\documentclass[pdflatex,sn-mathphys-num]{sn-jnl}

\usepackage{graphicx}%
\usepackage{multirow}%
\usepackage{amsmath,amssymb,amsfonts}%
\usepackage{amsthm}%
\usepackage{titlesec}
\usepackage{mathrsfs}%
\usepackage[title]{appendix}%
\usepackage{xcolor}%
\usepackage{textcomp}%
\usepackage{manyfoot}%
\usepackage{booktabs}%
\usepackage{algorithm}%
\usepackage{algorithmicx}%
\usepackage{algpseudocode}%
\usepackage{algpseudocodex}
\usepackage{listings}
\usepackage{subcaption}
\usepackage{float}
\usepackage{tikz}
\usetikzlibrary{positioning, fit, calc}
\usepackage{hyperref}
\usepackage[abs]{overpic}
\usepackage{float}
\usepackage{pifont} 
\usepackage{xcolor}
\usepackage{enumitem}
\usepackage{rotating}
\usepackage{cuted}
\usepackage{caption}
\usepackage{afterpage}
\usepackage{stfloats}
\usepackage{siunitx}
\usepackage{longtable}
\usepackage{tabularx}

\theoremstyle{thmstyleone}%
\theoremstyle{thmstyletwo}%
\definecolor{goldenyellow}{HTML}{C0C000}
\theoremstyle{thmstylethree}%

\begin{document}

\title[Article Title]{Identifying sensitivity-dominant parameters via active subspaces in reduced-order modeling of fluid dynamics}


\author[1]{\fnm{Dewu} \sur{Yang}}

\author[1]{\fnm{Rui} \sur{Wang}}
\author[1]{\fnm{Pengyu} \sur{Lai}}
\author[1]{\fnm{Junjie} \sur{Wang}}
\author[1]{\fnm{Feng} \sur{Wang}}

\author*[1]{\fnm{Hui} \sur{Xu}}\email{dr.hxu@sjtu.edu.cn}

\affil*[1]{\orgdiv{School of Aeronautics and Astronautics}, \orgname{Shanghai Jiao Tong University}, \orgaddress{\city{Shanghai}, \postcode{200240}, \country{China}}}


\abstract{Reduced-order models (ROMs) are widely employed to describe complex system dynamics when simulations with full-order models (FOMs) are computationally prohibitive. This study presents {POD--AS--PRS}, a novel model-reduction framework based on the active subspaces (AS) technique, which performs dimensionality reduction in both the state and parameter spaces, enabling efficient and high-fidelity approximations of quantities of interest (QoI). The approach employs proper orthogonal decomposition (POD) to extract low-dimensional coefficients from computational fluid dynamics (CFD) snapshots, which are inputs to a residual neural network (ResNet) with linear layers to learn their nonlinear mapping to QoI. Reverse-mode automatic differentiation (AD) is utilized to compute gradients with respect to the coefficients, enabling AS analysis to identify influential modes by shifting the analysis to the POD coefficient space, thereby achieving a dual-stage dimensionality reduction driven by QoI sensitivity rather than modal energy. A surrogate model is subsequently constructed using a polynomial response surface (PRS) based on AS-derived active variables, retaining only the highly influential POD coefficients to ensure accurate and efficient QoI reconstruction. The framework is validated on periodic and chaotic bluff-body flows, demonstrating high accuracy with few influential parameters, while AD-based gradients achieve a two-order-of-magnitude speed-up over finite-difference approximations. Sensitivity analysis further reveals that the influential coefficients are not necessarily proportional to modal energy, highlighting the critical flow structures. Consequently, POD--AS--PRS identifies a low-dimensional manifold of sensitivity-dominant parameters that govern the QoI, elucidating the essential flow structures and their coupling with control parameters, thereby enabling efficient and accurate QoI reconstruction.}




\maketitle

\section{Introduction}\label{sec1}

High-fidelity computational fluid dynamics (CFD) plays an indispensable role in engineering applications for the analysis and design of complex aerodynamic and turbulent flow systems~\cite{zhang2018direct,gimenez2025multiscale,li2022assessment}. However, the \emph{curse of dimensionality} in parametric studies imposes a prohibitive computational burden~\cite{bellman1957dynamic,bellman1959mathematical}, rendering direct high-fidelity simulations impractical for real-time analysis, design optimization, and uncertainty quantification (UQ)~\cite{duan2024non,amsallem2015design,xu2025reduced}. To reduce these costs, reduced-order models (ROMs) have emerged as a prevalent strategy. In this context, there is a growing preference for \emph{non-intrusive} approaches which rely solely on simulation data ~\cite{ghattas2021learning, DAVE2025118045} and avoid the detailed knowledge of governing equations required by traditional \emph{intrusive} methods~\cite{bai2002krylov,haasdonk2017reduced,chaturantabut2010nonlinear}. Among these non-intrusive techniques, proper orthogonal decomposition (POD) is a powerful tool for state-space reduction~\cite{loeve1977elementary}, capable of compressing complex flow fields into an optimal set of orthogonal modes based on energy content~\cite{berkooz1993proper,chatterjee2000introduction}. Nevertheless, a critical limitation remains: POD is kinematic in nature and does not inherently reveal the underlying dynamics~\cite{prakhar2025modal}. Consequently, interpreting individual modes and their influence on specific quantities of interest (QoI) can be challenging~\cite{mendez2019multi}. In particular, the contribution of each mode to integral quantities such as aerodynamic forces is not directly evident~\cite{prakhar2025modal}. To overcome these limitations, it is advantageous to exploit low-dimensional structures in the parameter space—where the QoI varies predominantly along a few dominant directions.

To identify and exploit these crucial low-dimensional structures in the input space, we turn to the \emph{active subspace} (AS) method~\cite{russi2010uncertainty,constantine2014active,constantine2015active}. AS provides a rigorous, gradient-based framework for global sensitivity analysis by determining the directions of maximal output variation, thus linking input changes directly to the most influential changes in the QoI. Specifically, AS analysis computes the second-moment matrix of the model's gradient with respect to input parameters, and its eigen-decomposition identifies the dominant eigenvectors, or active directions, along which the output exhibits the largest variance~\cite{constantine2014active,tezzele2018combined,activesubspaceshape2014}. Projecting the input space onto these low-dimensional active directions captures the principal sensitivity structure of the system, which has been successfully leveraged in engineering analysis~\cite{constantine2017time,yu2025novel,jefferson2017exploring} and Bayesian inverse problems~\cite{constantine2016accelerating}. However, in complex fluid dynamics systems, evaluating high-dimensional gradients remains computationally expensive, which limits the direct applicability of traditional AS analysis for large-scale or highly nonlinear models.

Once the active subspace is identified, the input space can be projected onto a reduced set of \emph{active variables}, each representing a linear combination of input parameters. This projection mitigates the curse of dimensionality in surrogate modeling. Response surfaces (RS) are then constructed to approximate the QoI as a function of these active variables, using limited evaluations of the underlying model. Such surrogate models have proven successful in forward~\cite{giunta2006promise,li2010evaluation} and inverse UQ problems~\cite{marzouk2007stochastic,bliznyuk2008bayesian,constantine2011surrogate}, as well as in optimization~\cite{jones2001taxonomy,wild2008orbit}. However, most RS techniques still suffer from exponential scaling with the number of inputs. A common remedy is to perform global sensitivity analysis~\cite{constantine2014active} to identify and retain only the most influential parameters. Local sensitivity methods are inexpensive but unreliable across wide parameter domains, whereas global approaches, such as AS, provide robust measures of variability across the entire input space, capturing nonlinear interactions~\cite{constantine2017global}.

Despite its potential, there is a notable paucity of research on integrating POD with AS for reduced-order modeling. Furthermore, the few existing methodologies~\cite{demo2019non,tezzele2018combined} are subject to intrinsic limitations. Typically, such approaches employ AS strictly as a preprocessing tool to reduce the dimensionality of the physical or geometric parameter space. In this conventional context, AS serves as an ancillary tool for input-parameter reduction to support subsequent field reconstruction (e.g., via POD interpolation (PODI) or POD-Galerkin projection), without directly informing the selection of the reduced-order basis. Departing from this convention, the present work establishes a novel POD–AS integration paradigm. The core innovation lies in shifting the application of AS from the physical input space to the POD coefficient space, thereby enabling QoI-driven sensitivity analysis for modal screening. The objective is to identify the specific linear combinations of POD coefficients that dominate the variations of a targeted QoI. Empowered by efficient gradient computation via neural-network surrogates and reverse-mode AD, this approach allows for a QoI-oriented truncation strategy: POD modes are retained based on their sensitivity impact on the QoI, rather than their energy content. Consequently, the proposed {POD--AS--PRS} framework offers a gradient-informed, dual-stage dimensionality reduction methodology, uniquely integrating state-space compression with AS-driven analysis in the latent coefficient space.

It should be noted that a direct performance comparison with conventional ROM frameworks is not appropriate in the present context, since the objectives and evaluation criteria are fundamentally different. Traditional ROM methods aim at minimizing full-field reconstruction or surrogate prediction errors in the state space, whereas the proposed framework is designed for identifying and interpreting the most influential modal directions governing QoI variations and enabling accurate QoI reconstruction from limited training samples, rather than for improving flow-field state reconstruction performance. Therefore, standard ROM benchmarking metrics based on field-level reconstruction errors do not provide a meaningful basis for comparison. Instead, the effectiveness of the present methodology is demonstrated through sensitivity consistency, stability of AS eigenspectra, and surrogate reconstruction accuracy for the targeted QoI, which constitute the appropriate validation measures for the objectives of this study.

By performing simultaneous reductions in both the state and parameter spaces, the framework enables the identification of sensitivity-dominant POD coefficients that drive the variation of the QoI. This facilitates the construction of a scalable and interpretable surrogate model based on the identified active variables. The integrated POD–AS paradigm provides not only enhanced computational efficiency but also improved physical insight into the flow structures that dominate QoI sensitivity, and is suitable for parametric studies, design optimization, and UQ in complex fluid systems.

The remainder of this paper is organized as follows. Section~\ref{sec:mb} presents the methodology, including POD decomposition, ResNet training, AS analysis, and response surface construction. Section~\ref{sec:HyROM} details the algorithmic implementation of the POD--AS--PRS framework. Section~\ref{sec:results} demonstrates its application to incompressible Navier--Stokes flows, including periodic cylinder and chaotic airfoil configurations. Finally, Section~\ref{sec:conclusion} summarizes the findings and outlines future research directions.

\section{Mathematical preliminaries}\label{sec:mb}

This section provides the foundational mathematical tools underpinning the proposed framework. We begin this section by introducing the POD method that extracts low-dimensional representations of high-dimensional flow fields and provides the basis for state-space reduction, followed by a description of the ResNet architecture for nonlinear mapping from POD coefficients while enabling efficient gradient computation via backpropagation. The AS method for gradient-based parameter-space dimensionality reduction is formulated in Section~\ref{subsec:AS}, before presenting the construction of response surfaces for efficient surrogate modeling in Section~\ref{subsec:RS}.

\subsection{Proper orthogonal decomposition}
\label{POD}

The purpose of the POD method is to extract an energy-optimal orthogonal basis from high-dimensional snapshot data in the least-squares sense. 
Consider a set of solution snapshots representing data at different times $t_k$ and spatial points $\boldsymbol{x_i}$, sampled uniformly with temporal discretization $\{t_k = (k-1)\Delta t\}_{k=1}^{n_t}$, where $\Delta t$ is the time step. Spatially, the data is defined on a regular Cartesian grid with $n_x \times n_y$. For matrix representation, these spatial points are mapped to a single linear index $\boldsymbol{i} \in [1, \dots, n_s]$, where $n_s = n_x \times n_y$ is the total number of spatial degrees of freedom.

These snapshots are organized into a matrix $\boldsymbol{S}\in\mathbb{R}^{n_t\times n_s}$. First, the snapshot matrix is mean-centered by subtracting its mean vector,
\begin{equation}
    \tilde{\boldsymbol{S}}=\boldsymbol{S}-\bar{\boldsymbol{S}},
\end{equation}
where $\bar{\boldsymbol{S}}\in\mathbb{R}^{1\times n_s}$ is the mean vector. Then, performing singular value decomposition (SVD)~\cite{kunisch1999control},
\begin{equation}
\label{eq:SVD}
\tilde{\boldsymbol{S}}=\boldsymbol{\Psi}\boldsymbol{\Sigma}\boldsymbol{\Phi}^T,
\end{equation}
where $\boldsymbol{\Psi}=[\boldsymbol{\psi}_1,\dots,\boldsymbol{\psi}_{r(\tilde{\boldsymbol{S}})}] \in \mathbb{C}^{n_t \times r(\tilde{\boldsymbol{S}})}$ and $\boldsymbol{\Phi}=[\boldsymbol{\phi}_1,\dots,\boldsymbol{\phi}_{r(\tilde{\boldsymbol{S}})}] \in \mathbb{C}^{n_s \times r(\tilde{\boldsymbol{S}})}$ collect the orthonormal temporal and spatial modes, respectively, and $\boldsymbol{\Sigma}=\operatorname{diag}(\sigma_1,\dots,\sigma_{r(\tilde{\boldsymbol{S}})})$ contains the corresponding singular values, with $r(\tilde{\boldsymbol{S}}) = \min(n_t, n_s)$ being the rank of $\tilde{\boldsymbol{S}}$.

The temporal basis matrix $\boldsymbol{\Psi}$ and the corresponding matrix of POD coefficients (modal amplitudes) $\boldsymbol{A}$ are obtained through the relationship:
\begin{equation}
    \boldsymbol{\Psi} \boldsymbol{\Sigma} = \tilde{\boldsymbol{S}} (\boldsymbol{\Phi}^T)^{-1} = \boldsymbol{A},\quad \text{which implies} \quad \boldsymbol{\Psi} = \boldsymbol{A} \boldsymbol{\Sigma}^{-1},\label{eq:pod_coeff}
\end{equation}
where $\sigma_r = \|\boldsymbol{A}_r\|_2$ and $\boldsymbol{A}_r$ denotes $r$-th column of the matrix $\boldsymbol{A}$.

The spatial modes $\boldsymbol{\Phi}$ constitute the POD basis, providing the best rank-$\hat{r}$ approximation of $\tilde{\boldsymbol{S}}$ in the least-squares sense. As established by~\cite{kunisch2006}, for a low-rank approximation, the corresponding approximation error, quantified by the Frobenius norm $|\cdot|_F$, is directly related to the singular values $\sigma_i$:

\begin{equation}
	\|\tilde{\boldsymbol{S}} - \hat{\boldsymbol{\Psi}} \hat{\boldsymbol{\Psi}}^T \tilde{\boldsymbol{S}}\|_F^2 = \sum_{i=\hat{r}+1}^{r(\tilde{\boldsymbol{S}})} \sigma_i^2,
\end{equation}
where $\hat{\boldsymbol{\Psi}}\in \mathbb{C}^{n_t \times \hat{r}}$ contains of the first $\hat{r}$ columns of $\boldsymbol{\Psi}$. This relationship explicitly demonstrates that the singular values $\sigma_i$ quantify both mode energy and truncation error. 

The error decays to zero as $\hat{r} \rightarrow r(\tilde{\boldsymbol{S}})$, ensuring the strongest possible convergence. If the snapshot matrix $\tilde{\boldsymbol{S}}$ captures the dynamics of the parametrized problem adequately, the decay of singular values serves as a surrogate for the Kolmogorov $m$-width. Many problems exhibit exponential singular value decay, enabling accurate approximations in a low-dimensional linear reduced space.

\subsection{Residual neural network with linear layers}
\label{ResNet}

In order to address the challenge of rapidly and accurately approximating the complex, nonlinear mapping between the input POD coefficients and the QoI in fluid dynamics, a deep ResNet architecture is employed. This framework is specifically chosen for its ability to mitigate the vanishing gradient problem inherent in deep neural networks. Furthermore, the inherent differentiability of the network, facilitated by skip connections that ensure highly stable backpropagation, allows for the precise and computationally efficient calculation of the QoI gradient with respect to the input POD coefficients via AD. By utilizing the ResNet as a differentiable bridge, we eliminate the need for an explicit analytical formulation of the flow physics, thereby enabling the AD engine to rapidly provide the gradient information required for the subsequent active subspace identification. This architectural advantage leads to faster convergence, enhancing the model's capacity to learn and approximate intricate functional relationships with high fidelity, ensuring a robust and efficient approximation of the QoI. It is important to emphasize that in this work, the deep ResNet architecture is employed not as a standalone predictive model, but strictly as a differentiable surrogate to facilitate the efficient computation of gradients via AD within the training distribution.

\subsubsection{Network architecture}

The canonical ResNet architecture, originally designed for image processing tasks, employs residual blocks constructed around convolutional operations~\cite{he2016deep}. In its standard configuration, each residual block contains two sequential $3\times3$ convolutional layers, each followed by batch normalization (BN) and a Rectified Linear Unit (ReLU) activation function. A fundamental characteristic of this architecture is the identity skip connection, which performs an element-wise addition of the block's input to the output of the second convolutional layer, after which a final ReLU activation is applied. This structural formulation transforms the learning objective from approximating a complete transformation $\mathcal{H}(\boldsymbol{a})$ to learning a residual function $\mathcal{F}(\mathbf{\boldsymbol{a}}) = \mathcal{H}(\mathbf{\boldsymbol{a}}) - \mathbf{\boldsymbol{a}}$, thereby simplifying the optimization landscape and promoting more effective training of deep networks.

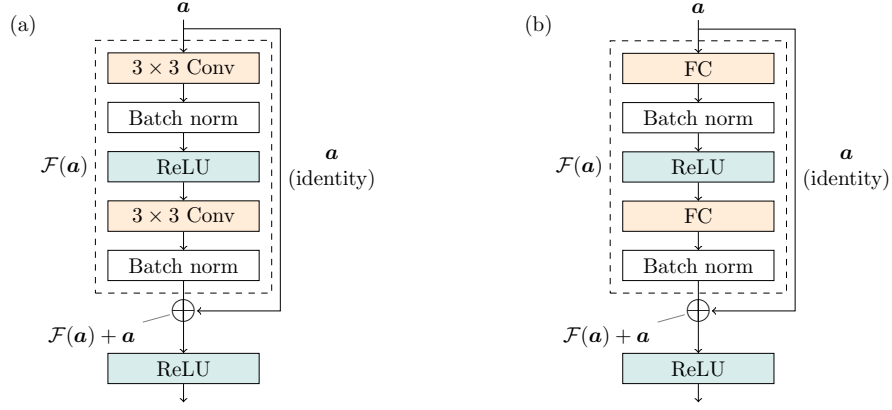
\begin{figure}[t]
\centering
\scalebox{0.8}{
\begin{subfigure}{0.48\textwidth}
\centering
\begin{tikzpicture}
    \tikzset{
        orange box/.style={rectangle, draw=black, fill=orange!15, minimum width=2.5cm, minimum height=0.5cm},
        white box/.style={rectangle, draw=black, minimum width=2.5cm, minimum height=0.5cm},
        teal box/.style={rectangle, draw=black, fill=teal!15, minimum width=2.5cm, minimum height=0.5cm},
        plus circle/.style={circle, draw=black, minimum size=0.5cm, inner sep=0pt},
    }
  \node[anchor=north west] at (-3, 1) {(a)};
  \node[orange box] (conv1) {$3\times3$ Conv};
  \node[white box, below=0.3cm of conv1] (bn1) {Batch norm};
  \node[teal box, below=0.3cm of bn1] (act1) {ReLU};
  \node[orange box, below=0.3cm of act1] (conv2) {$3\times3$ Conv};
  \node[white box, below=0.3cm of conv2] (bn2) {Batch norm};
  \node[below=0.3cm of bn2, font=\Large, inner sep=0, pin={190:$\mathcal F(\boldsymbol{a}) + \boldsymbol{a}$}] (add) {$\oplus$};
  \node[teal box, below=0.5cm of add] (act2) {ReLU};

  \draw[->] (conv1) -- (bn1);
  \draw[->] (bn1) -- (act1);
  \draw[->] (act1) -- (conv2);
  \draw[->] (conv2) -- (bn2);
  \draw[<-] (conv1) -- ++(0,0.8) node[above, pos=1] {$\boldsymbol{a}$};
  \draw[->] (bn2) -- (act2);
  \coordinate[below=0.3cm of act2] (output);
  \draw[->] (act2) -- (output);

  \node[fit=(conv1)(bn2), draw, dashed, inner sep=0.2cm] (fit_node) {};
  \node[xshift=-0.5cm] at (fit_node.west) {$\mathcal F(\boldsymbol{a})$};
  \draw[->] ($(conv1)-(0,-0.65)$) -- ++(1.6,0)-- node[right, align=center] {$\boldsymbol{a}$\\(identity)} node[right] {} ($(add)+(1.6,0)$) -- (add);

\end{tikzpicture}
\label{fig:convRes}
\end{subfigure}
\hspace{2cm}
\begin{subfigure}{0.48\textwidth}
\centering
\begin{tikzpicture}
    \tikzset{
        orange box/.style={rectangle, draw=black, fill=orange!15, minimum width=2.5cm, minimum height=0.5cm},
        white box/.style={rectangle, draw=black, minimum width=2.5cm, minimum height=0.5cm},
        teal box/.style={rectangle, draw=black, fill=teal!15, minimum width=2.5cm, minimum height=0.5cm},
        plus circle/.style={circle, draw=black, minimum size=0.5cm, inner sep=0pt},
    }
  \node[anchor=north west] at (-3, 1) {(b)};
  \node[orange box] (fc1) {FC};
  \node[white box, below=0.3cm of fc1] (bn1) {Batch norm};
  \node[teal box, below=0.3cm of bn1] (act1) {ReLU};
  \node[orange box, below=0.3cm of act1] (fc2) {FC};
  \node[white box, below=0.3cm of fc2] (bn2) {Batch norm};
  \node[below=0.3cm of bn2, font=\Large, inner sep=0, pin={190:$\mathcal F(\boldsymbol{a}) + \boldsymbol{a}$}] (add) {$\oplus$};
  \node[teal box, below=0.5cm of add] (act2) {ReLU};

  \draw[->] (fc1) -- (bn1);
  \draw[->] (bn1) -- (act1);
  \draw[->] (act1) -- (fc2);
  \draw[->] (fc2) -- (bn2);
  \draw[<-] (fc1) -- ++(0,0.8) node[above, pos=1] {$\boldsymbol{a}$};
  \draw[->] (bn2) -- (act2);
  \coordinate[below=0.3cm of act2] (output);
  \draw[->] (act2) -- (output);

  \node[fit=(fc1)(bn2), draw, dashed, inner sep=0.2cm] (fit_node) {};
  \node[xshift=-0.5cm] at (fit_node.west) {$\mathcal F(\boldsymbol{a})$};
  \draw[->] ($(fc1)-(0,-0.65)$) -- ++(1.6,0)-- node[right, align=center] {$\boldsymbol{a}$\\(identity)} node[right] {} ($(add)+(1.6,0)$) -- (add);

\end{tikzpicture}
\label{fig:fcRes}
\end{subfigure}
}
\caption{In a convolutional residual block (a), the portion within the dotted-line box needs to
learn the residual mapping, making the identity mapping easier to learn. In the fully connected residual block (b), which is a variant used in our framework, the convolutional layers are replaced with fully connected layers to adapt to the input.} 
\label{fig:ResNet_Block}
\end{figure}

In the present context, the input data consists of one-dimensional vectors of POD coefficients, rather than the multi-channel image data for which convolutional networks are naturally suited. Consequently, we adapt the architecture by replacing the convolutional layers with fully connected (FC) linear layers. This modification preserves the essential residual learning paradigm while accommodating the structural characteristics of the input data. As schematically illustrated in Fig.~\ref{fig:ResNet_Block}, the conventional convolutional residual block is reconfigured into its fully-connected counterpart. Within this formulation, $\mathcal{F}(\boldsymbol{a})$ represents the residual mapping learned by the two linear layers within the block, and $\boldsymbol{a}=[a_1,\dots,a_m]^T\in\mathbb{R}^m$ denotes the input vector of POD coefficients extracted from a vorticity field, where $m$ corresponds to the input dimension. The identity shortcut connection enables the direct summation of the input $\boldsymbol{a}$ with the output of the residual mapping, producing the block output $\mathcal{F}(\boldsymbol{a}) + \boldsymbol{a}$. This mechanism allows the network to focus on learning deviations from the identity mapping, which has been demonstrated to enhance training stability and performance in deep architectures.

The implemented ResNet variant is designed with a flexible architecture, comprising a configurable number of fully connected residual blocks whose depth and width are determined according to the complexity of the target task and the characteristics of the dataset. The number of hidden units per layer is a tunable hyperparameter; a default configuration of 128 hidden units is adopted in this study. The network accepts an input dimension equal to the number of retained POD modes and outputs a single scalar prediction corresponding to the QoI.

Training is performed by minimizing the mean squared error (MSE) loss using the \texttt{Adam} optimizer with an initial learning rate of $1\times10^{-3}$. To accelerate convergence and mitigate overfitting, a \texttt{ReduceLRonPlateau} learning rate scheduler is employed. This strategy halves the learning rate if the validation loss does not improve by at least $1\times10^{-4}$ over 10 consecutive epochs, and was found to converge faster while achieving comparable or superior validation-loss performance compared to \texttt{OneCycleLR}~\cite{al2022scheduling}. Additionally, an early stopping mechanism is applied to prevent unnecessary computation, halting training and restoring the model weights associated with the minimum validation loss if no significant improvement is observed within a predefined number of epochs.

By exploiting the residual learning paradigm, the network effectively addresses the vanishing-gradient problem that often impedes the training of deep architectures, thereby enabling robust and efficient learning of complex nonlinear mappings between the POD coefficient inputs and the target QoI without reliance on explicit theoretical formulations or analytical expressions. This capability makes the framework particularly well-suited for reduced-order modeling of fluid dynamics systems. By enabling the accurate prediction of key QoI, such as drag or lift, directly from the POD coefficients without reconstructing the full-order flow field, the proposed approach facilitates real-time inference and comprehensive exploration of high-dimensional parameter spaces while significantly reducing overall computational cost.

\subsubsection{Gradient approximation via reverse-mode automatic differentiation}

We propose a systematic framework for computing gradients of the QoI with respect to POD coefficients in fluid dynamics simulations, leveraging reverse-mode AD as implemented in \texttt{PyTorch}~\cite{paszke2019pytorch}. This framework addresses the limitations of conventional numerical differentiation techniques in high-dimensional settings.

We abstractly represent the map from inputs to the QoI by a function $f: \mathscr{X} \rightarrow \mathbb{R}$, where $\mathscr{X}\subseteq\mathbb{R}^m$ denotes the bounded input domain of interest and $f$ corresponds to the ResNet output. A probability density $\rho(\mathbf{x})$ that is strictly positive for $\mathbf{x}\in\mathscr{X}$ and zero otherwise. We further assume that  $\rho$ and $\mathscr{X}$ are defined such that the components of $\mathbf{x}$ are independent, have zero mean, and are similarly scaled (e.g., by range or variance). This normalization, achieved through rotation, shifting, or scaling, removes units and ensures each input contributes equally to parameter studies. For simplicity, we treat $f$ as incorporating any necessary transformations to map the normalized $\mathbf{x}$ to the simulation’s input range.

The primary objective is to compute the gradient $\nabla_{\mathbf{x}}f$, which characterizes the sensitivities of the QoI with respect to the POD coefficients. The function $f$ is formulated as a composition comprising an initial projection layer, $l$ residual blocks, followed by a final linear layer that produces the scalar output. Mathematically, this gradient corresponds to the Jacobian matrix that encodes sensitivities propagated through the network. Reverse-mode AD is employed for its efficiency and scalability. It enables the precise and efficient evaluation of this Jacobian for the scalar-valued QoI by performing only a single forward and backward pass through the network. In contrast, forward-mode AD requires $m$ forward passes for an $m$-dimensional input, which is computationally prohibitive for large $m$~\cite{griewank2008evaluating}. Consequently, reverse-mode AD exhibits scalability that is effectively independent of the input dimension, with a computational complexity at most four times that of evaluating $f$. This makes it particularly efficient for scalar-valued outputs from high-dimensional inputs. Subsequent numerical experiments confirm that gradients computed via reverse-mode AD agree with FD approximations to machine precision, while avoiding the latter’s inherent drawbacks, such as sensitivity to step-size selection and the linearly scaling computational cost with respect to $m$. Although computation time in practice increases with the number of input samples, batch size, or network depth due to memory usage and matrix operations, reverse-mode AD provides both high accuracy and computational efficiency, making it particularly suitable for high-dimensional POD inputs in gradient-based analyses. These results substantiate the suitability of reverse-mode AD for high-dimensional sensitivity analysis within the proposed ResNet-based reduced-order modeling framework. A detailed derivation, including the Jacobian formulation and adjoint propagation, is provided in Appendix~\ref{app:ad_derivation}.

The calculated gradient, $\nabla_{\mathbf{x}}f$, is the fundamental quantity for subsequent analysis of parameter sensitivities. For the subsequent rigorous mathematical treatment of this sensitivity analysis, it is necessary to explicitly define the properties of the function $f(\mathbf{x})$. Assuming the function $f(\mathbf{x})$ is differentiable and Lipschitz continuous with constant $L$, the gradient is oriented as a column $m$-vector, and Lipschitz continuity implies that the gradient's norm is bounded. Consequently, the gradient obtained through reverse-mode AD can be written as:
\begin{equation}
\begin{split}
\nabla_{\mathbf{x}} f(\mathbf{x}) &= \left(\frac{\partial f(\mathbf{x})}{\partial x_1}, \cdots, \frac{\partial f(\mathbf{x})}{\partial x_m}\right)^T, \quad \left\|\nabla_{\mathbf{x}} f(\mathbf{x})\right\| \leq L, \quad \forall\mathbf{x} \in \mathscr{X},
\end{split}
\label{eq:gradient}
\end{equation}
where $\|\cdot\|$ is the standard Euclidean norm. 

\subsection{Active subspace}
\label{subsec:AS}

To overcome the challenges of efficiently and effectively analyzing parameter sensitivities in the high-dimensional input space of POD coefficients, the methodology of AS is employed. AS constitutes an efficient and systematic approach for functional analysis and dimension reduction, designed to identify a low-dimensional subspace of the input parameter space in which the function of interest exhibits the most significant variation. This methodology provides a principled framework for reducing dimensionality while retaining the essential features of the underlying mapping. Unlike conventional approaches that select a subset of input variables based solely on their individual importance, AS identifies a set of important directions within the entire input space. Each direction is defined by a set of weights that form a linear combination of the inputs. These directions are obtained as the leading eigenvectors of the uncentered covariance matrix of the function gradients with respect to the inputs. If the function’s output remains invariant as the inputs vary along a specific direction, that direction can be neglected in the reduced analysis without significant loss of accuracy~\cite{constantine2015active}. By projecting the high-dimensional input space onto this active subspace, one obtains a reduced representation that preserves the dominant sensitivities of the function, thereby enabling efficient parameter studies and reduced-order modeling.

\subsubsection{Computing the active subspace}

Consider a matrix, $\boldsymbol{C}$, defined as the average of the outer product of the gradient with itself, and the matrix can be approximated using sample-based estimation:
\begin{equation}
\boldsymbol{C}=\mathbb{E}\left[\left(\nabla_{\mathbf{x}} f(\mathbf{x})\right)\left(\nabla_{\mathbf{x}} f(\mathbf{x})\right)^T\right]=\int\left(\nabla_{\mathbf{x}} f(\mathbf{x})\right)\left(\nabla_{\mathbf{x}} f(\mathbf{x})\right)^T \rho d \mathbf{x}, \label{eq:as}
\end{equation}
where $f(\mathbf{x})$ is  a sufficiently smooth function such that the matrix $\boldsymbol{C}$ exists (i.e., the gradient's partial derivatives are integrable), and $\mathbb{E}[\cdot]$ is the expected value. $\boldsymbol{C}$ can be interpreted as the uncentered covariance matrix of the gradient vector. The eigendecomposition of the resulting symmetric positive definite matrix is then performed 
\begin{equation}
\begin{split}
    \boldsymbol{C}=\boldsymbol{W}\boldsymbol{\Lambda}\boldsymbol{W}^T, \quad \boldsymbol{\Lambda}=\text{diag}(\lambda_1,\cdots,\lambda_m),\quad \lambda_1\geq\cdots\geq\lambda_m\geq0,\label{eq:C}    
\end{split}
\end{equation}
where $\boldsymbol{W}$ includes all orthogonal eigenvectors and $\boldsymbol{\Lambda}$ are the eigenvalues. All the eigenvalues are nonnegative because $\boldsymbol{C}$ is positive semidefinite. The relationship between the gradient of $ f(\mathbf{x}) $ and the eigendecomposition of $ \boldsymbol{C} $ is rigorously quantified as follows: the mean-squared directional derivative of $f(\mathbf{x})$ with respect to the eigenvector $\boldsymbol{w}_i$, where $\boldsymbol{w}_i$ represents an individual orthogonal eigenvector from the set $\boldsymbol{W}$, equals the corresponding eigenvalue~\cite{constantine2015active} 
\begin{equation}
\begin{split}
    \mathbb{E}[((\nabla_{\mathbf{x}} f(\mathbf{x}))^T \boldsymbol{w}_i)^2] =\int\left(\left(\nabla_{\mathbf{x}} f(\mathbf{x})\right)^T \boldsymbol{w}_i\right)^2 \rho d \mathbf{x} = \lambda_i, 
    \quad i=1,\cdots,m.\label{eq:as_lambda}
\end{split}
    \end{equation}
as proven by the definition of $\boldsymbol{C}$. If the smallest eigenvalue $\lambda_m$ is exactly zero, then the mean-squared change in $f(\mathbf{x})$ along the eigenvectors $\boldsymbol{w}_m$ is zero. Since $f(\mathbf{x})$ is continuous, the directional derivative $(\nabla_{\mathbf{x}}f(\mathbf{x}))^T\boldsymbol{w}_m$ is zero everywhere in the domain $\mathscr{X}$. In other words, $f(\mathbf{x})$ is always constant along the direction defined by $\boldsymbol{w}_m$. Given a point $\mathbf{x}_1\in\mathscr{X}$, choose a scalar $z$ such that $\mathbf{x}_1+z\boldsymbol{w}_m=\mathbf{x}_2\in\mathscr{X}$. Then $f(\mathbf{x}_1)=f(\mathbf{x}_2)$, which implies that the value of $f(\mathbf{x})$ is invariant with respect to $z$. This is information that we can exploit for dimension reduction.

One can split the matrix $\boldsymbol{\Lambda}$ and $\boldsymbol{W}$ into two parts,
\begin{equation}
\mathbf{\Lambda} =   \begin{bmatrix} \mathbf{\Lambda}_1 & \\
                                     &
                                     \mathbf{\Lambda}_2\end{bmatrix},\quad
\boldsymbol{W} = \left [ \boldsymbol{W}_1 \quad \boldsymbol{W}_2 \right ],
\end{equation}
where $\boldsymbol{\Lambda}_1=\text{diag}(\lambda_1,\cdots,\lambda_n)$ with $n<m$, and $\boldsymbol{W}_1\in \mathbb{R}^{m\times n}$ contains the first $n$ eigenvectors.The rotated coordinates $\mathbf{y}$ and $\mathbf{z}$ are defined by
\begin{equation}
\mathbf{y}=\boldsymbol{W}_1^T \mathbf{x} \in \mathbb{R}^n, \quad \mathbf{z}=\boldsymbol{W}_2^T \mathbf{x} \in \mathbb{R}^{m-n} ,\label{eq:active_variables}
\end{equation}
which are the \textit{active variables} and \textit{inactive variables}, respectively.

Any $\mathbf{x}\in\mathbb{R}^m$ can be expressed in terms of $\mathbf{y}$ and $\mathbf{z}$,
\begin{equation}
\mathbf{x}=\underbrace{\boldsymbol{W} \boldsymbol{W}^T}_{\mathbb{I}} \mathbf{x} =\boldsymbol{W}_1 \boldsymbol{W}_1^T \mathbf{x}+\boldsymbol{W}_2 \boldsymbol{W}_2^T \mathbf{x}=\boldsymbol{W}_1 \mathbf{y}+\boldsymbol{W}_2 \mathbf{z} .\label{eq:inputdecomp}
\end{equation}

Then we can compute the gradient of $f(\mathbf{x})$ with respect to the decomposed variables $\mathbf{y}$ and $\mathbf{z}$, where $f(\mathbf{x}) = f(\boldsymbol{W}_1 \mathbf{y} + \boldsymbol{W}_2 \mathbf{z})$. Applying the chain rule, we obtain:
\begin{equation}
\begin{aligned}
\nabla_\mathbf{y} f(\mathbf{x})&=\nabla_\mathbf{y} f\left(\boldsymbol{W}_1 \mathbf{y}+\boldsymbol{W}_2 \mathbf{z}\right) =\boldsymbol{W}_1^T \nabla_{\mathbf{x}} f\left(\boldsymbol{W}_1 \mathbf{y}+\boldsymbol{W}_2 \mathbf{z}\right)=\boldsymbol{W}_1^T \nabla_{\mathbf{x}} f(\mathbf{x}), \\
\nabla_\mathbf{z} f(\mathbf{x})&=\nabla_\mathbf{z} f\left(\boldsymbol{W}_1 \mathbf{y}+\boldsymbol{W}_2 \mathbf{z}\right) =\boldsymbol{W}_2^T \nabla_{\mathbf{x}} f\left(\boldsymbol{W}_1 \mathbf{y}+\boldsymbol{W}_2 \mathbf{z}\right)=\boldsymbol{W}_2^T \nabla_{\mathbf{x}} f(\mathbf{x}).
\end{aligned}
\end{equation}

This decomposition reveals that $f(\mathbf{x})$ varies more on average along the directions defined by the columns of $\boldsymbol{W}_1$ than directions defined by the columns of $\boldsymbol{W}_2$. Following~\cite{constantine2015active}, this variation is quantified as:

	\begin{equation}
		\begin{aligned}
			\mathbb{E}\left[\left(\nabla_{\mathbf{y}} f(\mathbf{x})\right)^T\left(\nabla_{\mathbf{y}} f(\mathbf{x})\right)\right] =\int\left(\nabla_\mathbf{y} f(\mathbf{x})\right)^T\left(\nabla_\mathbf{y} f(\mathbf{x})\right) \rho \mathrm{d} \mathbf{x} =\lambda_1+\cdots+\lambda_n, \\
			\mathbb{E}\left[\left(\nabla_{\mathbf{z}} f(\mathbf{x})\right)^T\left(\nabla_{\mathbf{z}} f(\mathbf{x})\right)\right] =\int\left(\nabla_\mathbf{z} f(\mathbf{x})\right)^T\left(\nabla_\mathbf{z} f(\mathbf{x})\right) \rho \mathrm{d} \mathbf{x} =\lambda_{n+1}+\cdots+\lambda_m,
		\end{aligned}\label{eq:variation}
	\end{equation}
where the average inner product of the gradient with itself to the eigenvalues of $\boldsymbol{C}$.

The \textit{active subspace} is defined as the subspace spanned by the eigenvectors in $\boldsymbol{W}_1$, while the \textit{inactive subspace} is spanned by the remaining eigenvectors in $\boldsymbol{W}_2$. Eq.~\eqref{eq:variation} indicates that minuscule perturbations in the active variables $\mathbf{y}$ produce larger changes in $f$ than equivalent perturbations in the inactive variables $\mathbf{z}$. In other words, when $\lambda_{n+1}=\cdots=\lambda_m=0$, $\nabla_{\mathbf{z}}f(\mathbf{x})$ is zero everywhere, i.e., $f(\mathbf{x})$ is $\mathbf{z}$-invariant. The $\mathbf{z}$-invariant functions have both linear contours and linear isoclines~\cite{constantine2014active}. The gradient $\nabla_{\mathbf{z}}f(\mathbf{x})$ being zero everywhere in $\mathscr{X}$ implies that $f(\mathbf{x}_1)=f(\mathbf{x}_2)$. To show that the gradients are equal, assume that $\mathbf{x}_1$ and $\mathbf{x}_2$ are in the interior of $\mathscr{X}$. Then for arbitrary $\mathbf{c}\in\mathbb{R}^m$, define
\begin{equation}
\mathbf{x}_1^{\prime}=\mathbf{x}_1+\varepsilon \mathbf{c}, \quad \mathbf{x}_2^{\prime}=\mathbf{x}_2+\varepsilon \mathbf{c},
\end{equation}
where $\varepsilon>0$ is chosen so that $\mathbf{x}_1^{\prime}$ and $\mathbf{x}_2^{\prime}$ are in $\mathscr{X}$. Note that $\boldsymbol{W}_1^T\mathbf{x}_1^{\prime}=\boldsymbol{W}_2^T\mathbf{x}_2^{\prime}$ so $f(\mathbf{x}_1^{\prime})=f(\mathbf{x}_2^{\prime})$. Then
\begin{equation}
\begin{split}
     \mathbf{c}^T\left(\nabla_{\mathbf{x}} f\left(\mathbf{x}_1\right)-\nabla_{\mathbf{x}} f\left(\mathbf{x}_2\right)\right) =\lim _{\varepsilon \rightarrow 0} \frac{1}{\varepsilon}\left[\left(f\left(\mathbf{x}_1^{\prime}\right)-f\left(\mathbf{x}_1\right)\right)-\left(f\left(\mathbf{x}_2^{\prime}\right)-f\left(\mathbf{x}_2\right)\right)\right] =0.
\end{split}
\end{equation}
Simple limiting arguments can be used to extend these results to $\mathbf{x}_1$ or $\mathbf{x}_2$ on the boundary of $\mathscr{X}$.

\subsubsection{Determination of the active subspace dimension}
 The identification of an active subspace requires estimation of the eigenpairs $\boldsymbol{\Lambda}$ and $\boldsymbol{W}$, followed by an analysis of the corresponding eigenvalues to assess their significance. For problems with relatively low input dimensionality $m$, high-order numerical integration schemes, such as tensor-product Gauss–Legendre quadrature, can be employed to accurately approximate the covariance matrix $\boldsymbol{C}$. However, as $m$ increases, the computational cost of direct numerical integration grows prohibitively, rendering it impractical for high-dimensional settings. In such cases, the covariance matrix is approximated via a  sample average over $M$ gradient evaluations:
\begin{equation}
    \boldsymbol{C} \approx \frac{1}{M} \sum_{i=1}^{M} \left( \nabla_{\mathbf{x}} f(\mathbf{x}_i) \right) \left( \nabla_{\mathbf{x}} f(\mathbf{x}_i) \right)^T
    = \hat{\boldsymbol{W}} \hat{\boldsymbol{\Lambda}} \hat{\boldsymbol{W}}^T,
    \label{eq:approx_C}
\end{equation}
where $\{\mathbf{x}_i\}_{i=1}^M$ denotes the set of input samples. Here, $\hat{\boldsymbol{W}}$ and $\hat{\boldsymbol{\Lambda}}$ represent the estimated eigenvectors and eigenvalues of $\boldsymbol{C}$, respectively. The magnitude of the eigenvalues and the gaps between them in $\hat{\boldsymbol{\Lambda}}$ indicate the dimensionality of the active subspace, thereby guiding the subsequent reduction of the input parameter space.

The dimension of the active subspace can be determined \emph{a priori} for a given parameter study, such as an $n$-dimensional regression, or selected to satisfy a prescribed approximation accuracy defined by a tolerance $\epsilon$. This criterion can be formalized in terms of minimizing the residual energy associated with ridge approximation~\cite{constantine2015active}:
\begin{equation}
    \mathbb{E}_{\rho}\left[\lVert f(\mathbf{x}) - \mathbb{E}_{\rho}\left[f \mid \boldsymbol{W}_{1}^{T}\mathbf{x}\right] \rVert_{2}^{2}\right]
    \leq \sum_{i=n+1}^{m} \lambda_{i} \leq \epsilon^{2},
\end{equation}
where $\mathbb{E}_{\rho}[\cdot]$ denotes the expectation with respect to the input density $\rho$, and $\lambda_i$ are the eigenvalues of the uncentered covariance matrix $\boldsymbol{C}$. The first inequality is a direct consequence of the properties of conditional expectations, reflecting the error incurred when restricting the function to the identified subspace. The second inequality bounds the tail sum of the eigenvalues via Poincaré-type inequalities~\cite{Poincare1890} or alternative estimates, thereby providing a rigorous measure of the fidelity of the reduced representation.

To quantify the approximation error, consider the distance between the subspace defined by the range of $\boldsymbol{W}_1$ and the subspace defined by the range of $\hat{\boldsymbol{W}}_1$. Following~\cite{stewart1973error}, the distance between subspaces is defined as
\begin{equation}
\epsilon=\left\|\boldsymbol{W}_1 \boldsymbol{W}_1^T-\hat{\boldsymbol{W}}_1 \hat{\boldsymbol{W}}_1^T\right\|=\left\|\boldsymbol{W}_1^T \hat{\boldsymbol{W}}_2\right\| \leq\frac{4\lambda_{1}\omega}{\lambda_{n}-\lambda_{n+1}},\label{eq:distance}
\end{equation}
where $\omega \in (0, (\lambda_n - \lambda_{n+1}) / (5 \lambda_1))$ bounds the relative discretization error in approximating $\boldsymbol{C}$ via the sample average, assuming sufficient samples $M$. This bound highlights that a large eigenvalue gap $\lambda_n - \lambda_{n+1}$ ensures robust subspace estimation. 

The identification of eigenvalue gaps represents a distinct strategy compared to other dimension‐reduction techniques that rely on eigenvalue spectra to construct low‐dimensional representations. For example, POD in model reduction of dynamical systems and principal component analysis (PCA) for high‐dimensional data typically determine the reduced dimension by energy criteria, such as retaining a prescribed fraction of the total variance (i.e., the cumulative sum of eigenvalues)~\cite{loeve1977elementary, lumley1967structure, perry2006principal}. In contrast, the active subspace method emphasizes the accuracy of the subspace approximation itself, which is governed more directly by the presence and magnitude of eigenvalue gaps rather than by cumulative energy measures. This focus enables the identification of directions in parameter space that contribute most significantly to the variation of the QoI, thereby facilitating more efficient and targeted dimension reduction.

In practical applications, the eigenvalue spectrum provides a physical heuristic for identifying the sensitivity-dominant directions of the active subspace~\cite{constantine2015active}. While the largest spectral gap, defined as
\begin{equation}
N_{\text{gap}} = \arg \max_n \left\{ \lambda_n - \lambda_{n+1} \right\}, \label{eq:as_gap}
\end{equation}
offers a principled criterion for isolating the most significant directions, it may not always account for the numerical saturation required for high-fidelity surrogate modeling. Therefore, in this study, the spectral gap is used as a qualitative guide to confirm the low-dimensional nature of the flow physics, while the final operational dimension $N_{\text{AS}}$ is determined via a grid search to ensure the reconstruction fidelity meets the target threshold, as detailed in Sec.~\ref{subsec:rs}. This strategy ensures that the selected subspace is both physically representative and numerically sufficient, balancing computational efficiency with representational accuracy.

\subsection{Response surface}
\label{subsec:RS}
In this study, the high-dimensional input vector $\mathbf{x} \in \mathbb{R}^m$ can be projected onto a low-dimensional representation $\mathbf{y} = \hat{\boldsymbol{W}}_1^T \mathbf{x} \in \mathbb{R}^n$, thereby enabling the construction of a reduced response surface $g(\mathbf{y})$ to approximate the original mapping $f(\mathbf{x})$. Even in cases where $\hat{\lambda}_{n+1}$ is small but nonzero, such a reduced representation can still provide an accurate approximation of $f(\mathbf{x})$ with a rigorously bounded error.

For a fixed $\mathbf{y}$, the optimal reduced model is given by the conditional expectation of $f$ conditioned on $\mathbf{y}$~\cite{constantine2015active}, namely
\begin{equation}
\begin{split}
    g(\mathbf{y}) &= \mathbb{E}_{\mathbf{z}}\left[ f(\mathbf{x}) \mid \mathbf{y} \right]  = \int f\left( \hat{\boldsymbol{W}}_1 \mathbf{y} + \hat{\boldsymbol{W}}_2 \mathbf{z} \right) 
\rho(\mathbf{z} \mid \mathbf{y}) \, \mathrm{d} \mathbf{z}, \label{eq:rs}
\end{split}
\end{equation}
where $\rho(\mathbf{z} \mid \mathbf{y})$ denotes the conditional probability density function of the complementary variables $\mathbf{z}$ conditioned on $\mathbf{y}$. Accordingly, the original function $f(\mathbf{x})$ can be approximated as
\begin{equation}
f(\mathbf{x}) \approx g\left( \hat{\boldsymbol{W}}_1^T \mathbf{x} \right), \label{eq:rs_app}
\end{equation}
thereby reducing the effective dimensionality of the problem while preserving essential features of the response.

Based on the Poincar\'{e} inequality~\cite{Poincare1890} and following \cite[Theorem 4.3]{constantine2015active}, the approximation error incurred by this dimension reduction is bounded by the sum of the tail eigenvalues of the associated covariance matrix:
	\begin{equation}
		\int\left(f(\mathbf{x})-g(\boldsymbol{W}_{1}^T\mathbf{x})\right)^2\rho\mathrm{d}\mathbf{x}\leq \mathcal{O}\left(\lambda_{n+1}+\ldots+\lambda_m\right) .\label{eq:rsapp_error}
	\end{equation}
This bound provides a quantitative measure of the fidelity of the reduced model relative to the full-dimensional representation.

\section{The hybrid reduced order model (POD--AS--PRS)}
\label{sec:HyROM}

\begin{figure*}[h] 
  \centering
  \includegraphics[width=1\textwidth]{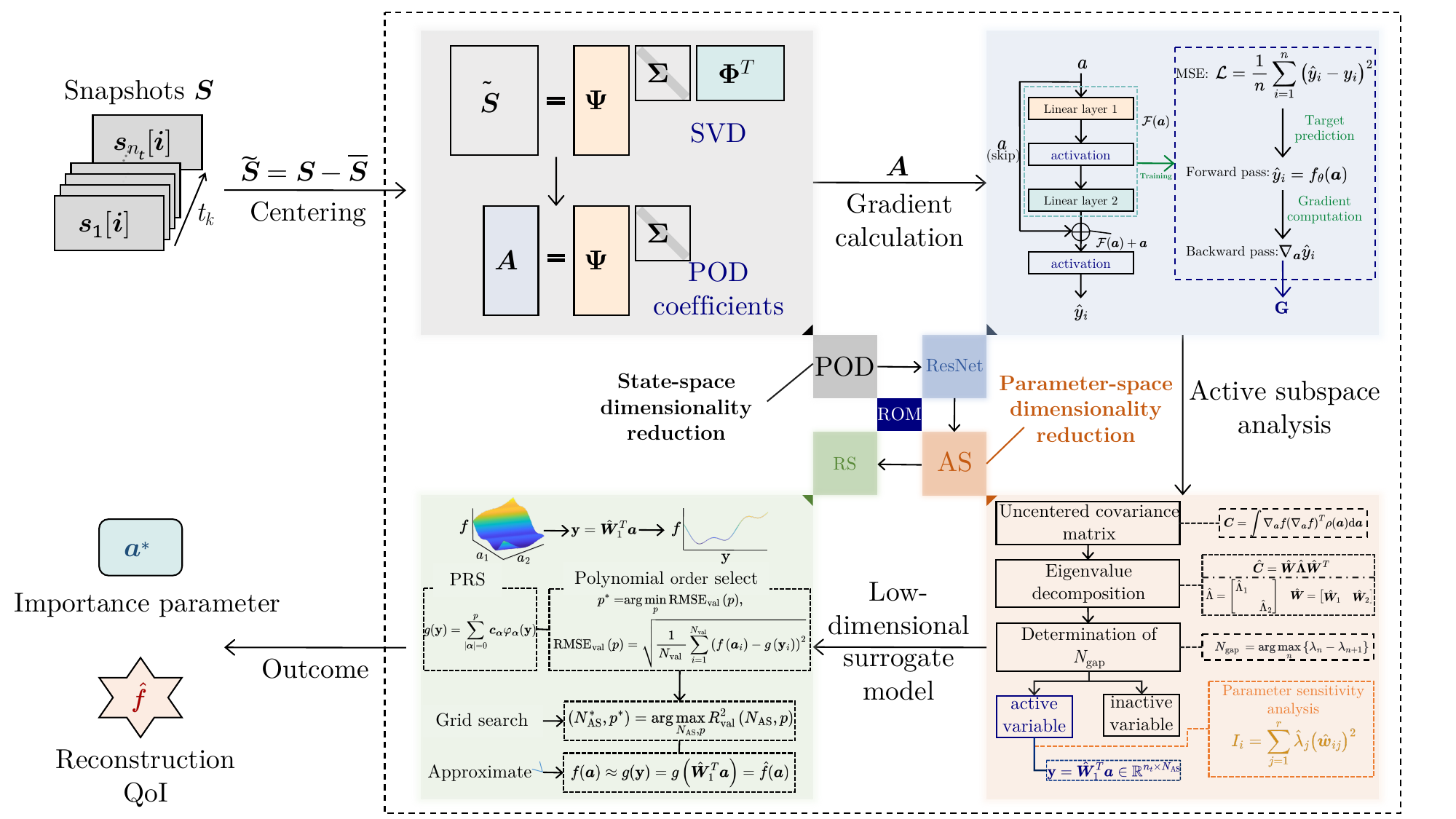} 
  \caption{POD--AS--PRS framework workflow.}
  \label{fig:pod_as_rom}
\end{figure*}

The proposed POD--AS--PRS framework constitutes a novel non-intrusive reduced-order modeling methodology based on POD, AS, and RS techniques for efficient and accurate QoI reconstruction in parametrized fluid dynamics simulations. Specifically, POD is employed for state-space dimensionality reduction by projecting the high-dimensional flow-field data onto a low-dimensional basis, thereby extracting a compact set of modal coefficients that encapsulate the dominant spatial and temporal structures of the system. This step isolates the essential dynamics while significantly reducing the computational complexity of subsequent analyses. The AS then performs parameter-space dimensionality reduction by identifying directions in the POD coefficient space that exert the greatest influence on the QoI. A key innovation of this work lies in the computation of the necessary gradients for AS via reverse-mode AD of a trained ResNet, ensuring high accuracy and computational efficiency even in high-dimensional settings. Finally, a response surface is constructed to map the AS-reduced coordinates directly to the QoI. This surrogate model enables rapid and reliable online evaluations without requiring the reconstruction of the full-order field, thereby dramatically accelerating reconstruction while maintaining fidelity. This integrated POD--AS--PRS strategy is particularly advantageous for systems characterized by low Kolmogorov width, where substantial reductions in computational cost can be achieved without compromising reconstruction accuracy. The overall structure and computational workflow of the proposed non-intrusive POD--AS--PRS framework are illustrated in Fig.~\ref{fig:pod_as_rom}, which serves as a schematic roadmap for the detailed exposition presented in the following sections.

\subsection{Data preparation and state-space dimensionality reduction}
\label{subsec:data_pod}

The POD--AS--PRS framework commences with the preparation of CFD data and the application of POD to reduce the state-space dimensionality of high-dimensional flow field snapshots. This subsection elucidates the computational pipeline for acquiring, preprocessing, decomposing, and processing CFD data to yield a low-dimensional representation in the form of POD coefficients, which serve as inputs to the subsequent ResNet surrogate model.

\subsubsection{Data acquisition and preprocessing}

The CFD snapshots of relevant flow quantities are obtained from time-resolved unsteady simulations conducted under fixed operating conditions, and are represented on a unified Cartesian grid, ensuring consistent spatial discretization and matrix dimensions for subsequent processing. The resulting multidimensional data are reshaped into a snapshot matrix $\boldsymbol{S} \in \mathbb{R}^{n_t \times n_s}$, where $n_t$ denotes the number of snapshots and $n_s$ the flattened spatial degrees of freedom. The mean snapshot $\bar{\boldsymbol{S}} \in \mathbb{R}^{n_s}$ is then computed and subtracted from $\boldsymbol{S}$ to obtain the mean-subtracted matrix $\tilde{\boldsymbol{S}} = \boldsymbol{S} - \bar{\boldsymbol{S}}$, thereby isolating the dynamic fluctuations about the mean flow field. Prior to decomposition, the matrix $\tilde{\boldsymbol{S}}$ is numerically conditioned by addressing undefined values and clipping small magnitude entries to ensure a well-posed input for subsequent reduced-order modeling procedures.

\subsubsection{POD reduction}

The POD reduction employs SVD to decompose the mean-subtracted snapshot matrix $\tilde{\boldsymbol{S}}$, as formulated in Sec.~\ref{POD}, producing a low-dimensional representation of the flow field. The energy contribution of the $k$-th mode is quantified as
\begin{equation}
    \frac{\sigma_k^2 }{\sum_{k=1}^{n_t} \sigma_k^2},\label{eq:pod_energy}
\end{equation}
where $\sigma_k$ is the $k$-th singular value. The number of retained modes $r$ is determined by a cumulative energy threshold, satisfying
\begin{equation}
    \frac{\sum_{j=1}^r \sigma_j^2 }{\sum_{k=1}^{n_t} \sigma_k^2} \geq \theta,\label{eq:pod_energy_norm}
\end{equation}
with $\theta$ chosen to be close to one, ensuring a balance between accuracy and dimensionality reduction. 

The POD coefficients are computed as in Eq.~\eqref{eq:pod_coeff}, and collecting all coefficient vectors as rows yields the POD coefficient matrix
\begin{equation}
\boldsymbol{A}=\left[\begin{array}{c}
\boldsymbol{a}_1 \\
\boldsymbol{a}_2 \\
\vdots \\
\boldsymbol{a}_{n_t}
\end{array}\right] \in \mathbb{R}^{n_t \times r}
\end{equation}
yielding $\boldsymbol{a}_i \in \mathbb{R}^r$ for each snapshot $i$. Hence, the $i$-th row of $\boldsymbol{A}$ corresponds to the POD coefficients of the $i$-th snapshot
\begin{equation}
\boldsymbol{a_i}=\left[a_i^{(j)}\right],\quad{{{i=1, \ldots, n_t},j=1, \ldots, r}}
\end{equation}
This enables the flow field to be approximated as 
\begin{equation}
    \boldsymbol{s}[k,\boldsymbol{i}] \approx \bar{\boldsymbol{s}}[\boldsymbol{i}]+\sum_{j=1}^{r}a_j(t_k)\phi_j(\boldsymbol{x_i}).
\end{equation}

The resulting POD coefficients serve as the low-dimensional input vector $\mathbf{x}$ for the subsequent ResNet-based surrogate model (Sec.~\ref{subsec:resnet}).

\subsection{ResNet model construction and training}
\label{subsec:resnet}

The POD coefficients $\boldsymbol{A}$ are mapped to the QoI using the ResNet described in Sec.~\ref{ResNet}, which additionally facilitates the computation of gradients required for subsequent AS analysis. This subsection delineates the computational workflow for constructing, training, and computing gradients from the ResNet model, ensuring both efficient QoI prediction and seamless integration with the subsequent AS analysis detailed in Sec.~\ref{subsec:as}.

\subsubsection{ResNet architecture implementation}

The ResNet architecture is initiated with an \emph{input projection layer} that maps the $r$-dimensional POD coefficient vector into a higher-dimensional hidden feature space. This projection is followed by a series of \emph{residual blocks} and optional dropout layers to enhance representational capacity while mitigating overfitting.  This design achieves a balance between representational capacity and computational efficiency. The network culminates in an \emph{output head} that reduces the high-dimensional hidden features to a scalar QoI through two FC layers, augmented with BN, ReLU activations, and dropout for additional regularization. This architectural design ensures both robust predictive performance and precise gradient computation required for AS analysis.

\subsubsection{Training}

Training is performed on a time-ordered dataset of paired POD coefficients and QoI, partitioned sequentially into training (80\%), validation (10\%), and test (10\%) sets to preserve temporal ordering and ensure realistic evaluation. The model is optimized by minimizing the MSE loss function~\cite{hastie2009elements}, defined as
\begin{equation}
    \mathcal{L} = \frac{1}{n} \sum_{i=1}^n (\hat{y}_i - y_i)^2, \label{eq:loss}
\end{equation}
where $\hat{y}_i$ denotes the predicted QoI and $y_i$ is the corresponding ground-truth value. Optimization employs the \texttt{Adam} optimizer~\cite{kingma2014adam} with an initial learning rate of $1\times10^{-3}$. To adapt the learning rate dynamically, a \texttt{ReduceLROnPlateau} scheduler is utilized, with a patience of 10 epochs and a reduction factor of 0.5, i.e., $P_{\text{LR}}=10$, $f_{\text{LR}}=0.5$, triggered when the validation loss plateaus. 

To mitigate overfitting and enhance generalization, early stopping is implemented~\cite{prechelt2002early}. This regularization technique monitors the validation loss and terminates training when no significant improvement is observed. Specifically, for each training epoch $N_{\text{ep}}$, the current validation loss $\mathcal{L}_{\text{val}}^{(N_{\text{ep}})}$ is compared to the historical minimum $\mathcal{L}_{\text{val}}^{\text{best}}$ using the criterion:
\begin{equation}
\begin{cases}\text { Reset counter, } & \text { if } \mathcal{L}_{\text {val }}^{(N_{\text{ep}})}<\mathcal{L}_{\text {val }}^{\text {best }}-\delta \\ \text { Counter }+=1, & \text { otherwise }\end{cases},\label{eq:early_stopping}
\end{equation}
where $\delta=1\times10^{-4}$ accounts for numerical fluctuations. Training halts if the counter reaches the patience value, and the model parameters corresponding to the lowest validation loss are restored. 

Mini-batch gradient descent is applied, with batch sizes ranging from 32 to 128 selected, facilitate efficient gradient updates. Hyperparameters, including the number of residual blocks, hidden layer dimensions, and dropout rates, are tuned systematically by minimizing the validation loss, optimizing performance for complex CFD applications.

To facilitate AS analysis (Sec.~\ref{subsec:as}), gradients of the QoI with respect to the POD coefficients, $\nabla_{\boldsymbol{a}} {f}$, are computed via reverse-mode AD. These gradients are processed in batches to accommodate large datasets and mitigate memory demands, subsequently aggregated into a matrix $\mathbf{G} \in \mathbb{R}^{n_t \times r}$ for covariance computation. This procedure yields derivatives accurate to machine precision while managing memory consumption effectively.

\subsection{Parameter-space dimensionality reduction through AS}
\label{subsec:as}

AS is performed in the POD coefficient space using gradients computed by the ResNet model. The pipeline ensures both scalability and numerical robustness in complex fluid dynamics applications.

The uncentered covariance matrix is estimated as:
\begin{equation}
\begin{split}
    \boldsymbol{C} = \int \nabla_{\boldsymbol{a}} f \, (\nabla_{\boldsymbol{a}} f)^T \, \rho(\boldsymbol{a}) \, \mathrm{d}\boldsymbol{a} \approx  \frac{1}{n_t} \sum_{i=1}^{n_t} \nabla_{\boldsymbol{a}} f_i \, (\nabla_{\boldsymbol{a}} f_i)^T = \hat{\boldsymbol{C}},\label{eq:active_subspace}    
\end{split}
\end{equation}
where $\rho(\boldsymbol{a})$ denotes the probability density of the POD coefficients. As a symmetric, positive semi-definite matrix, $\hat{\boldsymbol{C}}$ admits an eigenvalue decomposition $\hat{\boldsymbol{C}} = \hat{\boldsymbol{W}} \hat{\boldsymbol{\Lambda}} \hat{\boldsymbol{W}}^T$, with eigenvalues $\hat{\lambda}_i$ arranged in descending order and corresponding eigenvectors $\hat{\boldsymbol{W}}$ normalized to ensure consistent orientation.

The active subspace dimension $N_{\text{AS}}$ is determined via a grid search (Eq.~\eqref{eq:n_as_n_p}). The active subspace basis $\hat{\boldsymbol{W}}_1 \in \mathbb{R}^{r \times N_{\text{AS}}}$, comprising the first $N_{\text{AS}}$ eigenvectors, projects the normalized coefficients onto active variables $\mathbf{y} = \hat{\boldsymbol{W}}_1^T \boldsymbol{a} \in \mathbb{R}^{n_t \times N_{\text{AS}}}$, thereby reducing the dimensionality from $r$ to $N_{\text{AS}}$. Specifically, for unsteady flows, the time-dependent active variables $\mathbf{y}(t)$ are obtained by projecting 
the POD coefficients $\boldsymbol{a}(t)$—extracted from the high-fidelity FOM snapshots at each corresponding time instant—onto the active subspace. This projection is applied to both the training and test snapshots, such that the trained PRS can subsequently be evaluated on active variables derived from time instants not used during model construction. A nonparametric bootstrap analysis~\cite{tibshirani1993introduction}, involving 1000 iterations of gradient resampling, is performed to estimate confidence intervals for the eigenvalues and to assess subspace stability via the subspace distance metric.

The relative contribution of each POD mode to the active subspace is quantified using \emph{activity scores}~\cite{constantine2017global}:
\begin{equation}
   I_i = \sum_{j=1}^{r}\hat{\lambda}_j  (\hat{\boldsymbol{w}}_{ij})^2 , \quad i = 1,\cdots, n_t,\label{eq:as_score}
\end{equation}
where $\hat{\boldsymbol{w}}_{ij}$ denotes the $(i,j)$-th element of $\hat{\boldsymbol{W}}$, representing the projection of the $i$-th POD mode onto the $j$-th active direction, and $\hat{\lambda}_j$ weights the importance of that direction. Normalized activity scores are computed as
\begin{equation}
    \hat{I}_i = \frac{I_i}{\sum_{i=1}^{n_t} I_i},\label{eq:norm_as_score}
\end{equation}
providing a relative measure of each POD mode’s influence on the QoI’s variability within the active subspace. To identify which POD modes should be retained, a cumulative contribution measure is introduced to determine the subset of modes whose combined influence exceeds a prescribed threshold, and is defined as
\begin{equation}
    \frac{\sum_{\hat{i}=1}^{r_p}{\hat{I}_{\hat{i}}}}{\sum_{i=1}^{n_t}{\hat{I}}_i}=\sum_{\hat{i}=1}^{r_p}{\hat{I}_{\hat{i}}}\geq\epsilon,\label{eq:cumulative_as_score}
\end{equation}
where $r_p$ denotes the number of POD modes retained based on their cumulative contribution and used to construct the active variables, and  $\epsilon$ is a threshold close to unity that determines the retention criterion.

These scores quantify the relative influence of individual POD modes on the variability of the QoI. The structure of the AS provides a principled basis for interpreting these scores: the eigenvector $\hat{\boldsymbol{w}}_1$ identifies the dominant direction in the  POD coefficient space, along which perturbations of $\mathbf{a}$ induce the significant variations in $f$ (see Eq.~\eqref{eq:as_lambda}).  The magnitude and sign of the components of $\hat{\boldsymbol{w}}_1$ indicate the relative sensitivity of each POD coefficient along this most influential direction. The second most important direction is captured by $\hat{\boldsymbol{w}}_2$, whose significance relative to $\hat{\boldsymbol{w}}_1$ is measured by the spectral gap $\hat{\lambda}_1 - \hat{\lambda}_2$. Accordingly, scaling each eigenvector by its corresponding eigenvalue, as in Eq.~\eqref{eq:as_score}, yields a robust global sensitivity metric. While squaring the eigenvector components eliminates sign information, this procedure results in a practical and interpretable measure that can be directly compared to established sensitivity metrics~\cite{constantine2017global}. This formulation thus facilitates a quantitative ranking of the POD modes according to their influence on the QoI within the reduced active subspace.

\subsection{Construction of low-dimensional surrogate model through RS}
\label{subsec:rs}

We proceed to construct a reduced-order surrogate model based on a polynomial response surface (PRS) to approximate the mapping from the low-dimensional active variables $\mathbf{y}$ to the QoI, in accordance with the mathematical formulation presented in Sec.~\ref{subsec:RS}.

The PRS is expressed as a linear combination of monomial basis functions:
\begin{equation}
	g(\mathbf{y}) = \sum_{|\boldsymbol{\alpha}|=0}^{p} \boldsymbol{c_{\alpha}} \varphi_{\boldsymbol{\alpha}}(\mathbf{y}),\quad |\boldsymbol{\alpha}|=\alpha_1+\cdots+\alpha_n,
\end{equation}
where $\boldsymbol{c_{\alpha}}$ denote the polynomial coefficients obtained via least-squares regression, $p$ is the polynomial order, and $\varphi_{\boldsymbol{\alpha}}(\mathbf{y})=y_{1}^{\alpha_1} y_{2}^{\alpha_2}\cdots y_{n}^{\alpha_n}$ represent the monomial basis functions. Here, $n$ corresponds to the input dimension, i.e., the number of active variables $N_{\text{AS}}$. The training dataset consists of the active variables $\mathbf{y} = \hat{\boldsymbol{W}}_1^T \boldsymbol{a}$ and the corresponding QoI values $f(\boldsymbol{a})$.

For a training set containing $N_{\text{train}}$ samples, the polynomial approximation is cast in matrix form:
\begin{equation}
    \boldsymbol{f} \approx \boldsymbol{B}\boldsymbol{c},
\end{equation}
where $\boldsymbol{f} \in \mathbb{R}^{N_{\text{train}}}$ contains the QoI values, $\boldsymbol{B} \in \mathbb{R}^{N_{\text{train}} \times N_p}$ is the polynomial basis matrix with $N_p = \binom{n+p}{p}$ terms, and $\boldsymbol{c} \in \mathbb{R}^{N_p}$ is the vector of polynomial coefficients. Each row of $\boldsymbol{B}$ evaluates all monomial basis functions at a given sample of active variables:
\begin{equation}
\begin{split}
     \boldsymbol{B}_{i,j} = \varphi_{\boldsymbol{\alpha}_j}(\mathbf{y}_i)=\prod_{k=1}^{N_{\text{AS}}} y_{i, k}^{\alpha_{j, k}}, i = 1,\cdots,N_{\text{train}}, j = 1,\cdots,N_p,
\end{split}
\end{equation}
where $\mathbf{y}_i \in \mathbb{R}^{N_{\text{AS}}}$ is the $i$-th sample, and $\boldsymbol{\alpha}_j = (\alpha_{j,1}, \alpha_{j,2}, \dots, \alpha_{j,N_{\text{AS}}})$ denotes the multi-index of the $j$-th monomial basis. The least-squares regression problem is then formulated as:
\begin{equation}
   \boldsymbol{c} = \arg\min_{\boldsymbol{c}} \|\boldsymbol{B}\boldsymbol{c} - \boldsymbol{f}\|_2^2,
\end{equation}
Formally, the analytical solution is given by:
\begin{equation}
    \boldsymbol{c} = (\boldsymbol{B}^T\boldsymbol{B})^{-1}\boldsymbol{B}^T\boldsymbol{f},
\end{equation}
with SVD employed in practice to ensure numerical stability.

To identify the optimal combination of active subspace dimension $N_{\text{AS}}$ and polynomial order $p$, a grid search is conducted over candidate $N_{\text{AS}}$ values. For each $N_{\text{AS}}$, the corresponding optimal polynomial order $p^*$ is determined by cross-validation, minimizing the root mean square error (RMSE) over the validation set:

\begin{equation}
\begin{split}
    &p^*= \arg\min_{p} \text{RMSE}_{\text{val}}(p),\\
    &\text{RMSE}_{\text{val}}(p) = \sqrt{\frac{1}{N_{\text{val}}} \sum_{i=1}^{N_{\text{val}}} \left(f(\boldsymbol{a}_i) - g(\mathbf{y}_i)\right)^2},\label{eq:cross_val}
\end{split}
\end{equation}
where $N_{\text{val}}$ denotes the size of the validation set, $f(\boldsymbol{a}_i)$ is the true QoI, and $g(\mathbf{y}_i)$ the corresponding PRS reconstruction. After obtaining $p^*$ for each $N_{\text{AS}}$, the optimal configuration is selected by maximizing the validation set coefficient of determination:
\begin{equation}
    (N_{\text{AS}}^*, p^*) = \arg\max_{N_{\text{AS}},p} R^2_{\text{val}}(N_{\text{AS}},p), \label{eq:n_as_n_p}
\end{equation}
where $R^2_{\text{val}}(N_{\text{AS}},p)$ is defined as 
\begin{equation}
\begin{split}
    &R^2_{\text{val}} =1 - \frac{\sum_{i=1}^{N_{\text{val}}} (f(\boldsymbol{a}_i) - g(\mathbf{y}_i))^2}{\sum_{i=1}^{N_{\text{val}}} (f(\boldsymbol{a}_i) - \bar{f}_{\text{val}})^2},\\
    &\bar{f}_{\text{val}}=\frac{1}{N_{\text{val}}}\sum_{i=1}^{N_{\text{val}}}f(\boldsymbol{a}_i), \label{eq:n_as_n_p_r2}
\end{split}
\end{equation}

This procedure ensures that the final surrogate model balances model complexity and reconstruction accuracy, leveraging the most informative active subspace dimension while avoiding overfitting in the polynomial response surface.

With the optimal polynomial order $p^*$ determined, the final PRS is trained on the combined training and validation datasets to maximize data utilization. Model performance is subsequently evaluated on an independent test set to assess generalization capabilities. Beyond its reconstructive accuracy, the PRS offers a degree of physical interpretability that is intrinsic to its polynomial structure. Each active variable $y_k$ represents a physically meaningful linear combination of POD coefficients that most strongly governs QoI variability, as identified through the AS analysis. The polynomial coefficients $\boldsymbol{c}_{\boldsymbol{\alpha}}$ directly quantify how the QoI depends on these sensitivity-dominant directions and their interactions up to degree $p^*$, providing a transparent and analytically accessible mapping from the active variables to the QoI. This PRS construction finalizes the POD--AS--PRS framework, yielding an efficient and accurate surrogate model for QoI reconstruction, well-suited for real-time applications in fluid dynamics. It should be emphasized that, for unsteady flows, this process does not involve a self-contained time-marching procedure. Specifically, the trained PRS maps the active variables $\mathbf{y}(t)$ to the QoI at each time step. To assess the generalization capability of the surrogate, the dataset is partitioned chronologically: the first 80\% of snapshots are used for training, while the remaining 20\% constitute an independent test set. For the training interval, the PRS performs a temporal reconstruction of the QoI within the known data range. For the independent test interval, the PRS maps previously unseen active variables — obtained by projecting the POD coefficients of the test snapshots onto the identified active subspace — to the corresponding QoI values, thereby demonstrating the surrogate's ability to generalize to new inputs within the active subspace. It is acknowledged that this evaluation still relies on high-fidelity snapshots to provide the active variables $\mathbf{y}(t)$, and does not constitute autonomous temporal forecasting; the primary contribution lies in demonstrating that the identified low-dimensional active subspace faithfully captures the governing physics of the QoI across both the training and test intervals.

\subsection{The POD--AS--PRS algorithm}
\label{subsec:integrated}

This section synthesizes the computational pipeline of the POD--AS--PRS framework, elucidating the synergistic integration of the preceding methodologies. This integration successfully demonstrates the framework’s efficacy in identifying the low-dimensional, active flow structures that critically influence the QoI and establishing a non-intrusive, compact surrogate model for rapid and accurate reconstruction within parametrized fluid dynamics simulations. The complete procedure is summarized in Algorithm~\ref{alg:pod_as_rom}.

High-dimensional CFD snapshots are first projected onto the POD basis, yielding a set of modal coefficients $\boldsymbol{A}$ (Eq.~\eqref{eq:pod_coeff}). A ResNet is then trained on $\boldsymbol{A}$ to provide highly accurate gradient information via reverse-mode AD. These gradients are subsequently exploited within the AS framework to identify a low-dimensional set of active variables $\mathbf{y}$ that dominantly governs the variability of the QoI. Finally, a PRS surrogate is constructed in the reduced $\mathbf{y}$-space, enabling rapid, high-fidelity, and interpretable online evaluations.

Driven by this integrated methodology and its non-intrusive, modular design, the POD--AS--PRS framework is systematically validated using rigorous performance metrics, including test set accuracy, eigenvalue spectra, and bootstrap confidence intervals to assess subspace stability. This modular design facilitates scalable QoI reconstruction in large-scale, high-fidelity CFD problems, achieving substantial computational efficiency by exploiting the low Kolmogorov width of the system. As illustrated in Fig.~\ref{fig:pod_as_rom}, this workflow provides a robust and interpretable methodology for rapid and reliable QoI evaluation, positioning the framework as a powerful tool for advancing computational fluid dynamics.

\begin{algorithm}
\caption{The POD--AS--PRS Framework.}
\label{alg:pod_as_rom}
\begin{algorithmic}[1]
\Require CFD snapshots $\boldsymbol{S}(t_k, \boldsymbol{x_i})$, QoI $f$.
\Ensure Surrogate model $g(\mathbf{y})$ for QoI reconstruction.

\State \textbf{Preprocess Data and Apply POD} 
\State Construct snapshot matrix $\boldsymbol{S}$, compute mean-subtracted $\tilde{\boldsymbol{S}}$.
\State Perform SVD: $\tilde{\boldsymbol{S}} = \boldsymbol{\Psi} \boldsymbol{\Sigma} \boldsymbol{\Phi}^T$, select $r$ modes (Eq.~\eqref{eq:pod_energy_norm}).
\State Compute POD coefficients $\boldsymbol{A} \in \mathbb{R}^{n_t \times r}$.

\State \textbf{Train ResNet} 
\State Train ResNet to map $\boldsymbol{A}$ to QoI using MSE loss (Eq.~\eqref{eq:loss}).
\State Compute gradient matrix $\mathbf{G} \in \mathbb{R}^{n_t \times r}$, where $\mathbf{G}_{i,:} = \nabla_{\boldsymbol{a}} f(\boldsymbol{a}_i)$ via reverse-mode AD.

\State \textbf{Perform Active Subspace Analysis} 
\State Compute covariance $\hat{\boldsymbol{C}}$ from gradients, decompose $\hat{\boldsymbol{C}} = \hat{\boldsymbol{W}} \hat{\boldsymbol{\Lambda}} \hat{\boldsymbol{W}}^T$ (Eq.~\eqref{eq:approx_C}).
\State Select active subspace basis $\hat{\boldsymbol{W}}_1 \in \mathbb{R}^{r \times N_{\text{AS}}}$ via eigenvalue gap.
\State Project coefficients: $\mathbf{y} = \hat{\boldsymbol{W}}_1^T \boldsymbol{a} \in \mathbb{R}^{n_t \times N_{\text{AS}}}$.

\State \textbf{Construct Response Surface} 
\State Fit polynomial $g(\mathbf{y}) = \sum_{|\boldsymbol{\alpha}| \leq p^*} c_{\boldsymbol{\alpha}} \varphi_{\boldsymbol{\alpha}}(\mathbf{y})$ using least-squares (Eq.~\eqref{eq:cross_val}).

\State \textbf{Output}: Surrogate model $g(\mathbf{y})$ for QoI reconstruction.
\State Evaluate $g(\mathbf{y})$ on test set for generalization.
\end{algorithmic}
\end{algorithm}

\section{Numerical investigations}
\label{sec:results}

This section presents numerical validation of the POD--AS--PRS framework through two canonical fluid dynamics problems exhibiting fundamentally different dynamical characteristics. The investigations commence with the well-established benchmark of flow past a circular cylinder, which exhibits strongly periodic dynamics. Subsequently, we examine the more complex case of flow over a NACA 4412 airfoil, characterized by intermittent and chaotic dynamics that pose significant challenges for reduced-order modeling. High-fidelity simulations provide the full-order model data essential for constructing and validating the reduced-order framework. The subsequent analysis strictly adheres to the computational pipeline established in the aforementioned methodology. Quantitative assessments of reconstruction accuracy, computational efficiency, and physical interpretability demonstrate the robustness in identifying sensitivity-dominant parameters and elucidating the underlying physics across different flow regimes.

\subsection{The incompressible Navier–Stokes equations}
\label{subsec:solver}
We consider incompressible Navier–Stokes equations with constant fluid properties, coupled with the continuity equation:
\begin{equation}
\frac{\partial \boldsymbol{u}}{\partial t}=-\boldsymbol{u} \cdot \nabla \boldsymbol{u}-\nabla p+\frac{1}{R e} \nabla^2 \boldsymbol{u}+\boldsymbol{f_b}, \label{eq:ns}
\end{equation}
\begin{equation}
    \nabla \cdot \boldsymbol{u}=0.\label{eq:continuity}
\end{equation}
where $\boldsymbol{u}=\{u, v\}^{T}$ is the velocity vector with components in the $x$ and $y$  directions, $p$ represents the pressure, and $\boldsymbol{f_b}$ is the body-force term. Eq.~\eqref{eq:ns} enforces momentum continuity, while Eq.~\eqref{eq:continuity} ensures mass continuity. These equations are integrated in time using the \texttt{Nek5000} code developed by Fischer, Lottes, and Kerkemeier~\cite{nek5000-web-page}. The system of equations is expressed in weak form, with the solution expanded using basis and test functions on each of these elements~\cite{offermans2020adaptive}. 

In the spectral element method (SEM)~\cite{patera1984spectral}, the physical domain is decomposed into spectral elements. Within each element, the local flow field is approximated as a sum of Lagrange interpolants constructed on an orthogonal basis of Legendre polynomials up to degree $N$. The same polynomial order $N$ is used for all spatial directions. Following the $\mathbb{P}_N-\mathbb{P}_{N-2}$ spatial discretisation~\cite{maday1989spectral}, $N + 1$ Gauss–Lobatto–Legendre (GLL) nodes are used to build the velocity Lagrange polynomial interpolants and $N - 1$ Gauss–Legendre (GL) nodes for the pressure Lagrange polynomial interpolants (two orders less than the velocity field) in every spectral element. The discrete Poisson equation for the pressure is solved using a variational multigrid GMRES solver, incorporating local overlapping Schwarz methods for element-based smoothing at resolution $N$ and approximately $N/2$, coupled with a global coarse-grid problem based on linear elements~\cite{fischer1997overlapping}. Here we adopt $N = 11$ for case~1 and $N=7$ for case~2. For temporal discretization, the equations are advanced in time using a third-order (case~1) and second-order (case~2) conditionally stable backward differentiation and extrapolation scheme (BDF$k$/EXT$k$)~\cite{fischer2003implementation}, employing an implicit treatment of the diffusion term and an explicit treatment of the advection term. \texttt{Nek5000} has been extensively validated for a variety of incompressible flows, including cases with flow past a cylinder and an airfoil~\cite{amor2023higher,citro2017efficient,shahriari2018control,rudy2022prediction}.

\subsection{Case~1: Cylinder flow at Re=100}
\label{subsec:cylinder}

This part presents the application and validation of the POD--AS--PRS framework to the canonical problem of flow past a circular cylinder. This benchmark configuration exhibits periodic vortex shedding and serves as an ideal test case for identifying sensitivity-dominant parameters with well-defined coherent structures. Comprehensive quantitative assessments demonstrate the framework's accuracy in reconstructing aerodynamic quantities while providing physical insights into the dominant mechanisms governing drag variability.

\subsubsection{Dataset description}
The present investigation considers the flow over a circular cylinder, which exhibits periodic vortex shedding, specifically employing a cylinder geometry with a diameter of $D$. The diameter-based Reynolds number is $Re=U_{\infty}D/\nu=100$ with $U_\infty$ as the incoming free-stream velocity and $\nu$ as the kinematic viscosity. Length and velocity are non-dimensionalised by cylinder diameter $D$ and $U_\infty$, respectively. Time refers to non-dimensional convective time normalised by $DU_\infty^{-1}$.

The computational extent is $(x,y)\in\left[-15,35\right]\times\left[-15,15\right]$ with the origin located at the cylinder's center. Within each element, flow quantities are resolved using high-order polynomial approximations implemented in the spectral code described in Sec.~\ref{subsec:solver}. For the cylinder case, a combination of 1472 elements and a polynomial order of 11 was selected. The operator-integrating-factor splitting method proposed in~\cite{maday1990operator} was employed to ensure stable time integration with a maximum timestep of $\Delta t=0.001$. 

\begin{figure*}[t]
\centering
\includegraphics[width=1\textwidth]{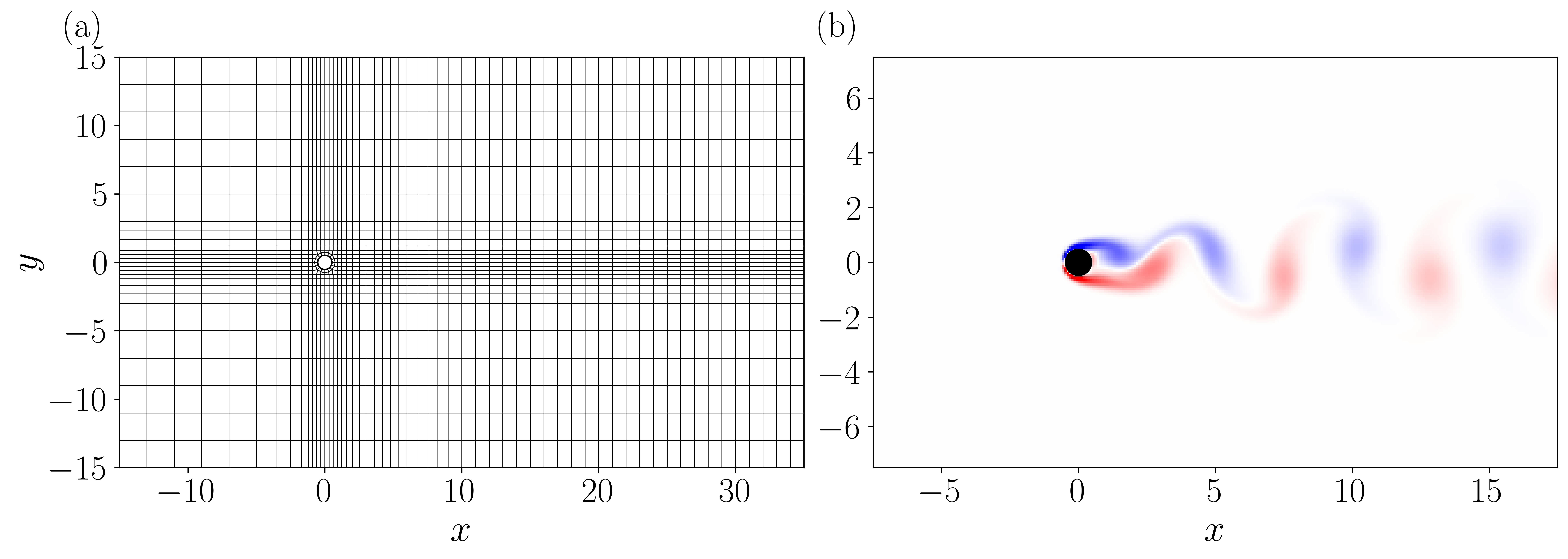}
\caption{(a) Computational domain for the flow past a circular cylinder showing outlines of spectral elements without internal interpolation points. (b) Snapshot of vorticity.}
\label{fig:cylinder_mesh}
\end{figure*}

The inflow boundary condition was prescribed as a constant velocity $\boldsymbol{u} = (1, 0)$ throughout the simulation. No-slip (Dirichlet) boundary conditions were imposed on the wall of the cylinder, while symmetry conditions were enforced along the upper and lower domain boundaries. An open boundary condition with zero pressure (Neumann) is applied at the exit section, allowing vortical structures to exit the computational domain smoothly. Fig.~\ref{fig:cylinder_mesh} shows the spectral element grid (without GLL interpolation points) along with a representative vorticity snapshot.

In this case, snapshots from $t=250$ to $t=300$ are collected every 50 steps, yielding a total of 1000 snapshots used as inputs.

\subsubsection{The quantity of interest}

\begin{figure*}[h]
    \centering
    \vspace{10pt}

    \color{black}
    \begin{subfigure}[t]{0.48\textwidth}
        \centering
        \begin{overpic}[height=4.5cm]{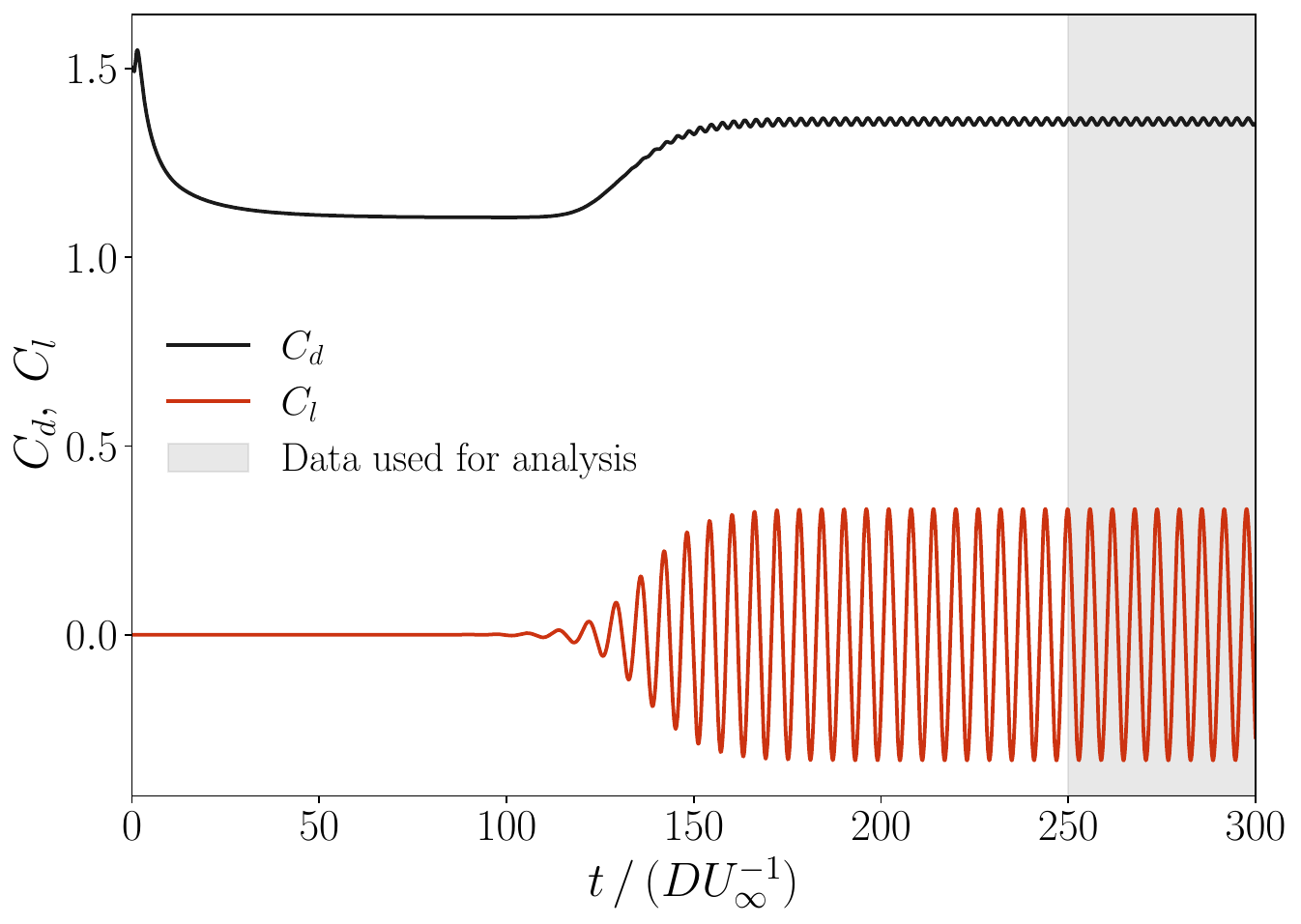}
            \put(2,130){\small{(a)}} 
        \end{overpic}
    \end{subfigure}
    \hfill
    \begin{subfigure}[t]{0.48\textwidth}
        \centering
        \begin{overpic}[height=4.5cm]{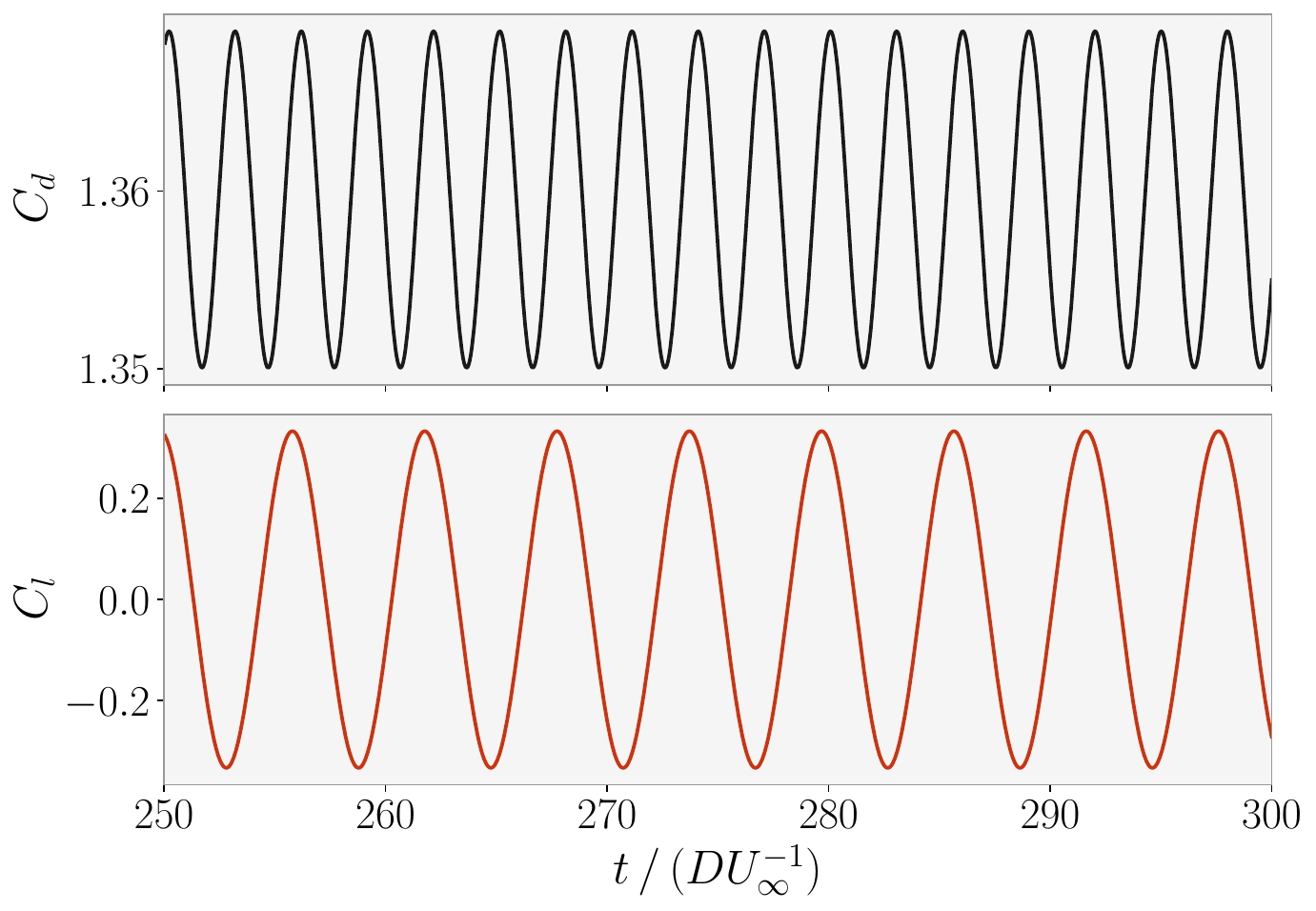}
            \put(2,130){\small{(b)}} 
        \end{overpic}
    \end{subfigure}

    \caption{(a)Temporal evolution of the drag coefficient $C_d$ and lift coefficient $C_l$, with the grey-shaded region denoting the snapshot sampling interval adopted for constructing the ROM; (b) zoomed view of $C_d$ (top) and $C_l$ (bottom) over the sampling interval.}
    \label{fig:case1_qoi}

\end{figure*}

The drag and lift coefficients ($C_d$ and $C_l$, respectively) are defined as
\begin{equation}
C_d = \dfrac{F_d}{\frac{1}{2}\rho U_\infty^2 D},\hspace{1 cm} C_l = \dfrac{F_l}{\frac{1}{2}\rho U_\infty^2 D}.
\label{eq:aerodynamic_coefficients}
\end{equation}
where $\rho$ is the fluid density, while $F_d$ and $F_l$ are the drag and lift forces per unit span length, respectively. In this case, both $C_d$ and $C_l$ are designated as the QoIs. The temporal evolution of the QoI is illustrated in Fig.~\ref{fig:case1_qoi}, where the grey-shaded region highlights the snapshot sampling interval employed in this case study.

\subsubsection{Results and discussion}

\paragraph{Dimensionality reduction of POD state-space}

The POD analysis is performed on fluctuating components to extract the dominant oscillatory dynamics from fully converged full-order models (FOMs) snapshots. As shown in Fig.~\ref{subfig:case1_pod_spatial} in Appendix~\ref{app:pod_figures}, the spatial modes emerge in pairs, reflecting the periodic nature of vortex shedding. These results reveal that modes emerge in pairs due to the periodic nature of the flow, capturing real and imaginary components that can be combined into complex modes as $\boldsymbol{u}_i^{'}=\boldsymbol{u}_i+i\boldsymbol{u}_{i+1}$. The first six spatial POD modes (left column) and their corresponding temporal coefficients (right column) show that paired modes (e.g., Mode 2 as a phase-shifted version of Mode 1) exhibit sinusoidal behavior in their coefficients, reflecting the dominant vortex-shedding frequency and its harmonics in higher-order modes. This structure aligns with typical bluff-body flows dominated by vortex shedding. The phase diagram in Fig.~\ref{subfig:case1_pod_phase} demonstrates that the $(a_1, a_2)$ evolution traces a nearly perfect circle, confirming the periodic nature of saturated vortex shedding. Additionally, the phase plots highlight that the third and fourth POD modes correspond to the second harmonics of the vortex shedding, while the fifth and sixth modes capture its third harmonics. Complementing this, the probability density functions (PDFs) of the first six POD mode time coefficients exhibit bimodal distributions, as shown in Fig.~\ref{subfig:case1_pod_phase}. The kernel density estimate (KDE) overlays highlight this continuous oscillation between positive and negative extrema, which is characteristic of coherent structures in vortex-dominated wakes. These analyses confirm that POD effectively extracts the key oscillatory dynamics, providing low-dimensional coefficients suitable as inputs for the subsequent ResNet-based QoI prediction and gradient computation in AS analysis.

\begin{figure}[t]
	\centering
	\begin{subfigure}[t]{0.48\columnwidth} 
    \centering
        \begin{overpic}[height=4.5cm]{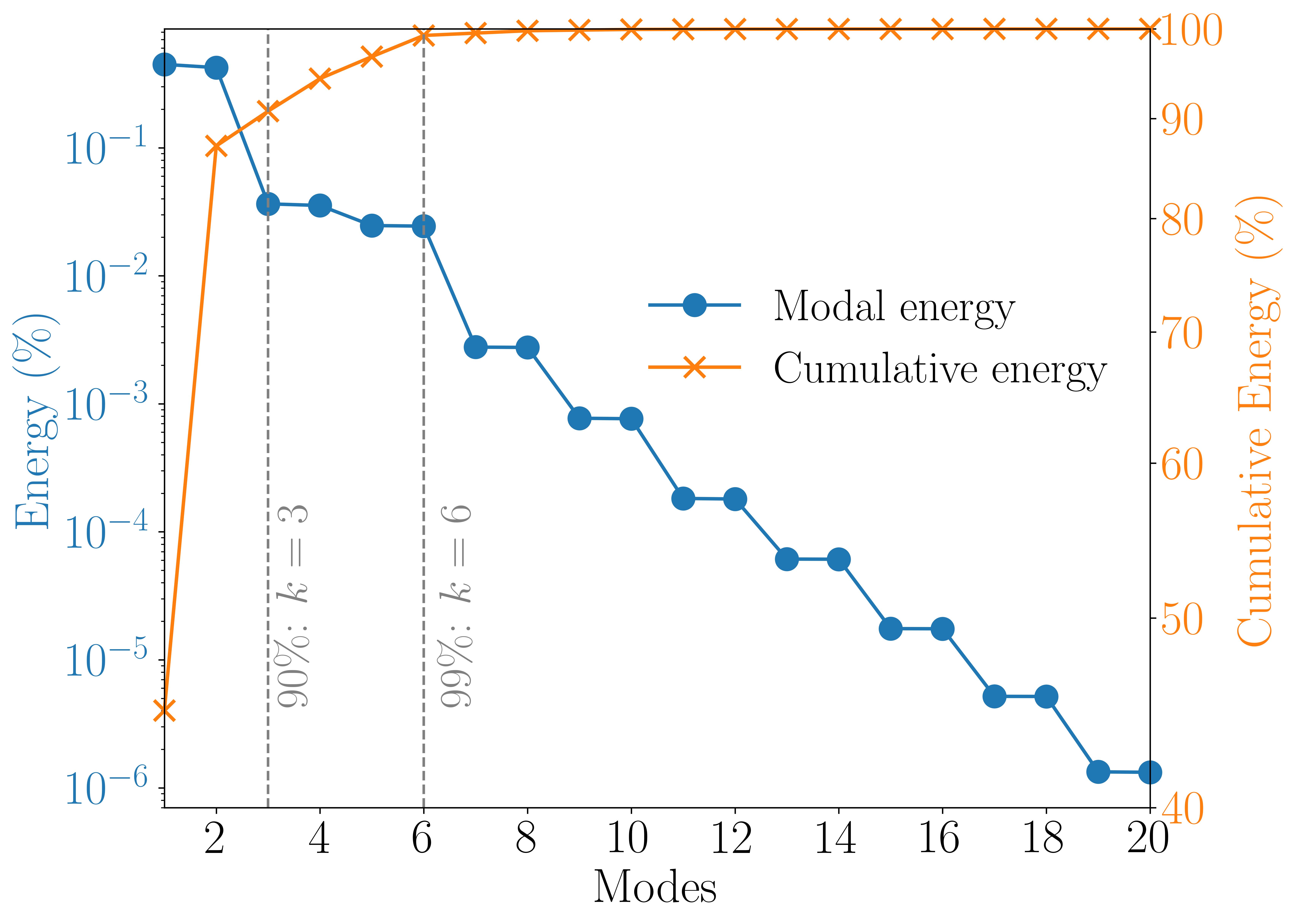}
            \put(2,130){\small{(a)}} 
        \end{overpic}
		\phantomcaption
		\label{subfig:case1_energy} 
	\end{subfigure}
	\hfill
	\begin{subfigure}[t]{0.48\columnwidth}
    \centering
        \begin{overpic}[height=4.5cm]{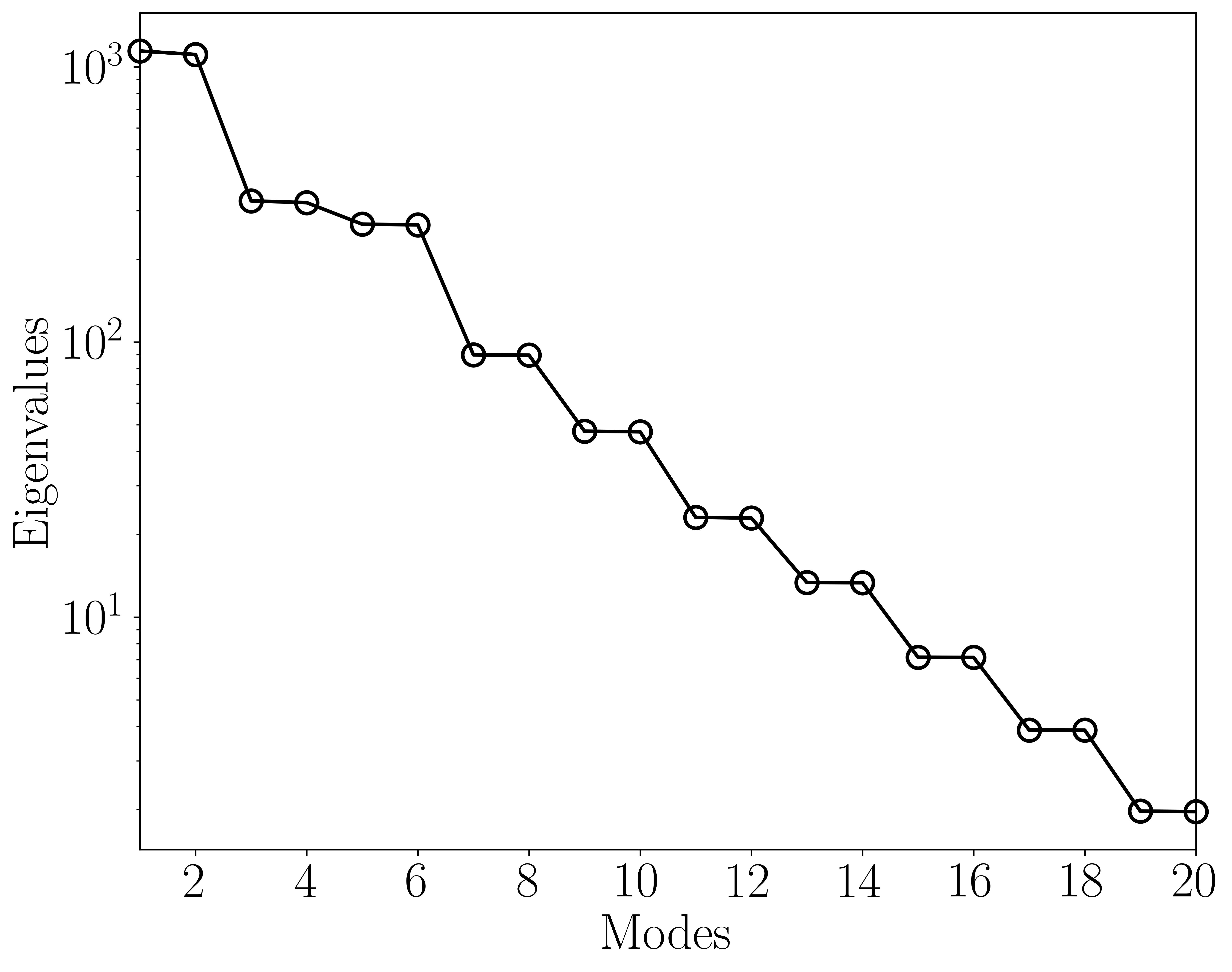}
            \put(2,130){\small{(b)}}  
        \end{overpic}
		\phantomcaption
		\label{subfig:case1_eigenval} 
	\end{subfigure}
	\caption{(a) The kinetic energy explained by each of the first 20 POD modes and the corresponding cumulative energy, and (b) the eigenvalue distribution of the first 20 modes for the test case~1.}
	\label{fig:case1_pod_energy}
\end{figure}

The kinetic energy $\sigma_{k}^{2}$ of each POD mode and the corresponding cumulative energy, represented in Fig.~\ref{subfig:case1_energy}, indicate strong dominance by the first two modes, which encapsulate the primary shedding frequency and account for the majority of the flow's kinetic energy. The first six modes collectively capture over 99\% of the total energy (Table~\ref{tab:case1_pod_energy_retained}), thereby justifying the truncation of the basis at a very low order. This is corroborated by the eigenvalue spectrum of the first 20 modes in Fig.~\ref{subfig:case1_eigenval}, which decays rapidly and quantifies the separation between dominant and subordinate modes. Such a distribution not only validates the energy capture but also guides mode selection for the POD--AS--PRS framework, ensuring computational efficiency in constructing response surface-based surrogate models from AS-derived active variables.

\begin{table}[h]
\centering
\small
\begin{tabularx}{\textwidth}{>{\centering\arraybackslash}X >{\centering\arraybackslash}X}
\toprule
Minimum retained energy [\%]&$N_{\text{POD}}$ \\
\midrule
90&3\\
94&4\\
96&5\\
99&6\\
99.5&7\\
99.9&9\\
99.99&13\\
\bottomrule
\end{tabularx}
\caption{Cumulative retained energy for different numbers of POD modes for the test case~1.}\label{tab:case1_pod_energy_retained}
\end{table}

\paragraph{Training of linear-layer ResNet and gradient computation} 

Following the POD decomposition, the extracted POD coefficients serve as low-dimensional inputs to learn the nonlinear mapping to the QoI, enabling efficient approximation and gradient computation through reverse-mode AD.

\begin{figure*}[h] 
\color{black}
	\centering

	\begin{subfigure}[t]{0.48\textwidth} 
		\centering
		\begin{overpic}[height=5cm]{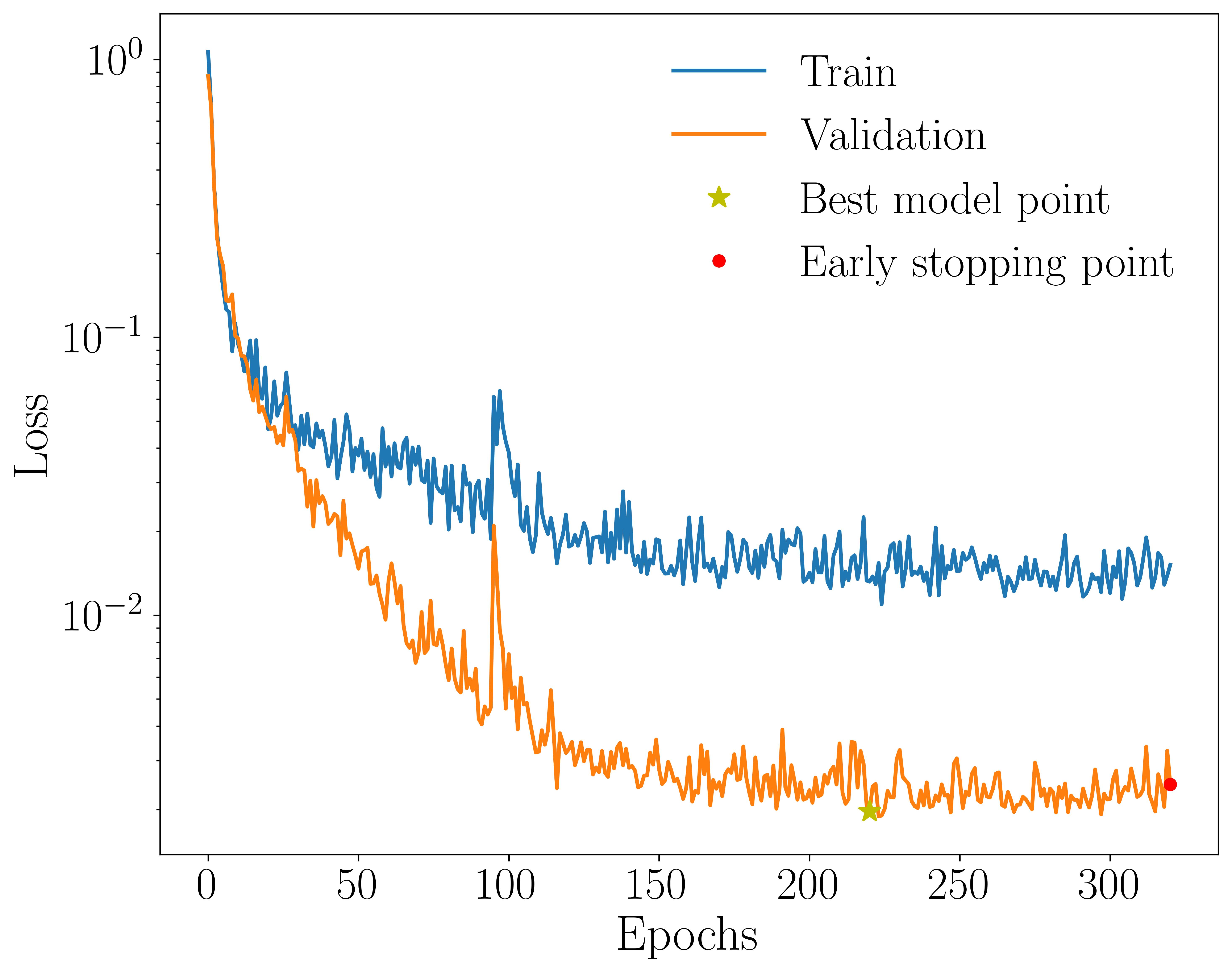}
			\put(2,140){\small{(a)}} 
		\end{overpic}
		\phantomcaption
		\label{subfig:case1_loss_cd} 
	\end{subfigure}
	\hfill
	\begin{subfigure}[t]{0.48\textwidth}
		\centering
		\begin{overpic}[height=5cm]{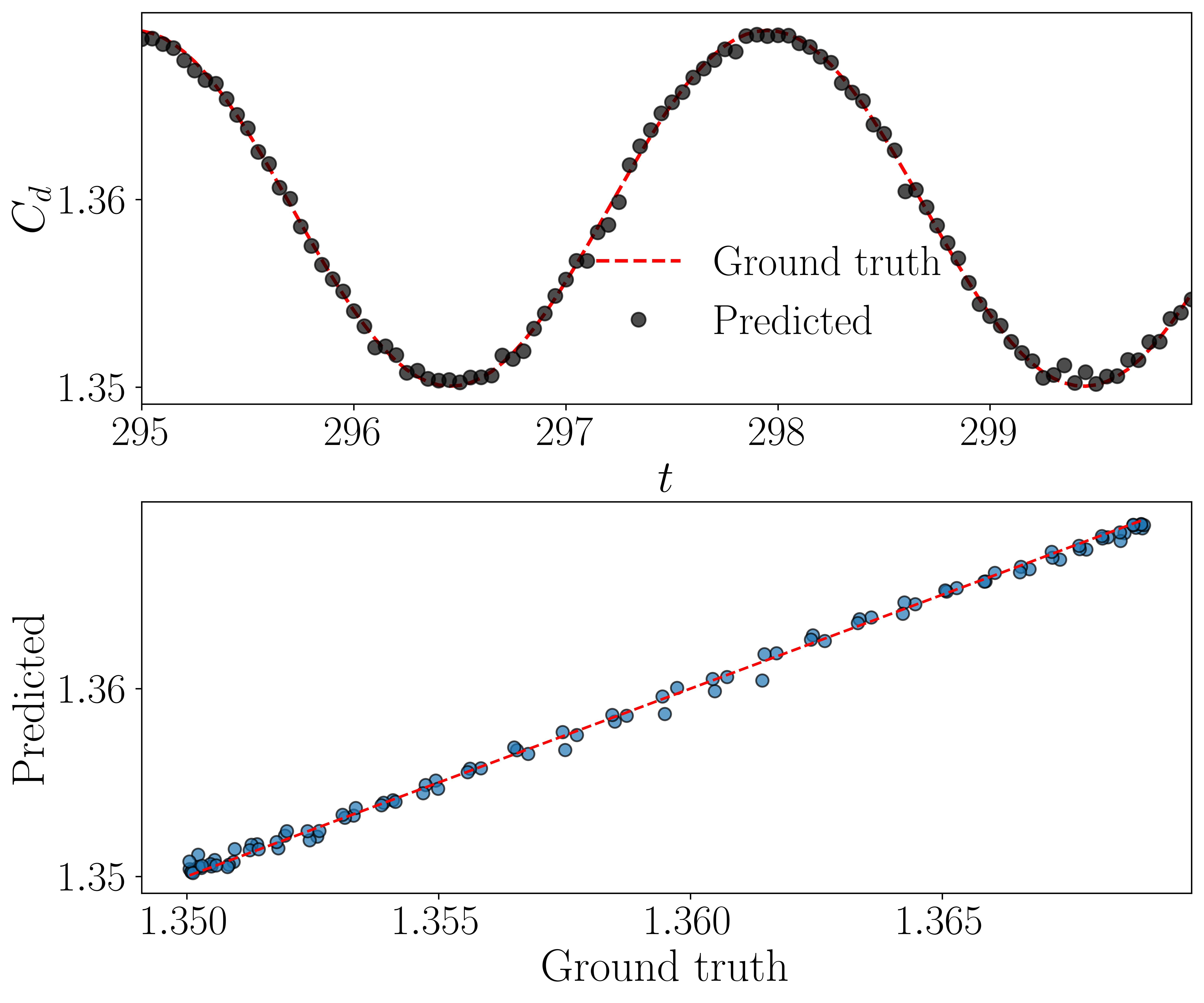}
			\put(2,140){\small{(b)}} 
		\end{overpic}
		\phantomcaption
		\label{subfig:case1_pred_cd} 
	\end{subfigure}

	\vspace{4mm} 

	\begin{subfigure}[t]{0.48\textwidth} 
		\centering
		\begin{overpic}[height=5cm]{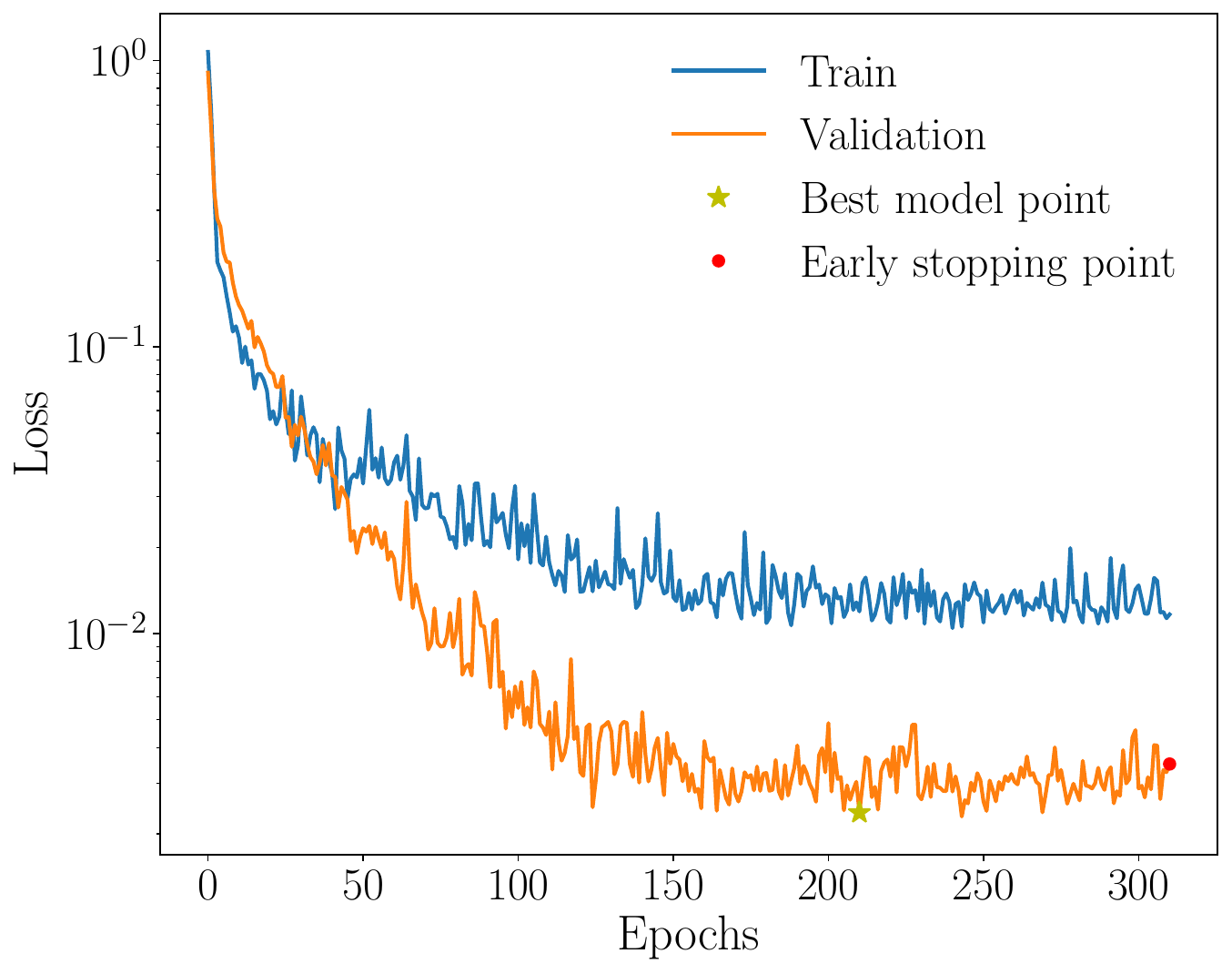}
			\put(2,140){\small{(c)}} 
		\end{overpic}
		\phantomcaption
		\label{subfig:case1_loss_cl} 
	\end{subfigure}
	\hfill
	\begin{subfigure}[t]{0.48\textwidth}
		\centering
		\begin{overpic}[height=5cm]{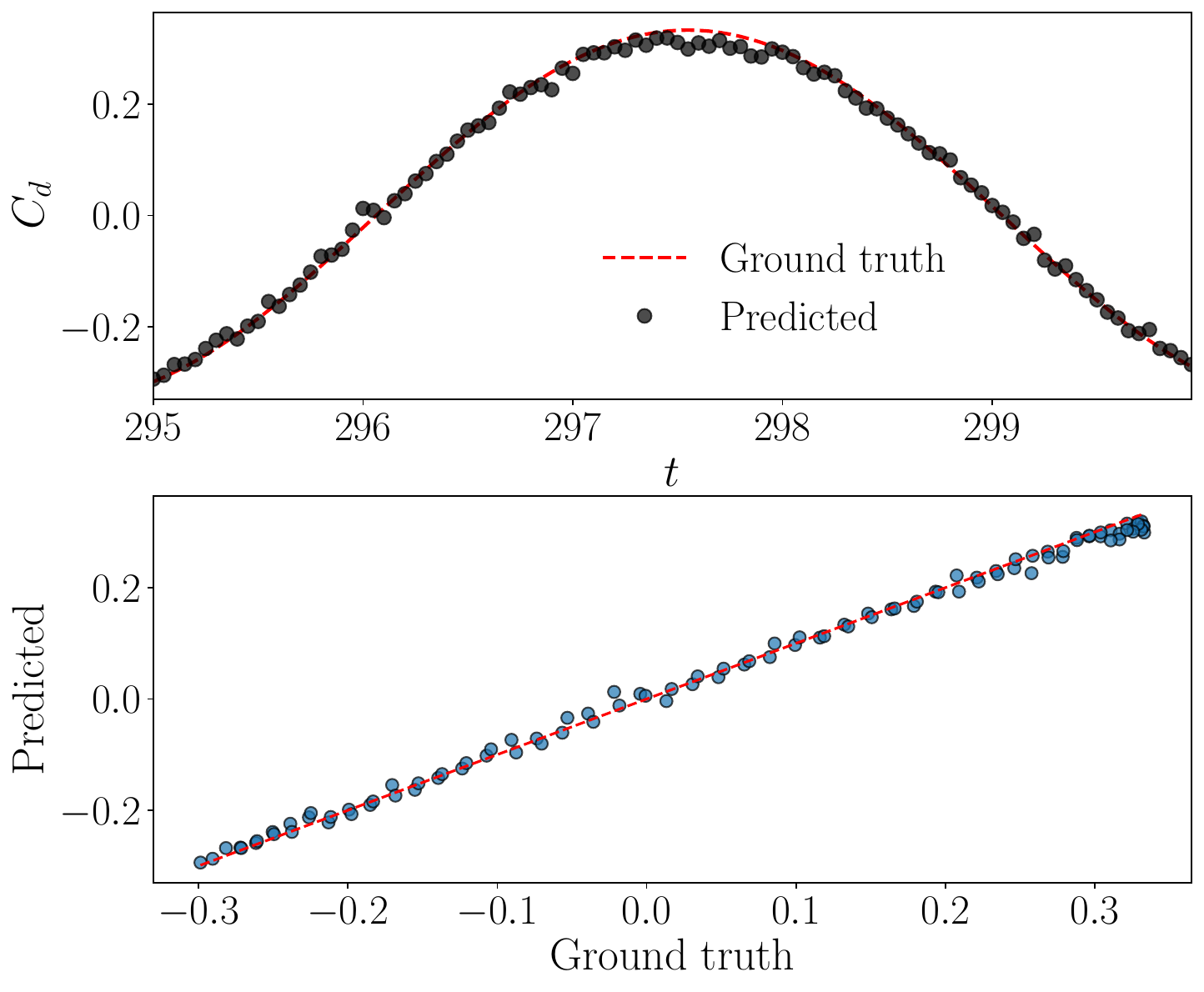}
			\put(2,140){\small{(d)}} 
		\end{overpic}
		\phantomcaption
		\label{subfig:case1_pred_cl} 
	\end{subfigure}

	\caption{Loss function evolution for (a) $C_d$ and (c) $C_l$; the yellow star (\textcolor{goldenyellow}{\ding{72}}) indicates the best model, while the red dot (\textcolor{red}{$\bullet$}) marks the early stopping point (Eq.~\eqref{eq:early_stopping}). Model performance on the test dataset for (b) $C_d$ and (d) $C_l$: the upper panels compare predicted and ground truth values over time, while the lower panels show the corresponding parity plots.}
	\label{fig:case1_resnet_training}
\end{figure*}

The training process exhibited robust convergence under the MSE loss function, with early stopping (Eq.~\eqref{eq:early_stopping}) to prevent  overfitting. Fig.~\ref{fig:case1_resnet_training} illustrates the training process and model performance for both QoIs. Figs.~\ref{subfig:case1_loss_cd} and~\ref{subfig:case1_loss_cl} show that the loss for both training and validation datasets rapidly decreases initially and then stabilizes for $C_d$ and $C_l$, respectively, indicating effective convergence without significant overfitting in either case. Figs.~\ref{subfig:case1_pred_cd} and~\ref{subfig:case1_pred_cl} evaluate the predictive accuracy of the model in the independent test dataset for $C_d$ and $C_l$, respectively. The time-series comparisons demonstrate close agreement with the ground truth across the temporal domain, capturing transient flow behaviors accurately. The parity plot further confirms high fidelity of the ResNet predictions, with predicted values closely clustering around the diagonal in both cases.

To quantify the model's performance more rigorously, Table~\ref{tab:case1_resnet_performance} reports key metrics evaluated across the training, validation, and test datasets for both $C_d$ and $C_l$, including the RMSE, mean absolute error (MAE), and coefficient of determination ($R^2$). For both quantities, the low RMSE and MAE values, combined with an $R^2$ approaching unity, affirm the strong predictive performance of the linear-layer ResNet in mapping the POD coefficients to the respective QoI. These results validate the model's suitability for subsequent gradient computation through AD, which forms the basis for the AS analysis in the POD--AS--PRS framework.

\begin{table*}[h]
\color{black}
\centering
\small
\begin{tabular}{ccccc}
\toprule
QoI & Metric & Training set & Validation set & Test set \\
\midrule
\multirow{3}{*}{$C_d$} 
    & RMSE & $1.78\times 10^{-4}$ & $3.32\times 10^{-4}$ & $3.16\times 10^{-4}$ \\
    & MAE  & $1.44\times 10^{-4}$ & $2.75\times 10^{-4}$ & $2.48\times 10^{-4}$ \\
    & $R^2$ & 0.999 & 0.998 & 0.998 \\
\midrule
\multirow{3}{*}{$C_l$} 
    & RMSE  & $6.17\times 10^{-3}$ & $1.39\times 10^{-2}$ & $1.24\times 10^{-2}$ \\
    & MAE  & $5.00\times 10^{-3}$ & $1.14\times 10^{-2}$ & $9.52\times 10^{-3}$ \\
    & $R^2$ & 0.999 & 0.997 & 0.996 \\
\bottomrule
\end{tabular}
\caption{Performance metrics of the linear layer ResNet for test case~1.}
\label{tab:case1_resnet_performance}
\end{table*}

\begin{figure*}[h]
\color{black}
    \centering

    \begin{subfigure}[t]{0.81\textwidth} 
        \centering
        \begin{overpic}[height=3.5cm]{figures/Figure_7a}
            \put(2,100){\small{(a)}} 
        \end{overpic}
        \phantomcaption
        \label{subfig:case1_grad_diff_cd} 
    \end{subfigure}
    \hfill
    \begin{subfigure}[t]{0.18\textwidth}
        \centering
        \begin{overpic}[height=3.5cm]{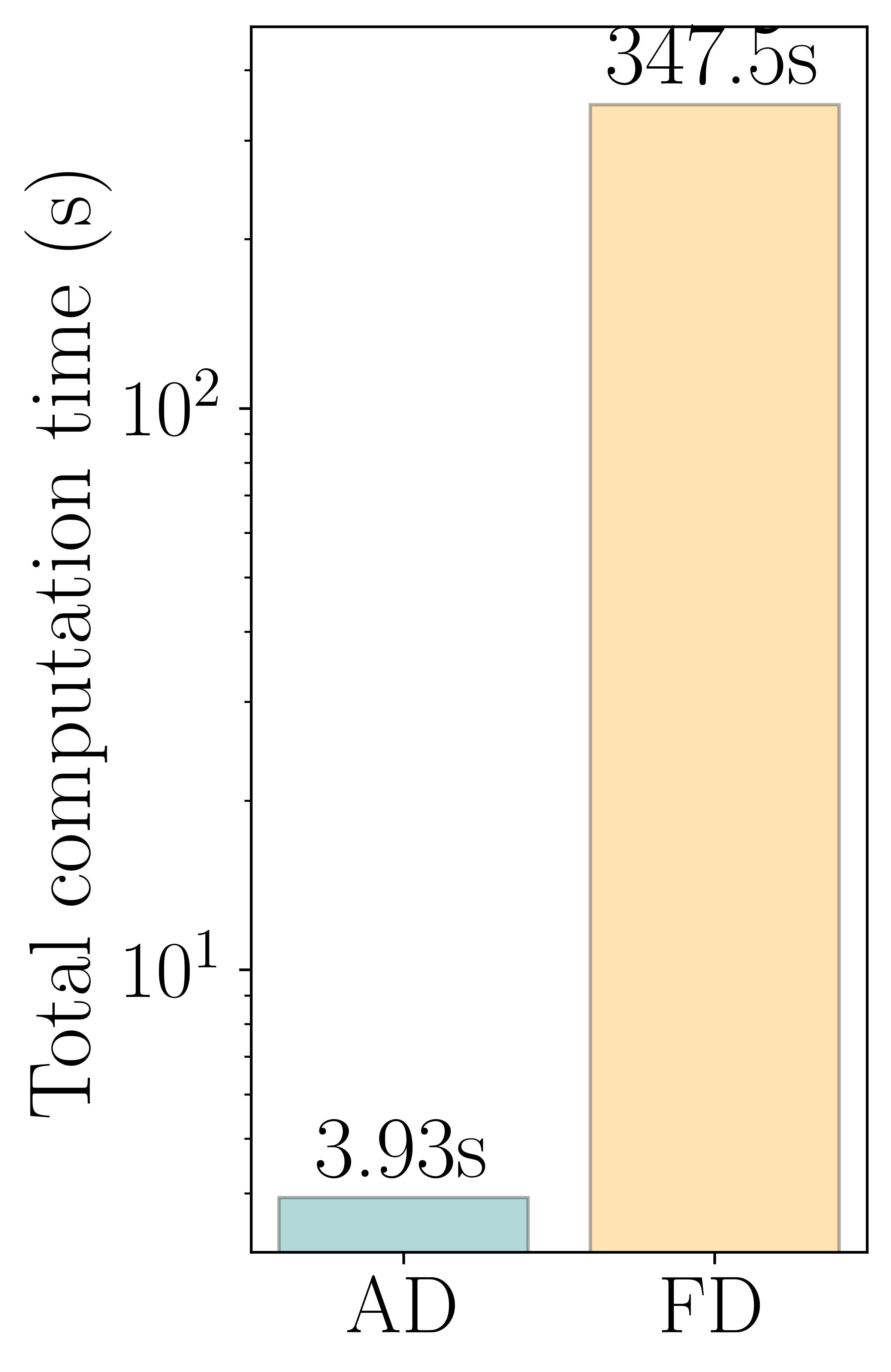}
            \put(2,100){\small{(b)}} 
        \end{overpic}
        \phantomcaption
        \label{subfig:case1_grad_time_cd} 
    \end{subfigure}

    \vspace{4mm}

    \begin{subfigure}[t]{0.81\textwidth} 
        \centering
        \begin{overpic}[height=3.5cm]{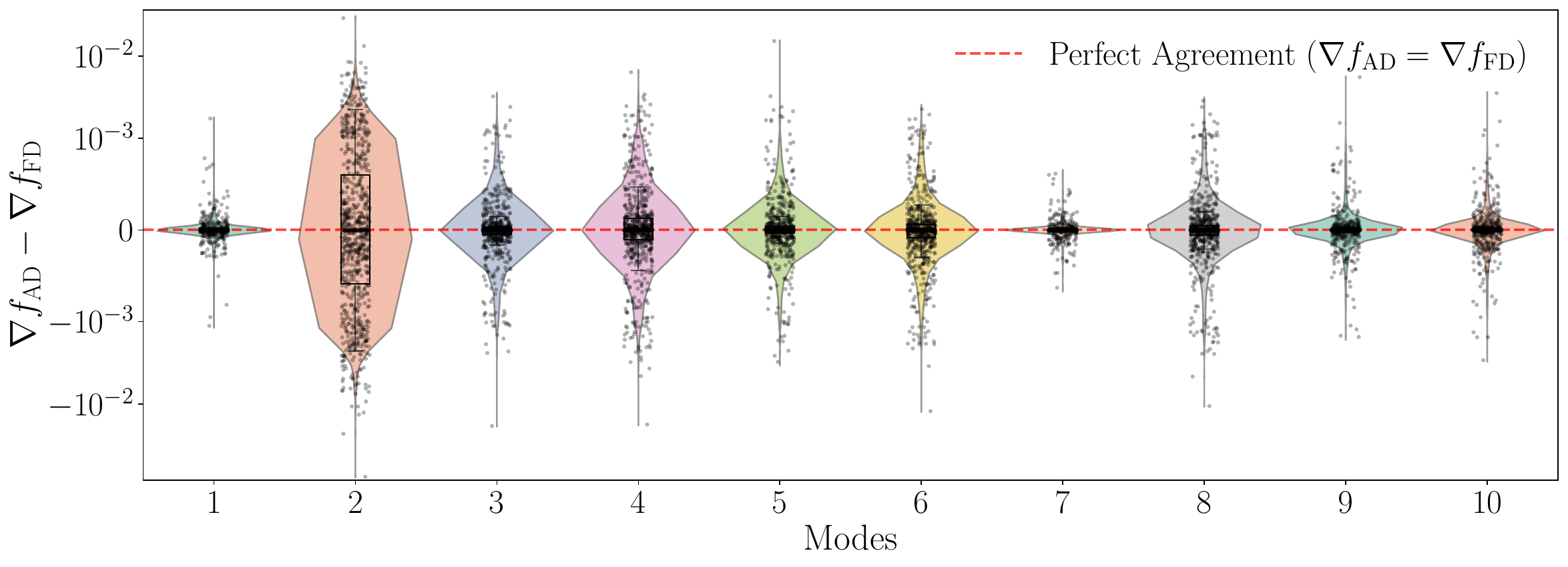} 
            \put(2,100){\small{(c)}} 
        \end{overpic}
        \phantomcaption
        \label{subfig:case1_grad_diff_cl} 
    \end{subfigure}
    \hfill
    \begin{subfigure}[t]{0.18\textwidth}
        \centering
        \begin{overpic}[height=3.5cm]{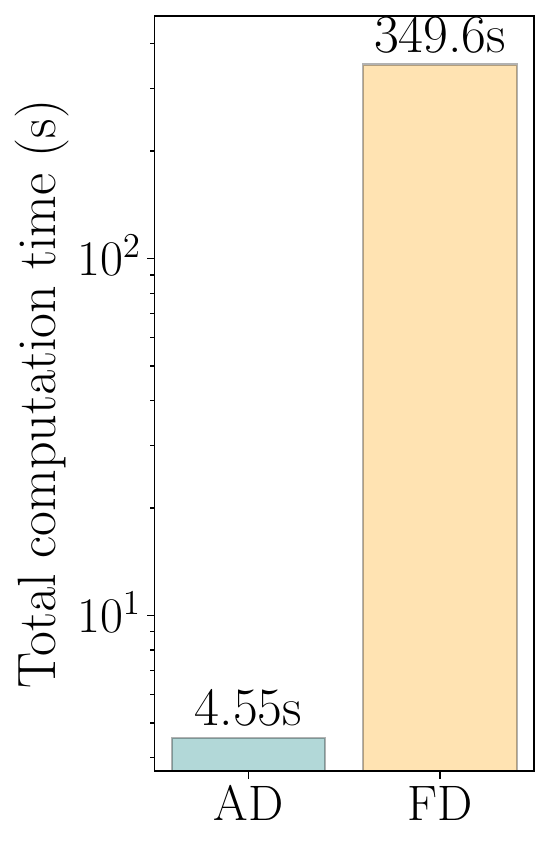} 
            \put(2,100){\small{(d)}} 
        \end{overpic}
        \phantomcaption
        \label{subfig:case1_grad_time_cl} 
    \end{subfigure}

    \caption{Gradient differences of the first 10 POD modes across all samples for (a) $C_d$ and (c) $C_l$, shown as violin plots with boxplot and strip overlays; (b, d) total gradient computation times for AD and FD for (b) $C_d$ and (d) $C_l$ in test case~1.}
    \label{fig:case1_gradient}
\end{figure*}

Gradients of the predicted QoI with respect to the POD coefficients were computed using reverse-mode AD through the trained network. In the gradient-based sensitivity analysis, the first 150 POD modes were retained as the input dimensions for each sample. This choice was made to capture sufficient system information and to provide a rich enough candidate space for the AS analysis to identify all potentially sensitive physical directions. To evaluate accuracy and efficiency, AD-based gradients were compared with FD approximations using central differences with step size $h = 1 \times 10^{-2}$. The gradient differences are presented in Figs.~\ref{subfig:case1_grad_diff_cd} and~\ref{subfig:case1_grad_diff_cl} as violin plots overlaid with boxplots and strip plots for $C_d$ and $C_l$, respectively, exhibiting symmetric distributions around the zero reference line, with medians close to zero and narrow interquartile ranges, indicating excellent agreement without systematic bias. Complementarily, the computational timings in Figs.~\ref{subfig:case1_grad_time_cd} and~\ref{subfig:case1_grad_time_cl} show that AD requires only 3.93s and 4.55s for $C_d$ and $C_l$ in total, respectively, whereas FD takes 347.5s and 349.6s, corresponding to a speedup of roughly two orders of magnitude in both cases. All computations were performed on a workstation equipped with an NVIDIA GeForce RTX 3090 GPU (compute capability 8.6, 24 GB memory). Beyond computational efficiency, AD inherently applies the chain rule across the computational graph, yielding derivatives with machine-precision accuracy and circumventing the truncation–round-off trade-off inherent to FD. For scalar outputs with high-dimensional inputs, reverse-mode AD computes the entire gradient at a cost comparable to a single function evaluation, in stark contrast to the $d+1$ evaluations required by FD (where $d$ is the input dimension)~\cite{baydin2018automatic}. This property renders AD particularly advantageous for large-scale gradient-based sensitivity analysis and model reduction.

\paragraph{Dimensionality reduction of active subspaces parameter-space} 

Using gradients of both $C_d$ and $C_l$ with respect to the POD coefficients obtained by reverse-mode AD, active subspace analyses were conducted independently for each QoI to identify low-dimensional structures capturing their respective variability.

\begin{figure*}[h]
\color{black}
    \centering
      \begin{subfigure}[t]{0.48\textwidth} 
        \centering
        \begin{overpic}[height=5cm]{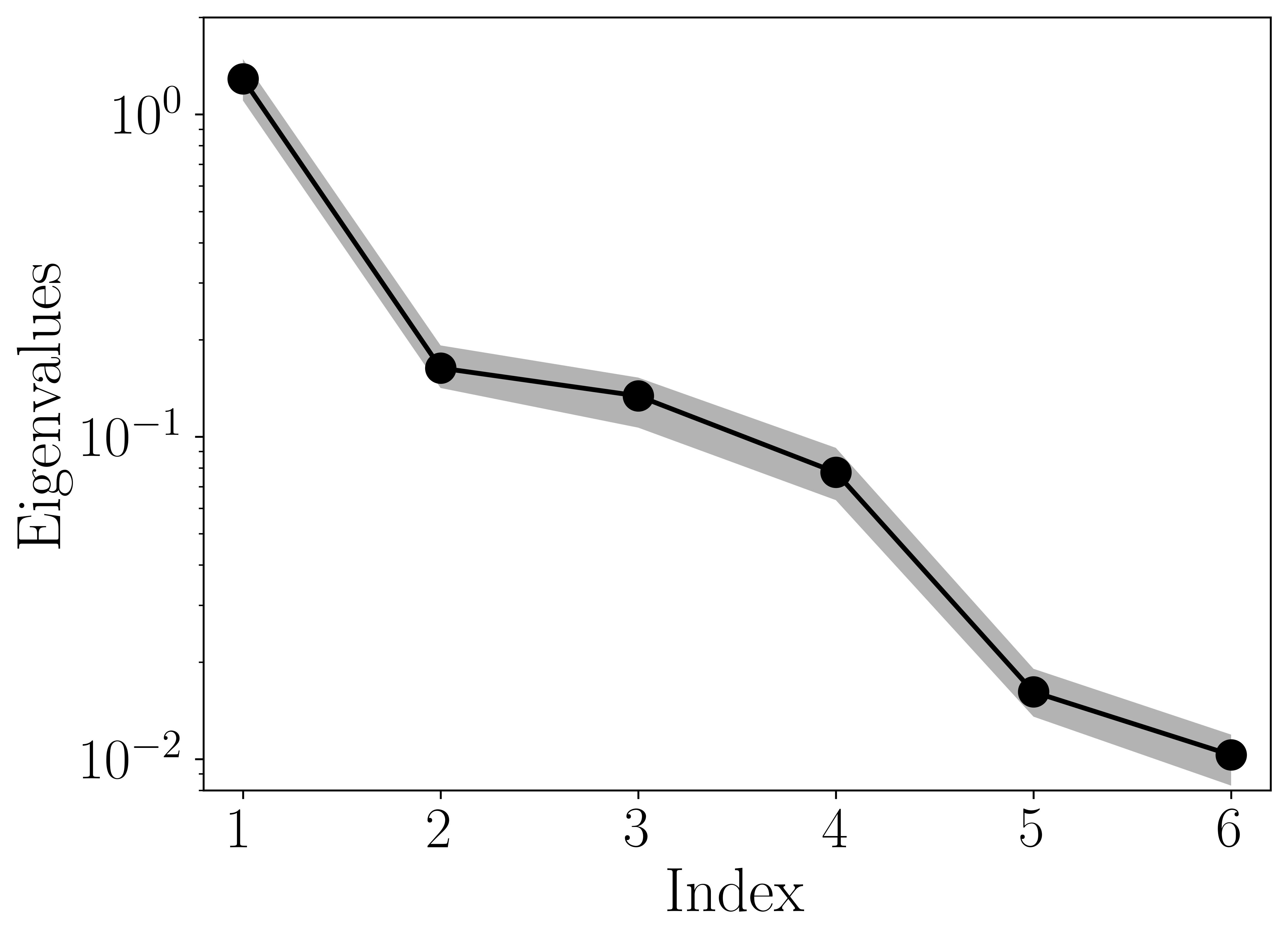}
            \put(2,140){\small (a)} 
        \end{overpic}
        \phantomcaption
        \label{subfig:case1_as_eigenval_cd} 
    \end{subfigure}
    \hfill
    \begin{subfigure}[t]{0.48\textwidth} 
        \centering
        \begin{overpic}[height=5cm]{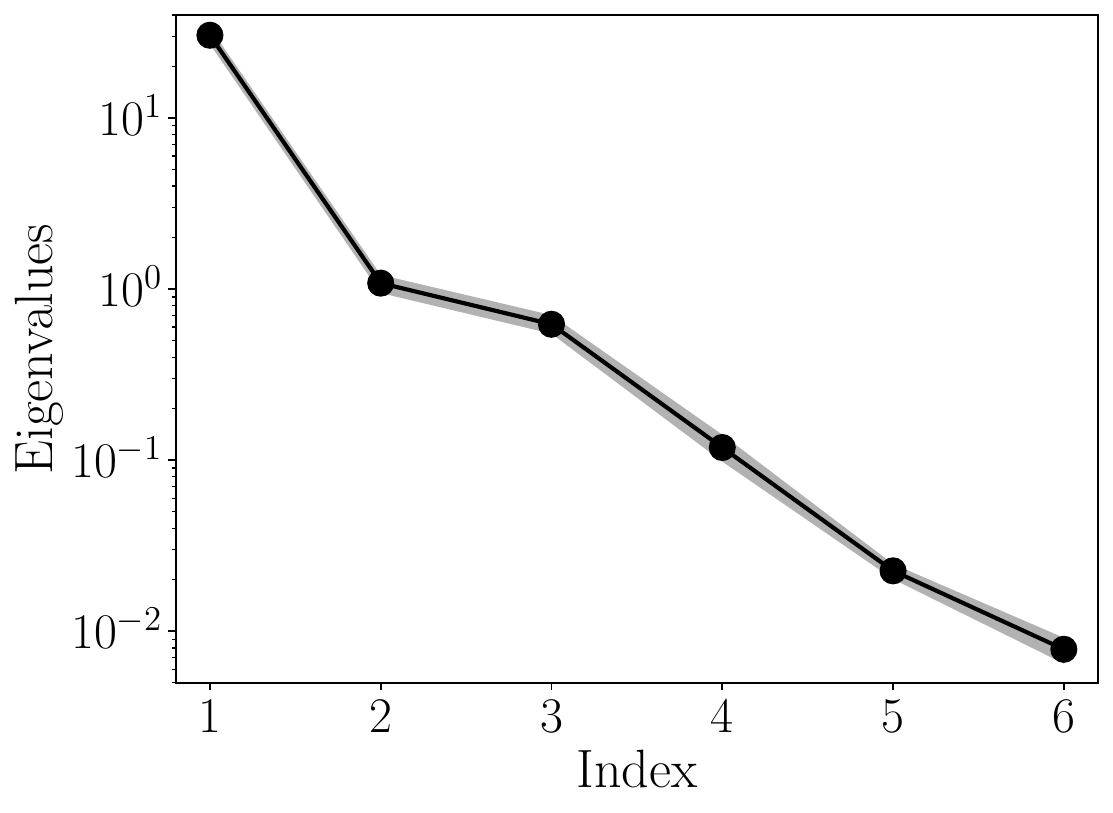} 
            \put(2,140){\small (b)}
        \end{overpic}
        \phantomcaption
        \label{subfig:case1_as_eigenval_cl} 
    \end{subfigure}

    \vspace{4mm}

    \begin{subfigure}[t]{0.48\textwidth} 
        \centering
        \begin{overpic}[height=5cm]{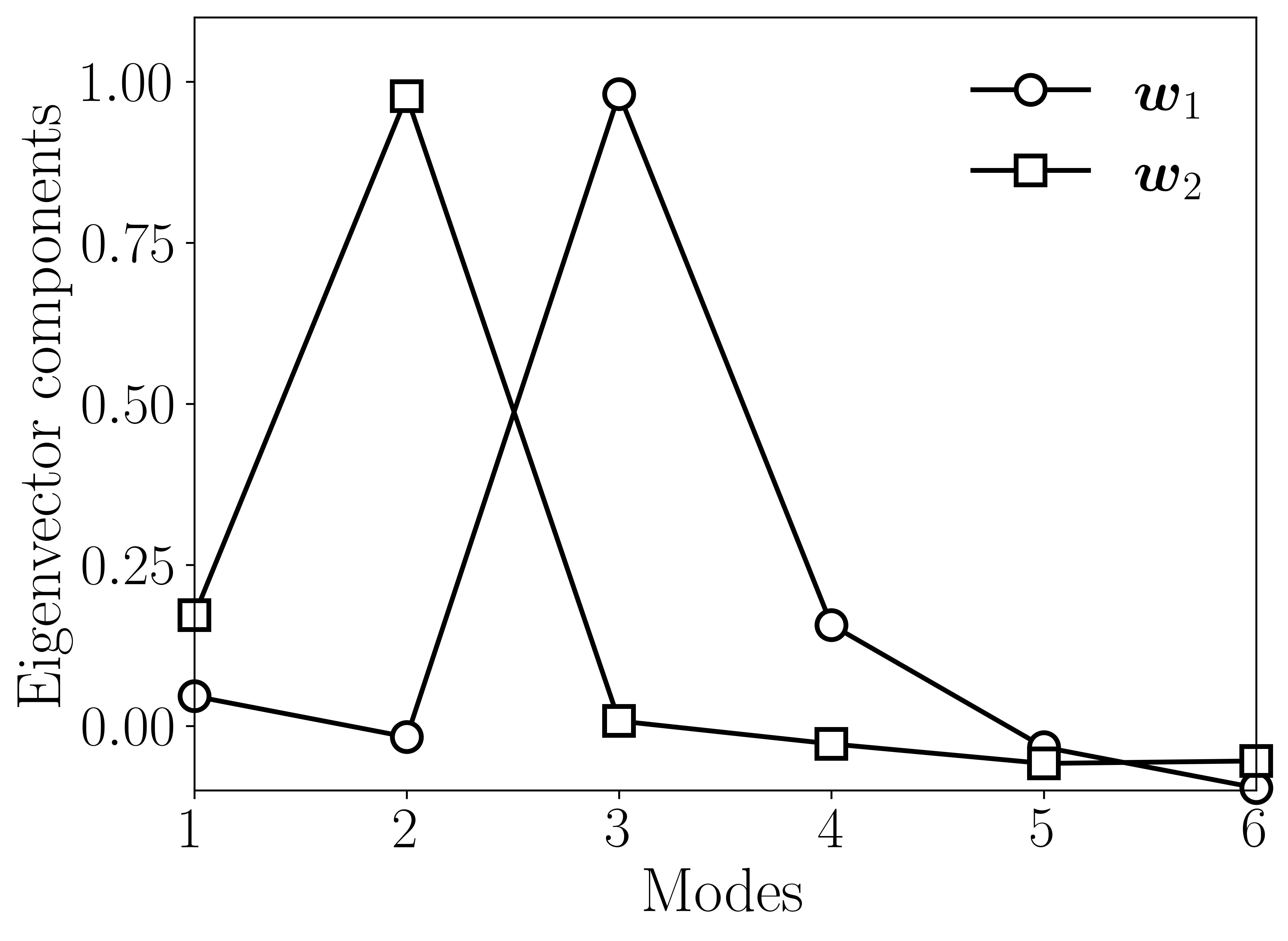} 
            \put(2,140){\small (c)} 
        \end{overpic}
        \phantomcaption
        \label{subfig:case1_as_eigenvec_cd} 
    \end{subfigure}
    \hfill
    \begin{subfigure}[t]{0.48\textwidth} 
        \centering
        \begin{overpic}[height=5cm]{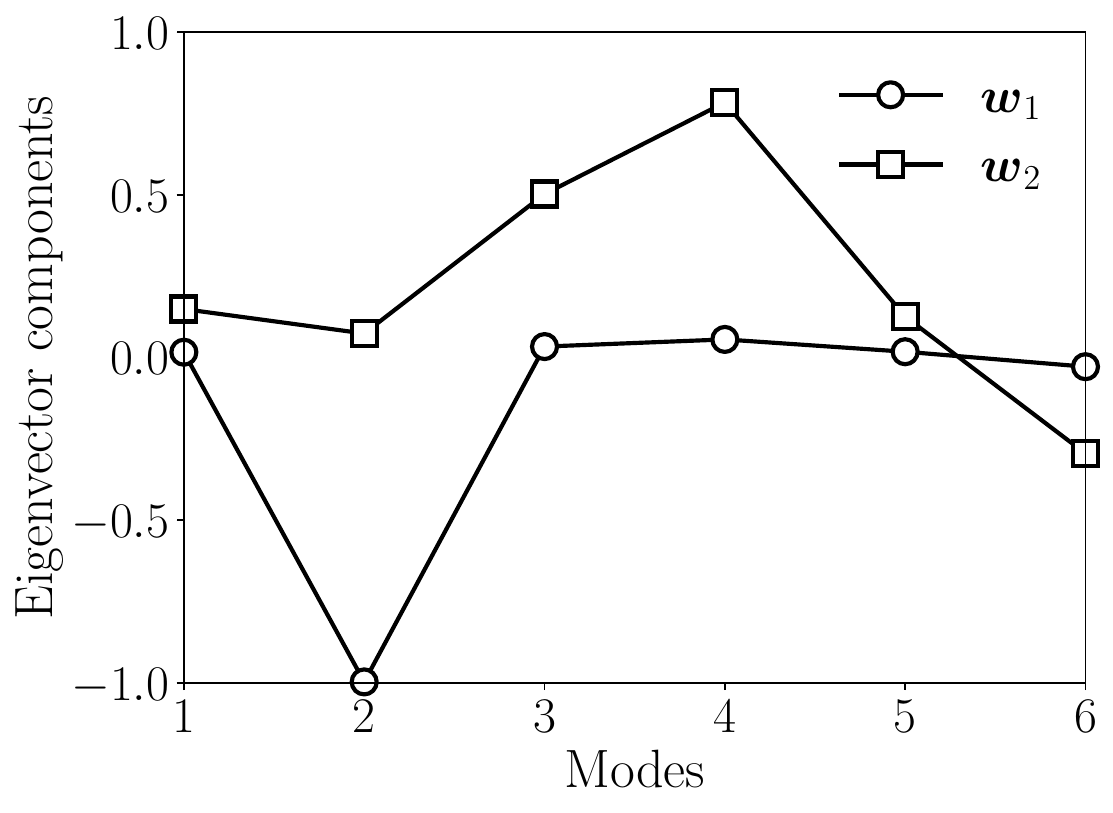}
            \put(2,140){\small (d)}
        \end{overpic}
        \phantomcaption
        \label{subfig:case1_as_eigenvec_cl} 
    \end{subfigure}

    \caption{The first six eigenvalues of the covariance matrix with bootstrap confidence intervals (grey regions) for (a) $C_d$ and (b) $C_l$ and components of the first two eigenvectors for (c) $C_d$ and (d) $C_l$ in test case~1.}
    \label{fig:case1_as_eigen}
\end{figure*}

Following the AS analysis described in Sec.~\ref{subsec:AS}, the eigenvalue spectrum of the covariance matrices $\hat{\boldsymbol{C}}$ for $C_d$ and $C_l$ are shown in Figs.~\ref{subfig:case1_as_eigenval_cd} and~\ref{subfig:case1_as_eigenval_cl}, respectively. For both quantities, pronounced order-of-magnitude gaps among the leading eigenvalues indicate the existence of a low-dimensional active subspace, with dominant contributions concentrated in the first few directions.

\begin{figure*}[h]
\color{black}
    \centering
      \begin{subfigure}[t]{0.48\textwidth} 
        \centering
        \begin{overpic}[height=5cm]{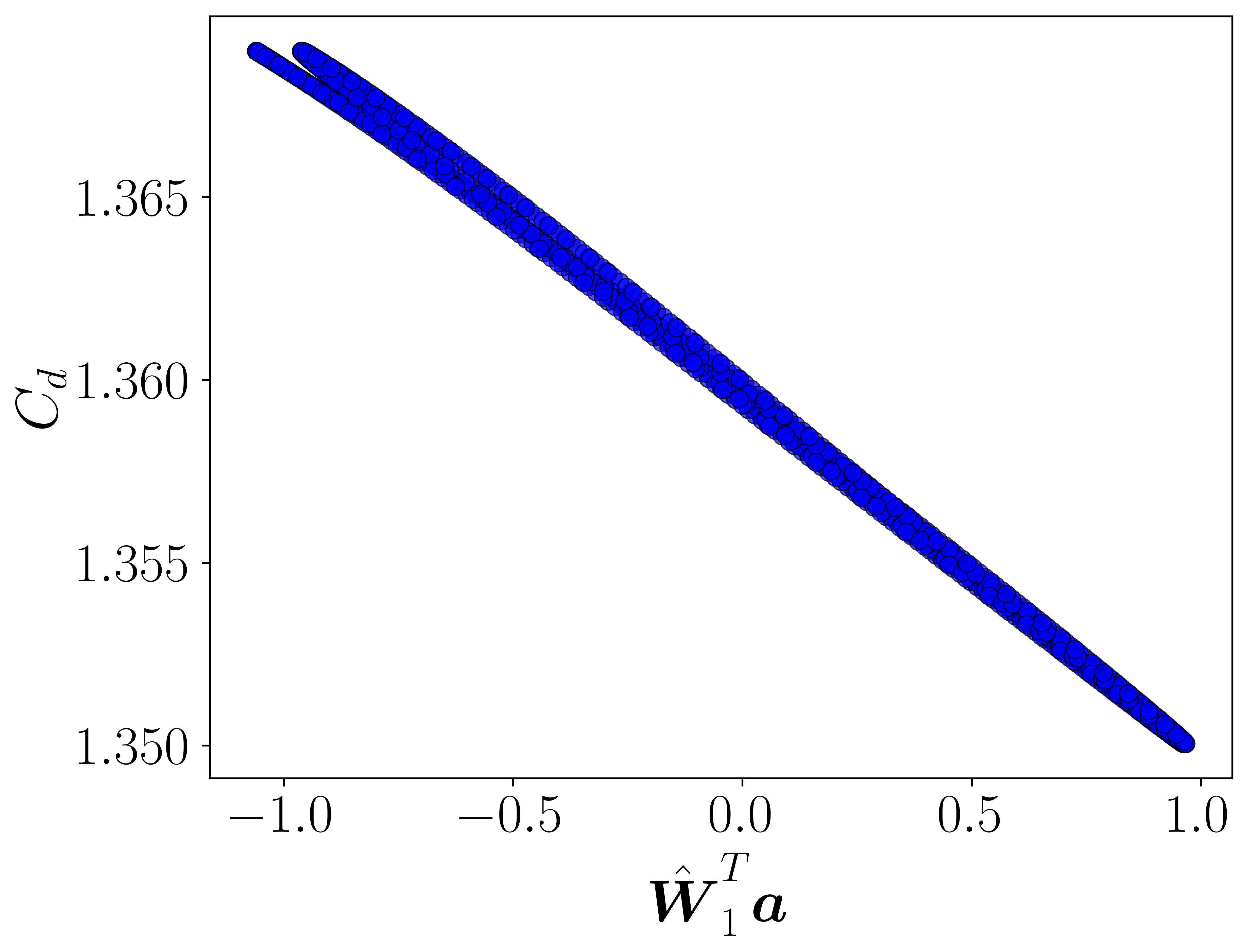}
            \put(2,140){\small (a)} 
        \end{overpic}
        \phantomcaption
        \label{subfig:case1_ssp1_cd} 
    \end{subfigure}
    \hfill
    \begin{subfigure}[t]{0.48\textwidth} 
        \centering
        \begin{overpic}[height=5cm]{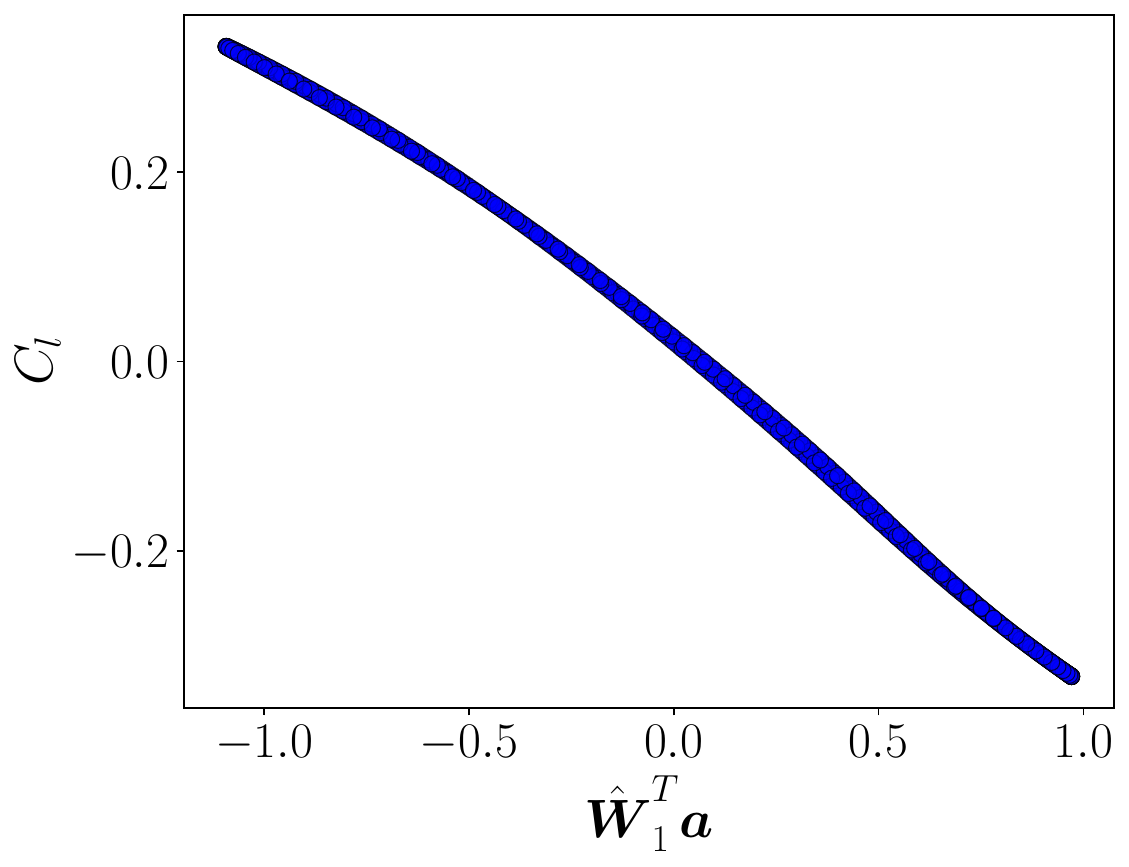} 
            \put(2,140){\small (b)}
        \end{overpic}
        \phantomcaption
        \label{subfig:case1_ssp1_cl} 
    \end{subfigure}

    \vspace{4mm}

    \begin{subfigure}[t]{0.48\textwidth} 
        \centering
        \begin{overpic}[height=5cm]{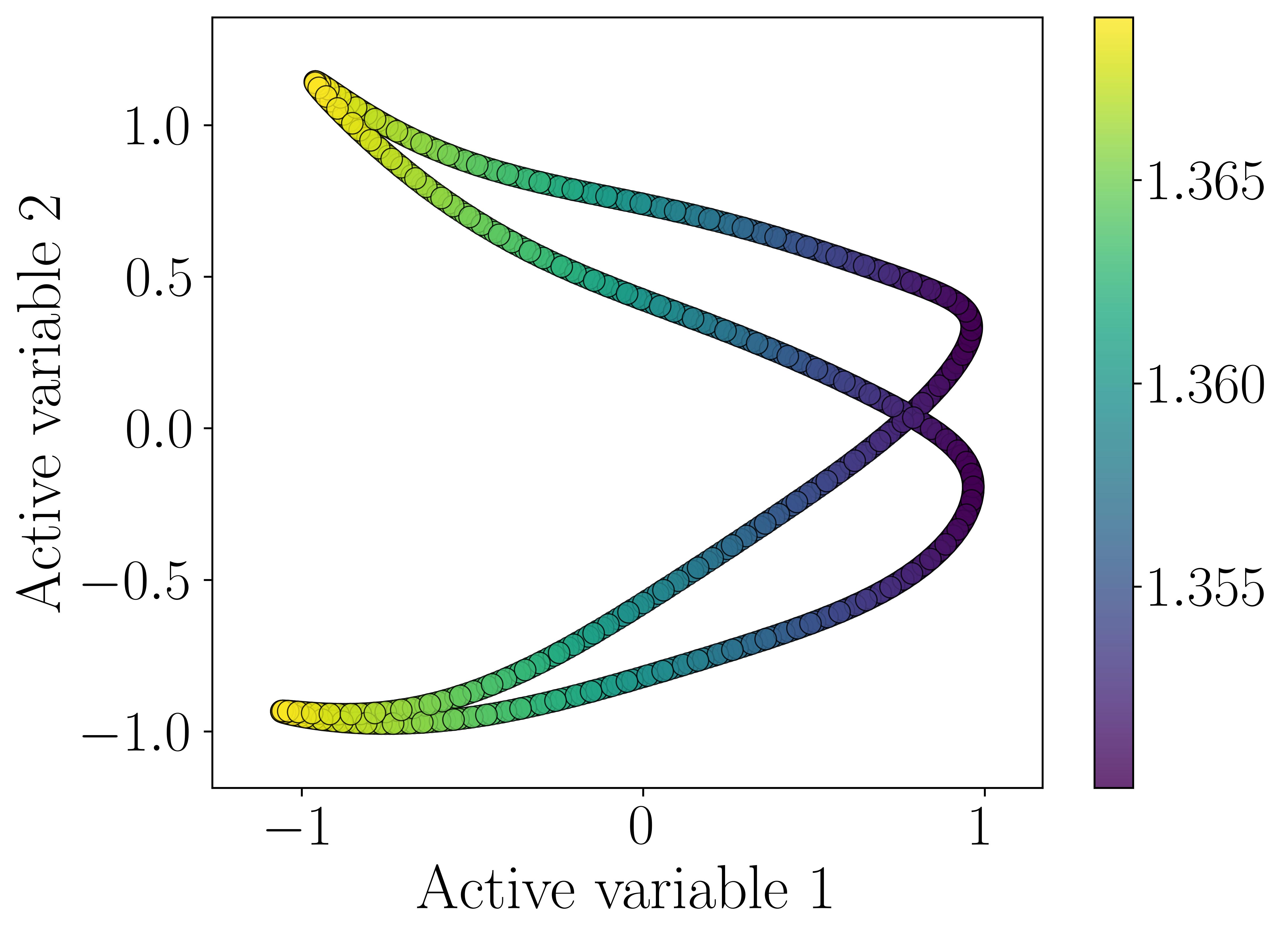} 
            \put(2,140){\small (c)} 
        \end{overpic}
        \phantomcaption
        \label{subfig:case1_ssp2_cd} 
    \end{subfigure}
    \hfill
    \begin{subfigure}[t]{0.48\textwidth} 
        \centering
        \begin{overpic}[height=5cm]{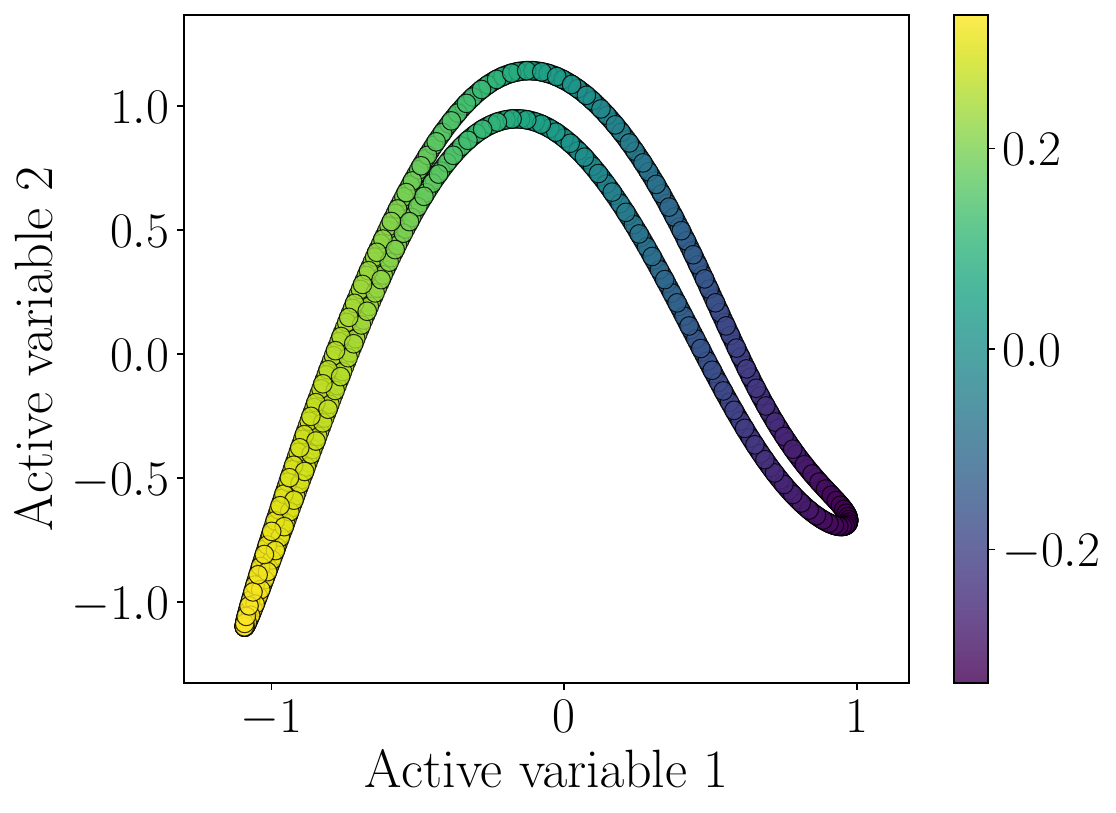}
            \put(2,140){\small (d)}
        \end{overpic}
        \phantomcaption
        \label{subfig:case1_ssp2_cl} 
    \end{subfigure}

    \caption{Sufficient summary plots for (a, c) $C_d$ and (b, d) $C_l$, using (a, b) one active variable and (c, d) two active variables for test case~1}
    \label{fig:fig:case1_ssp}
\end{figure*}

Figs.~\ref{subfig:case1_as_eigenvec_cd} and~\ref{subfig:case1_as_eigenvec_cl} present the components of the first two eigenvectors of $\hat{\boldsymbol{C}}$ for $C_d$ and $C_l$, respectively, which are utilized to generate one- and two-dimensional summary plots in Figs.~\ref{subfig:case1_ssp1_cd}--\ref{subfig:case1_ssp2_cl} to identify the appropriate dimension of the active subspace for each QoI. These summary plots, commonly referred to as sufficient summary plots\cite{cook2009regression}, employ scatter representations that incorporate all available regression information. As described in Eq.~\eqref{eq:active_variables}, Figs.~\ref{subfig:case1_ssp1_cd}--\ref{subfig:case1_ssp2_cl} depict $f(\boldsymbol{a})$ against active variable $\mathbf{y}=\hat{\boldsymbol{W}}_1^T\boldsymbol{a}$, where $\hat{\boldsymbol{W}}_1^T$ contains the leading one and two eigenvectors, for both $C_d$ and $C_l$. For both QoIs, the one-dimensional summary plots exhibit a gross monotonic trend in the output as a function of the ﬁrst active variable. Notably, this monotonic relationship is more distinct for $C_l$ than for $C_d$, suggesting that a one-dimensional active subspace may already be sufficient for representing the lift coefficient. Nonetheless, we prefer to retain more information about the QoI function by employing a two-dimensional active subspace for both QoIs. The two-dimensional summary plot captures more of the input/output relationship, where some curvature is apparent in the two-dimensional level sets. Consequently, an approximation of the form Eq.~\eqref{eq:rs_app} constructed with two eigenvectors may be more informative than one relying solely on the first eigenvector. This sort of qualitative reasoning is typical when the engineer is determining how to exploit the active subspace for a particular application.

\begin{figure*}[!ht]
\color{black}
	\centering

	\begin{subfigure}[t]{0.48\textwidth} 
		\centering
		\begin{overpic}[height=5.5cm]{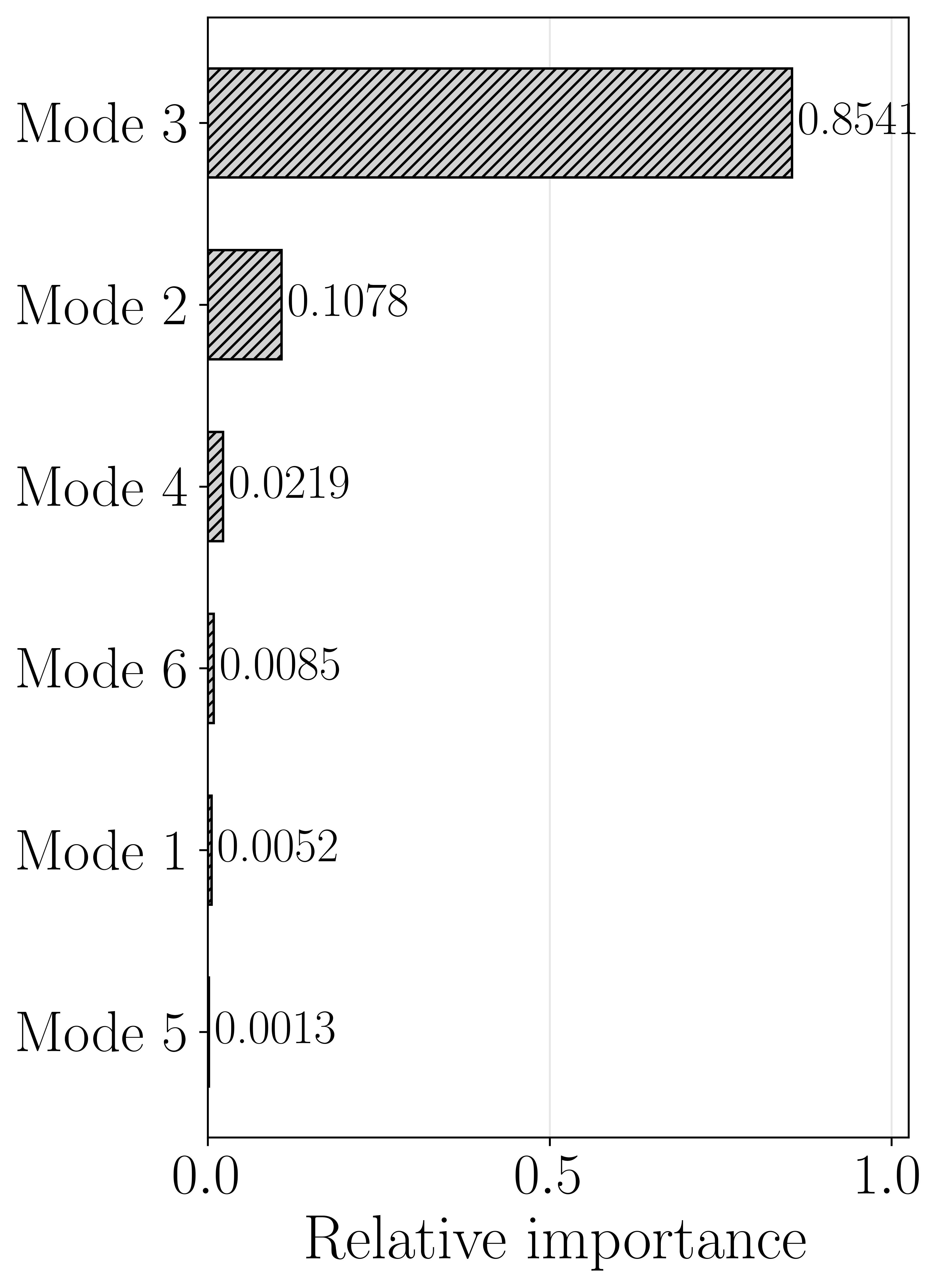}
			\put(2,155){\small (a)} 
		\end{overpic}
		\phantomcaption
		\label{subfig:case1_importance_cd} 
	\end{subfigure}
	\hfill
	\begin{subfigure}[t]{0.48\textwidth} 
		\centering
		\begin{overpic}[height=5.5cm]{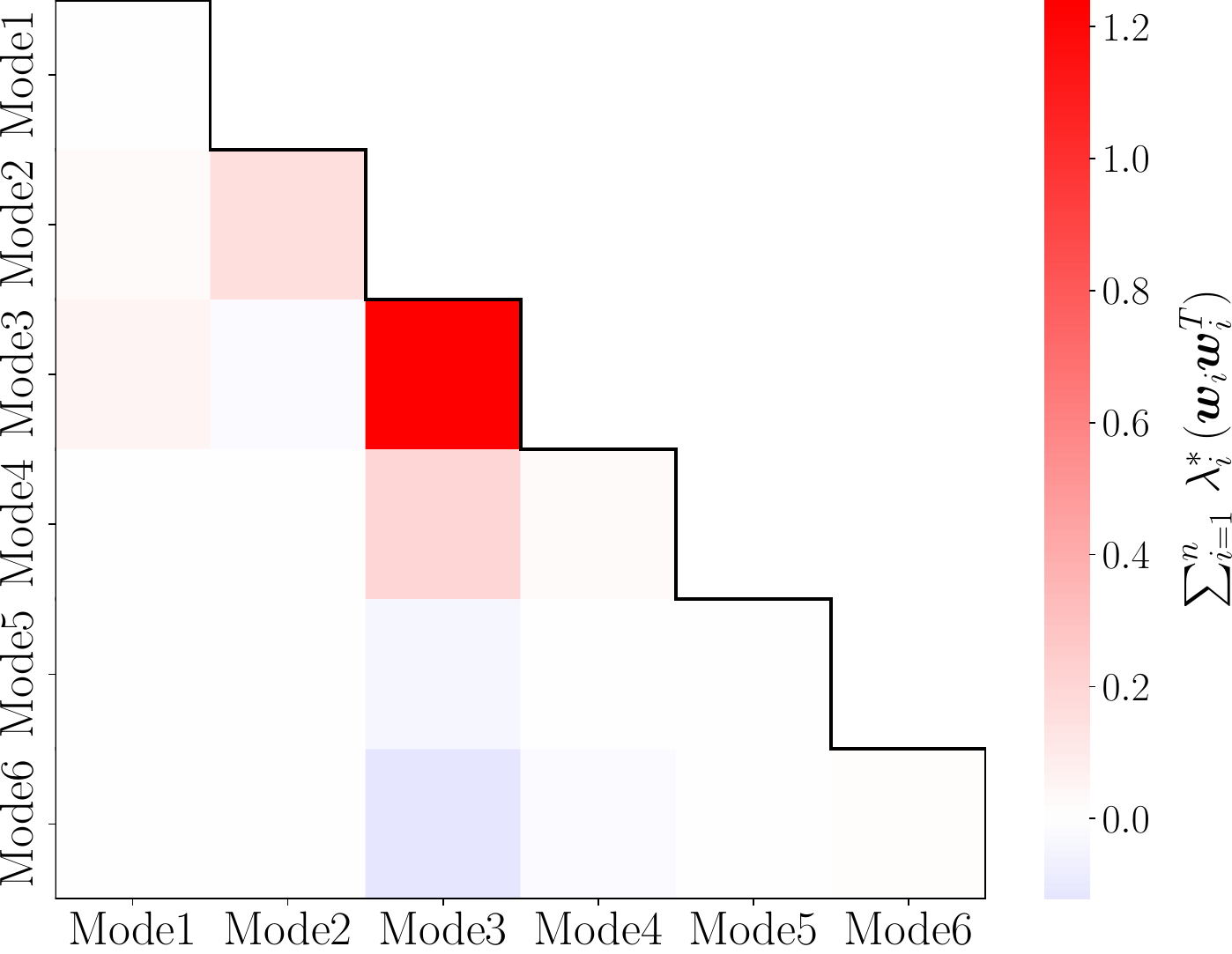}
			\put(-10,155){\small (b)}
		\end{overpic}
		\phantomcaption
		\label{subfig:case1_importance_heat_cd} 
	\end{subfigure}

	\vspace{4mm}

	\begin{subfigure}[t]{0.48\textwidth} 
		\centering
		\begin{overpic}[height=5.5cm]{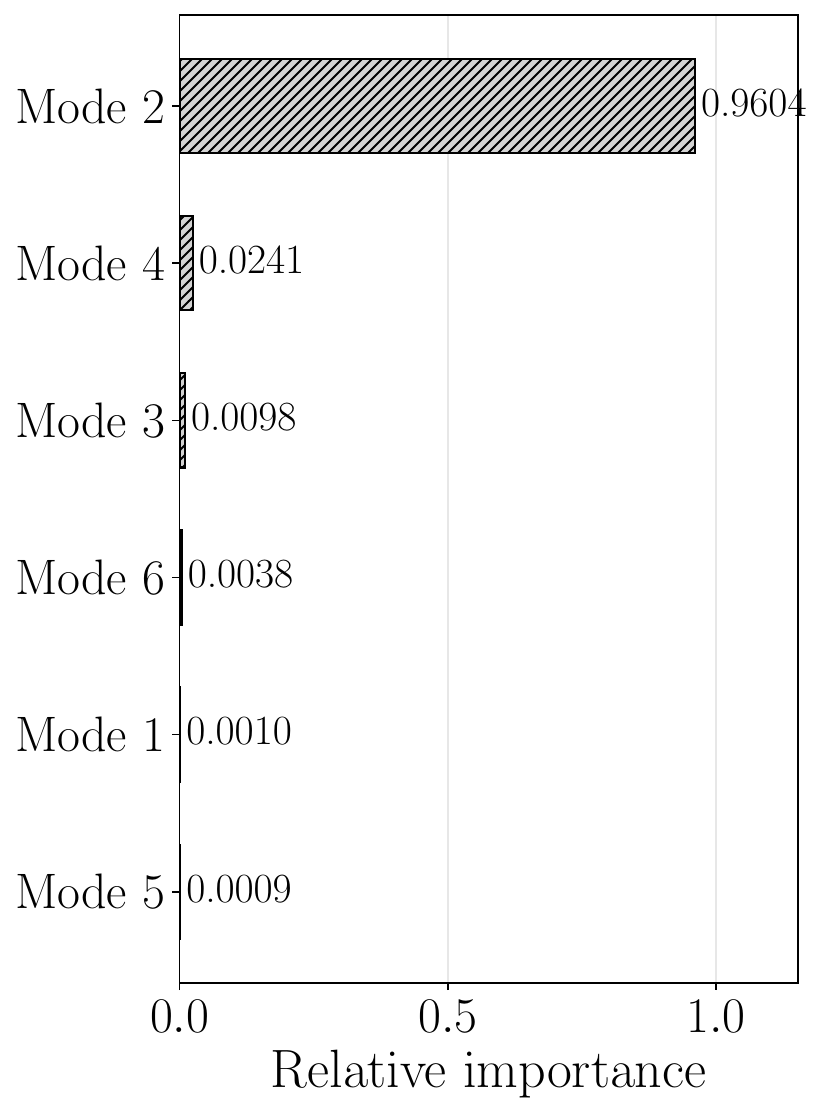} 
			\put(2,155){\small (c)} 
		\end{overpic}
		\phantomcaption
		\label{subfig:case1_importance_cl} 
	\end{subfigure}
	\hfill
	\begin{subfigure}[t]{0.48\textwidth} 
		\centering
		\begin{overpic}[height=5.5cm]{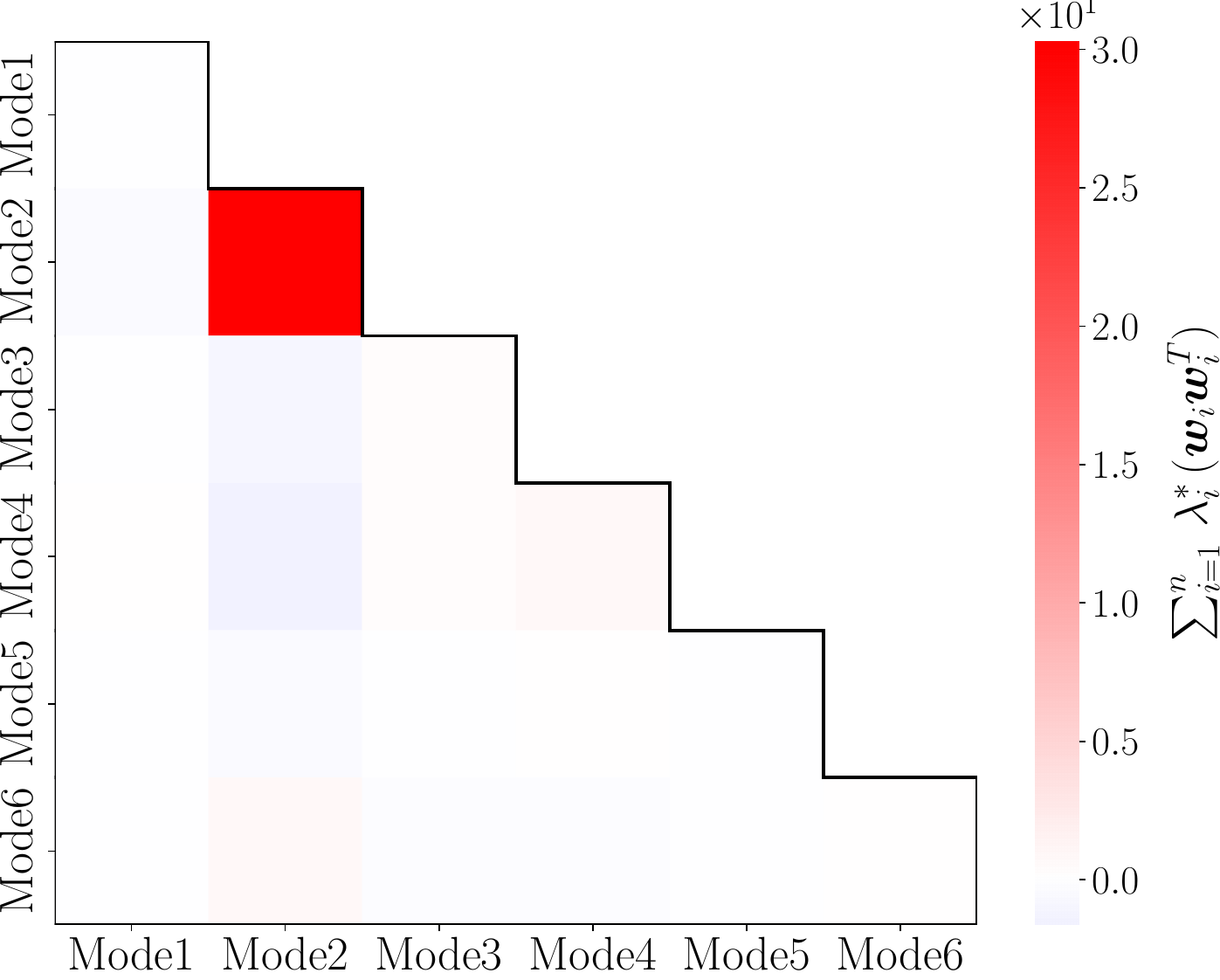} 
			\put(-10,155){\small (d)}
		\end{overpic}
		\phantomcaption
		\label{subfig:case1_importance_heat_cl} 
	\end{subfigure}

	\caption{Global sensitivity analysis of POD coefficients with respect to (a, b) $C_d$ and (c, d) $C_l$ for test case~1. (a, c) Normalized activity scores for the top six coefficients; (b, d) pairwise interaction matrices of the leading active subspace directions, where diagonal entries show individual sensitivities and off-diagonal entries indicate correlations (red for synergistic amplification, blue for antagonistic reduction).}
	\label{fig:case1_importance}
\end{figure*}

\begin{table*}[h]
\color{black}
	\centering
    \small
	\begin{tabular}{cc@{\hspace{1cm}}cc} 
		\toprule
		\multicolumn{2}{c}{$C_d$} & \multicolumn{2}{c}{$C_l$} \\
		\cmidrule(r){1-2} \cmidrule(l){3-4}
		Minimum retained influence [\%] & $N_{\text{POD}}$ & Minimum retained influence [\%] & $N_{\text{POD}}$ \\
		\midrule
		85    & 1 & 96     & 1 \\
		96    & 2 & 98     & 2 \\
		98    & 3 & 99     & 3 \\
		99    & 4 & 99.9   & 5 \\
		99.9  & 6 & 99.99  & 6 \\
		99.99 & 9 & 99.999 & 9 \\
		\bottomrule
	\end{tabular}
	\caption{Cumulative retained influence of POD modes on $C_d$ and $C_l$ for test case~1.}
    \label{tab:case1_as_influence_retained}
\end{table*}

To elucidate the global sensitivity of the POD coefficients to both $C_d$ and $C_l$, normalized activity scores were evaluated using a two-dimensional active subspace ($N_{\text{AS}} = 2$) for each QoI. It is worth noting that the activity scores adopted here as the sensitivity metric, rather than simpler alternatives such as Pearson correlation coefficients or standardized regression coefficients. Constantine \& Diaz~\cite{constantine2017global} have shown that activity score rankings are broadly consistent with those from standard sensitivity metrics for well-behaved engineering models, while different sensitivity metrics measure different characteristics of the function, discrepancies between metrics can arise when the mapping between the QoI and the input variables exhibits nonlinear or non-monotonic behaviour. This distinction is further substantiated by the quantitative ranking comparison presented in Appendix~\ref{app:correlation_metrics}. Beyond ranking, the AS framework additionally identifies directions of linear combinations of input parameters (i.e., active variables) that most strongly drive the model response. By exploiting these important directions, the dimension of the model's parameter space can be significantly reduced, facilitating the construction of high-fidelity surrogates and extensive parameter studies that are otherwise numerically prohibitive~\cite{constantine2017global}. This enables the AS framework to capture nonlinear and coupled contributions between POD coefficients and the QoI that are structurally inaccessible to univariate correlation or regression-based metrics, where each parameter is assessed independently without accounting for inter-parameter synergy. 
Figs.~\ref{subfig:case1_importance_cd} and~\ref{subfig:case1_importance_cl} depict these scores for the primary POD coefficients for $C_d$ and $C_l$, respectively. The total influence of the retained number of POD modes on $C_d$ and $C_l$ is summarised in Table~\ref{tab:case1_as_influence_retained}.
For $C_d$, Mode 3 emerges as the sensitivity-dominant parameter, contributing approximately $85\%$, followed by Mode 2 with about $11\%$, while higher-order modes have a negligible influence. For $C_l$, Mode 2 emerges as the most dominant parameter, accounting for approximately $96\%$ of the total sensitivity, with Mode 4 as the secondary contributor at approximately $2\%$, while the remaining modes have negligible influence. For both QoIs, the sensitivity ranking aligns well with the structure of the active subspace presented in Figs.~\ref{subfig:case1_as_eigenvec_cd} and~\ref{subfig:case1_as_eigenvec_cl}, where the largest components of the leading eigenvectors correspond to the most influential modes identified by the activity scores.
The dominance of Mode 3 further reflects its preferential alignment with the steepest descent direction of $C_d$ in parameter space, such that small perturbations in $a_3$ induce disproportionately large variations in drag. 
This underscores the physical significance of sensitivity-dominant modes over energetic ones, which encapsulates flow features such as localized vortical structures or shear instabilities that significantly modulate the pressure distribution and viscous stresses, thereby exerting a substantial influence on $C_d$. In contrast, more energetic modes such as Mode 1 contribute less, as their dominant structures correspond to quasi-steady or bulk oscillatory motions with a weaker impact on $C_d$. Mode 2, emphasized in $\boldsymbol{w}_2$, contributes as the secondary direction of sensitivity, consistent with its association to large-scale coherent motions that interact with the shear layer and modulate drag in a less dominant but still significant manner. The pronounced dominance of Mode 2 for $C_l$ is consistent with its well-established role as the primary carrier of the large-scale vortex-shedding dynamics, whose oscillatory motion is directly coupled to the periodic lift generation mechanism. This analysis highlights that the importance of POD modes does not scale monotonically with their energy content for either QoI. Instead, their sensitivity is governed by the specific coupling between modal structures and the respective force-producing mechanisms. It is worth noting that, under the same flow condition, $C_d$ and $C_l$ are simultaneously generated, yet the AS analysis reveals that their sensitivity-dominant modes are distinctly different: mode 3 governs $C_d$ while mode 2 governs $C_l$. This raises a practically important question regarding modal selection when both QoIs are of interest. In the present case, however, this question admits a remarkably compact answer: retaining only modes 2, 3, and 4 is sufficient to accurately reconstruct both QoIs simultaneously, since the active subspace for $C_d$ is spanned by modes 3 and 2, while that for $C_l$ is spanned by modes 2 and 4. The union of these two sets yields a combined basis of just three modes, demonstrating that the AS framework enables a highly efficient dual-QoI surrogate with minimal modal redundancy. This further underscores the advantage of sensitivity-driven modal selection over energy-based truncation, where a much larger number of modes would be retained without necessarily improving QoI reconstruction accuracy.

Figs.~\ref{subfig:case1_importance_heat_cd} and~\ref{subfig:case1_importance_heat_cl} delineate the pairwise interactions among POD coefficients encoded in the matrix $\mathbf{M}=\sum_{i=1}^{n}\hat{\lambda}_i (\hat{\boldsymbol{w}}_i \hat{\boldsymbol{w}}_i^T)$, where the diagonal elements recover the individual activity scores, consistent with Figs.~\ref{subfig:case1_importance_cd} and~\ref{subfig:case1_importance_cl}, while the off-diagonal entries quantify second-order modal couplings. Positive off-diagonal values (red) indicate synergistic interactions, whereby concurrent perturbations in two modes reinforce variations in the QoI, whereas negative values (blue) signify antagonistic interactions in which modal contributions counteract each other, thereby reducing the fluctuations of QoI. For $C_d$, Modes 3 and 4 exhibit a comparatively strong positive correlation, suggesting that in-phase variations of their associated flow structures synergistically enhance drag. Conversely,  Modes 3 and 6 display a pronounced negative correlation, indicating that the effect of Mode 6 tends to counterbalance the dominant drag contribution of Mode 3. These interaction patterns reinforce the single-mode sensitivity analysis, highlighting that Mode 3 is the dominant driver of $C_d$ variability, with its effects either amplified by Mode 4 or moderated by Mode 6. For $C_l$, the interaction matrix is markedly sparser, with Mode 2 overwhelmingly dominating the diagonal, and a pronounced antagonistic coupling observed between Modes 2 and 4, indicating that Mode 4 tends to attenuate the lift variations driven by Mode 2. Notably, in both cases, the QoI variations are governed by a sparse subset of sensitivity-dominant parameters and their nonlinear couplings, providing a high-fidelity basis for constructing reduced-order models that transcend simple modal energy considerations.

\paragraph{Construction of response surface surrogate models}

\begin{figure*}[h]
\color{black}
    \centering

    \begin{subfigure}[t]{0.3\textwidth} 
    \centering
        \begin{overpic}[height=3.5cm]{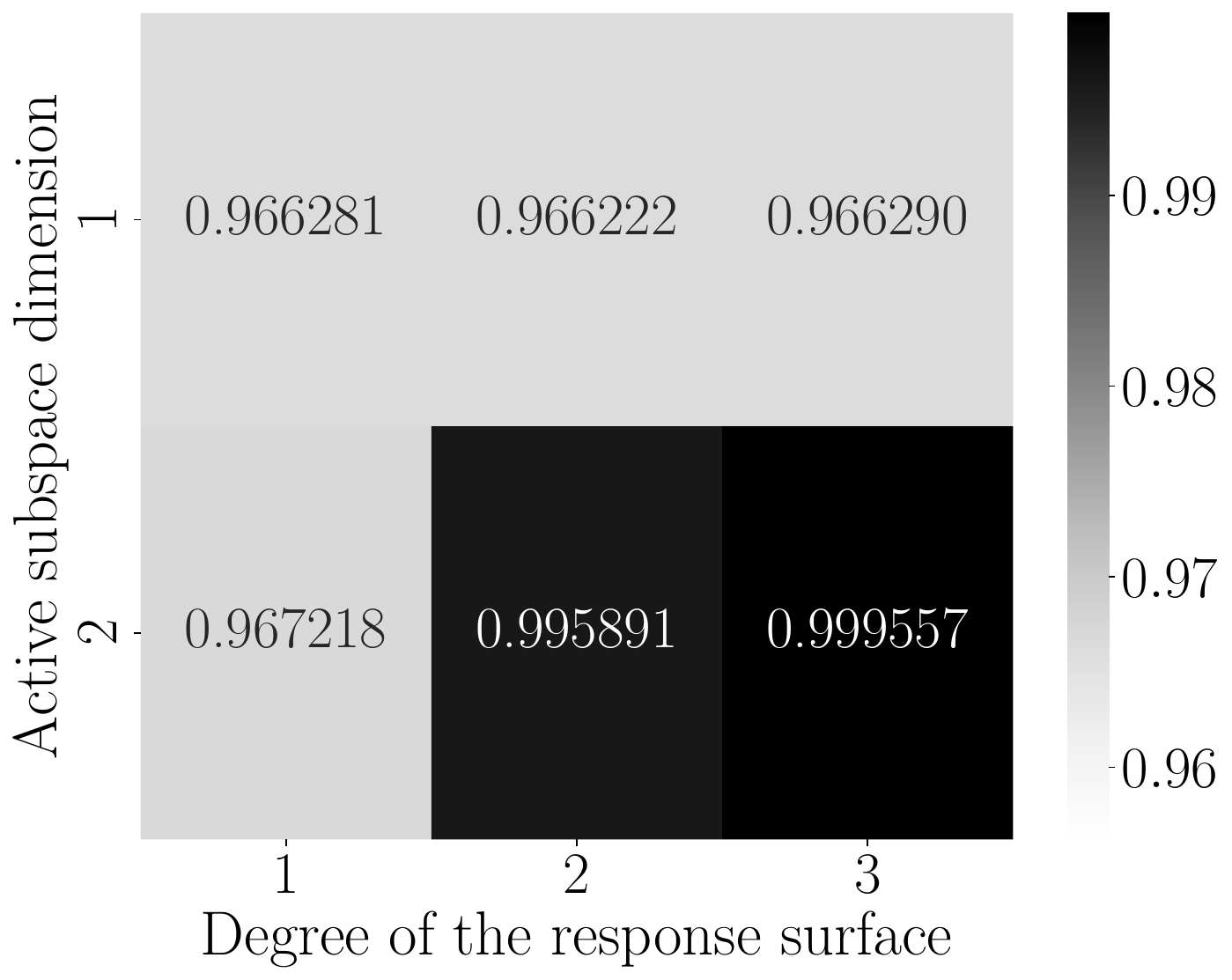}
            \put(2,100){\small (a)} 
        \end{overpic}
        \phantomcaption
        \label{subfig:case1_rs_heat_cd} 
    \end{subfigure}
    \hfill
    \begin{subfigure}[t]{0.3\textwidth} 
    \centering
        \begin{overpic}[height=3.5cm]{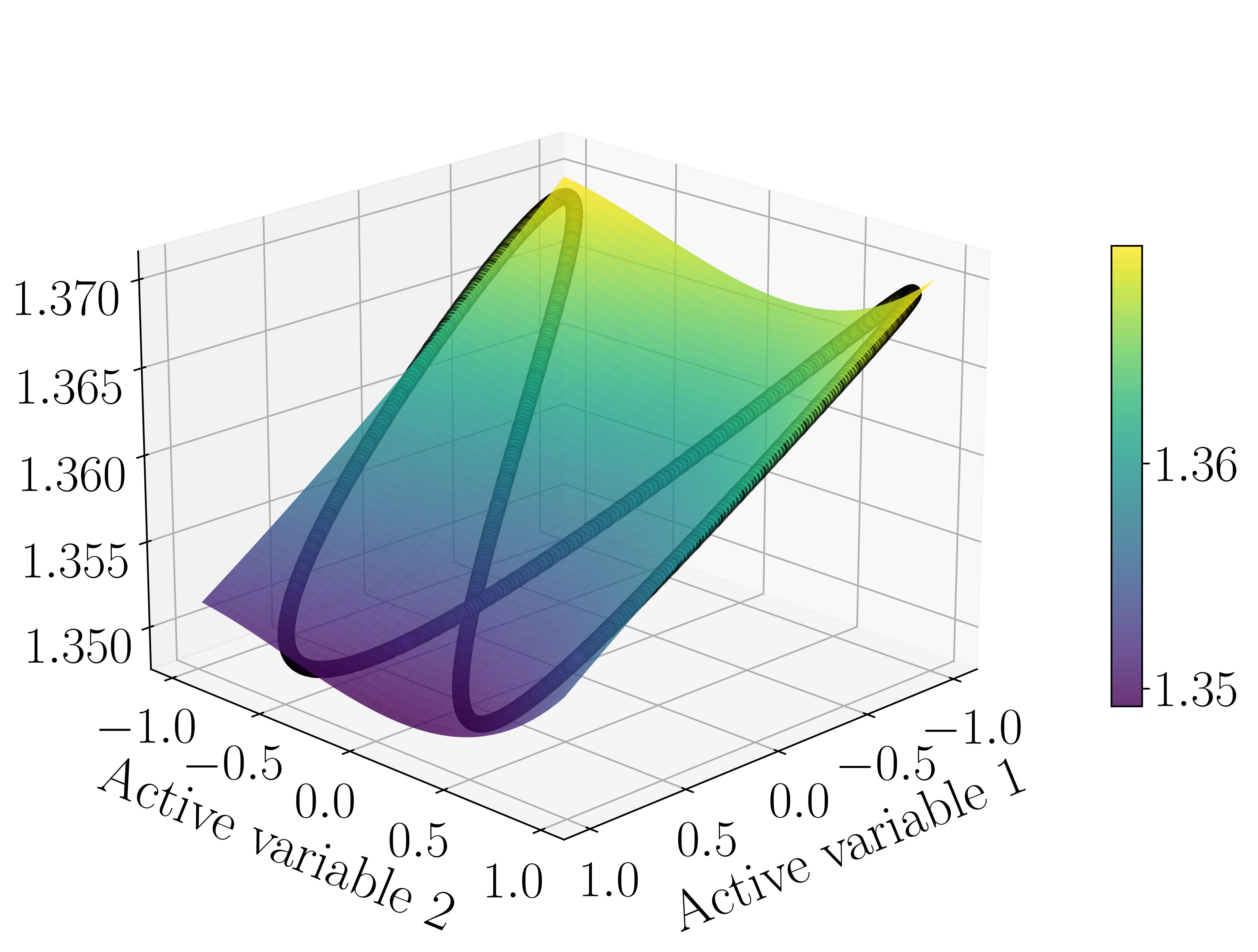}
            \put(2,100){\small (b)}
        \end{overpic}
        \phantomcaption
        \label{subfig:case1_rs_3d_cd} 
    \end{subfigure}
    \hfill
    \begin{subfigure}[t]{0.3\textwidth} 
    \centering
        \begin{overpic}[height=3.5cm]{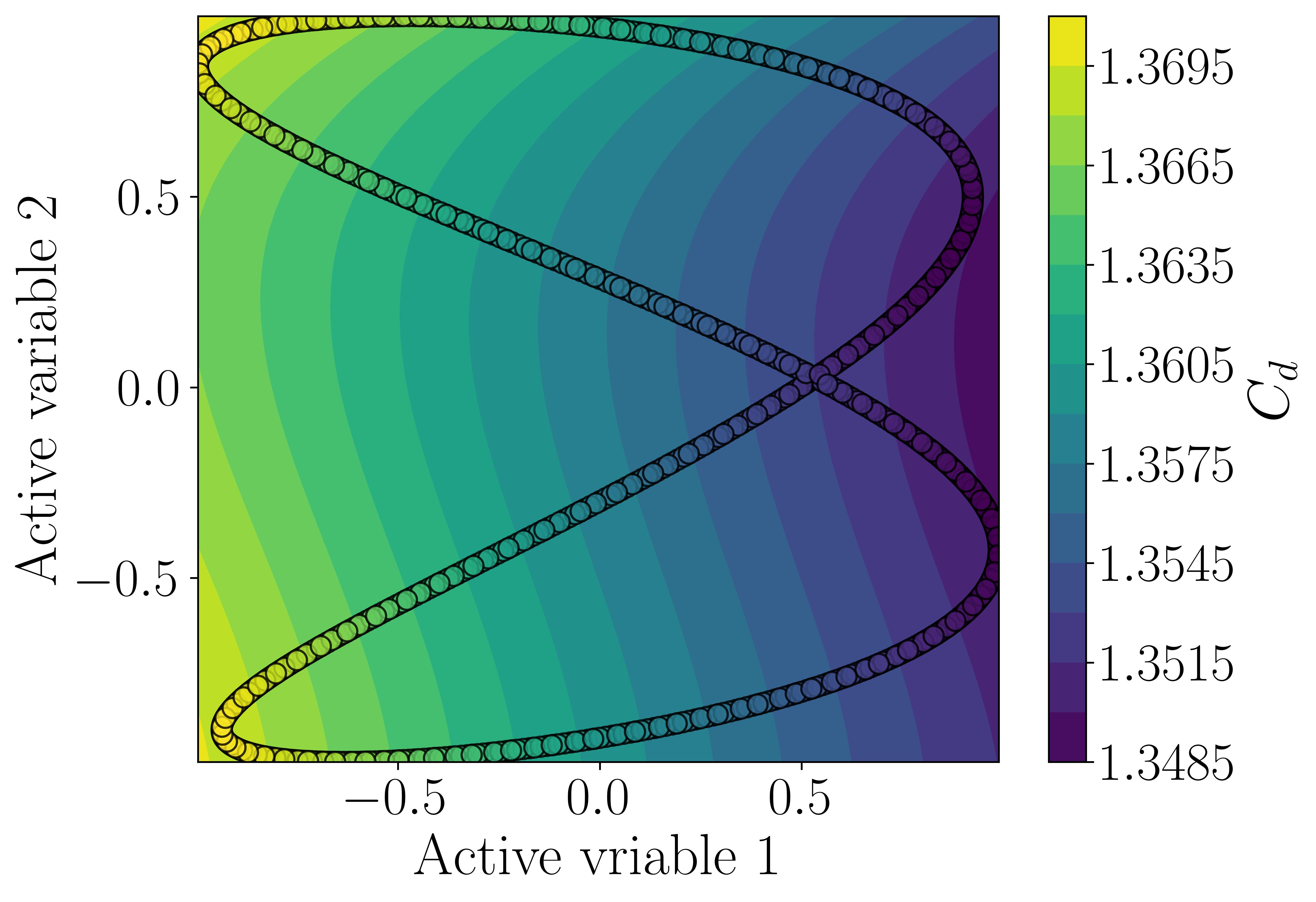}
            \put(2,100){\small (c)}
        \end{overpic}
        \phantomcaption
        \label{subfig:case1_rs_contour_cd} 
    \end{subfigure}

    \vspace{4mm} 
    
    \begin{subfigure}[t]{0.3\textwidth} 
    \centering
        \begin{overpic}[height=3.5cm]{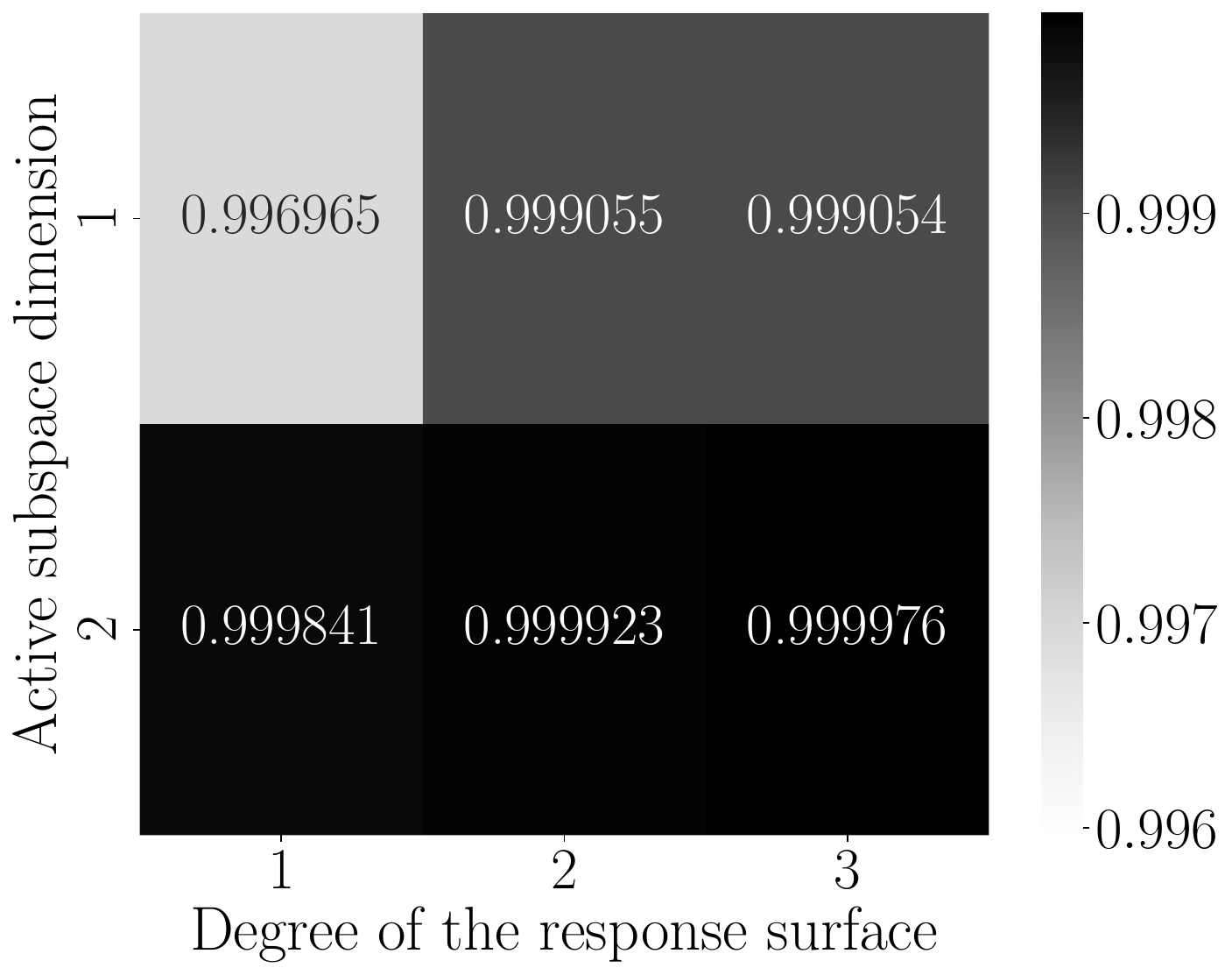} 
            \put(2,100){\small (d)} 
        \end{overpic}
        \phantomcaption
        \label{subfig:case1_rs_heat_cl} 
    \end{subfigure}
    \hfill
    \begin{subfigure}[t]{0.3\textwidth} 
    \centering
        \begin{overpic}[height=3.5cm]{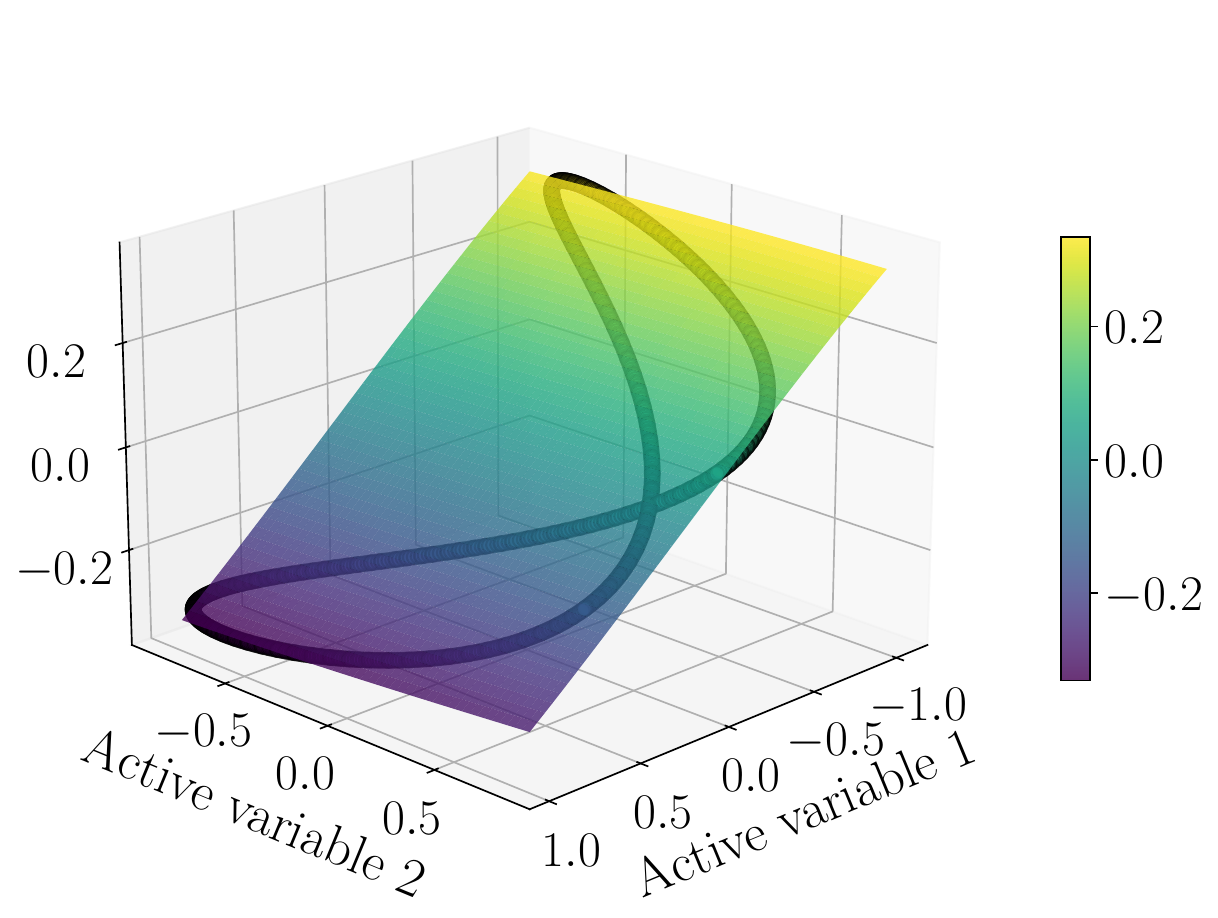}
            \put(2,100){\small (e)}
        \end{overpic}
        \phantomcaption
        \label{subfig:case1_rs_3d_cl} 
    \end{subfigure}
    \hfill
    \begin{subfigure}[t]{0.3\textwidth} 
    \centering
        \begin{overpic}[height=3.5cm]{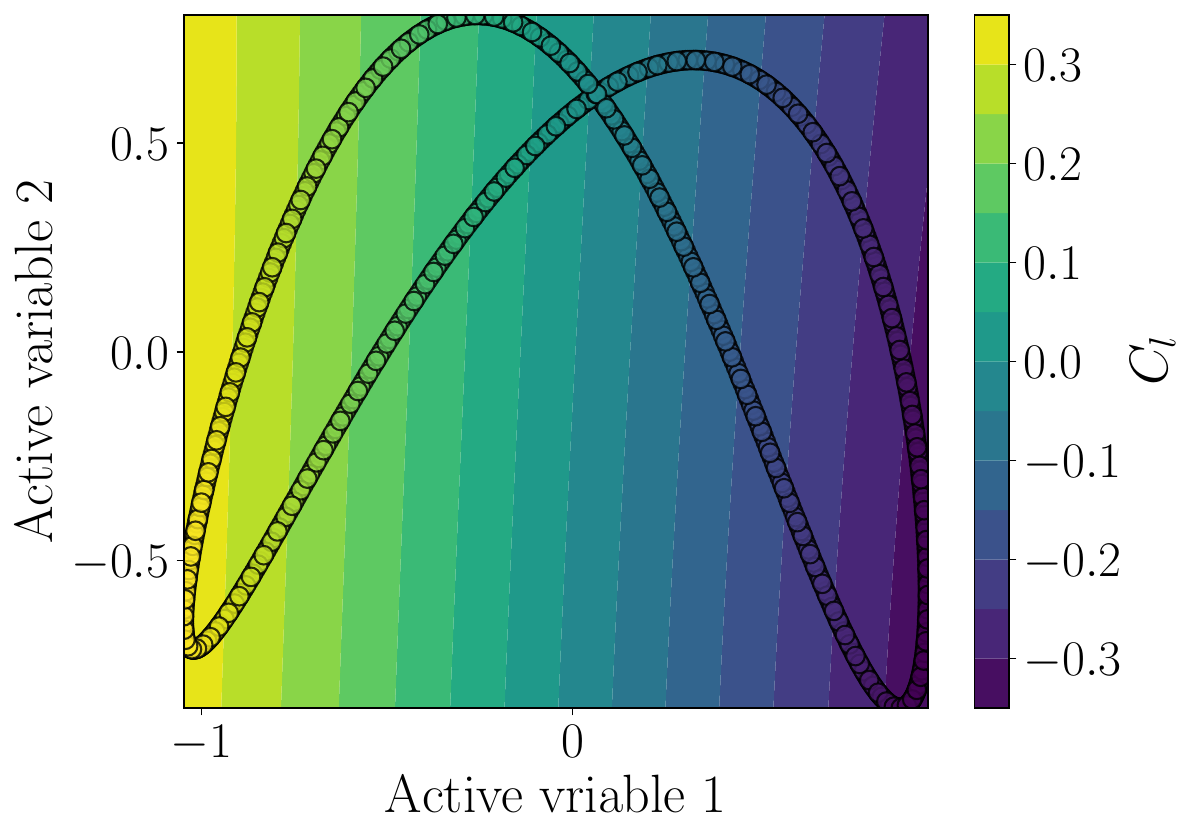}
            \put(2,100){\small (f)}
        \end{overpic}
        \phantomcaption
        \label{subfig:case1_rs_contour_cl} 
    \end{subfigure}

    \caption{Response surface construction and evaluation for reconstructing (a--c) $C_d$ and (d--f) $C_l$. (a, d) Heatmaps of $R^2$ values for different combinations of active subspace dimensions and polynomial orders; (b, e) three-dimensional response surfaces with respect to the two active variables; (c, f) corresponding two-dimensional contour plots (iso-contours).}
    \label{fig:case1_rs}
\end{figure*}

Based on the preceding AS analysis, the response surfaces for both $C_d$ and $C_l$ were constructed using active variables, derived from the respective sensitivity-dominant POD coefficients (Modes 3 and 2 for $C_d$ and Modes 2 and 4 for $C_l$). This strategy effectively reduces the input dimension while preserving the governing variability of the QoI. Following the procedure described in Eq.~\eqref{eq:n_as_n_p}, we perform a grid search over different active subspace dimensions and polynomial orders. The $R^2$ values reported in the heatmaps are evaluated on the independent chronological test set, reflecting the surrogate's generalization capability to active variables not seen during model construction. Figs.~\ref{subfig:case1_rs_heat_cd} and \ref{subfig:case1_rs_heat_cl} present the heatmaps of $R^2$ values obtained from the grid search. For both QoIs, even a single-dimensional active subspace ($N_{\text{AS}}=1$) achieves high reconstruction fidelity, confirming that the dominant physics of the cylinder flow can be represented by a very low-dimensional manifold. For $C_l$, the accuracy shows a slight improvement as the polynomial order increases from 1 to 2, while $C_d$ remains relatively stable. To further capture the remaining variance and achieve near-unity fidelity, the active subspace is extended to two dimensions ($N_{\text{AS}}=2$). This expansion, combined with a third-order polynomial, enhances the model's approximation accuracy, effectively capturing the complex variations of the QoI. Thus, the $(N_{\text{AS}},p)=(2,3)$ is identified, providing a superior balance between computational efficiency and reconstruction accuracy for the sensitivity-driven ROM.

 To further illustrate this result, Figs.~\ref{subfig:case1_rs_3d_cd}--\ref{subfig:case1_rs_contour_cl} depict the constructed response surface in the two-dimensional active subspace. Figs.~\ref{subfig:case1_rs_3d_cd} and \ref{subfig:case1_rs_3d_cl} show the three-dimensional surface of the reconstructed $C_d$ and $C_l$, respectively, with respect to the first and second active variables, while Figs.~\ref{subfig:case1_rs_contour_cd} and \ref{subfig:case1_rs_contour_cl} present the corresponding two-dimensional contour plot, showcasing projected iso-contours. These results collectively demonstrate that the surrogate model yields a smooth and high-fidelity representation of the aerodynamic coefficient variations, capturing the underlying manifold governed by the sensitivity-dominant parameters, thereby highlighting the efficacy of the active subspace in representing the input-output relationship.

\paragraph{Accuracy assessment of the POD--AS--PRS framework}

\begin{figure*}[h]

\color{black}
	\centering
	\begin{subfigure}[t]{0.57\textwidth} 
		\centering
		\begin{overpic}[height=4cm]{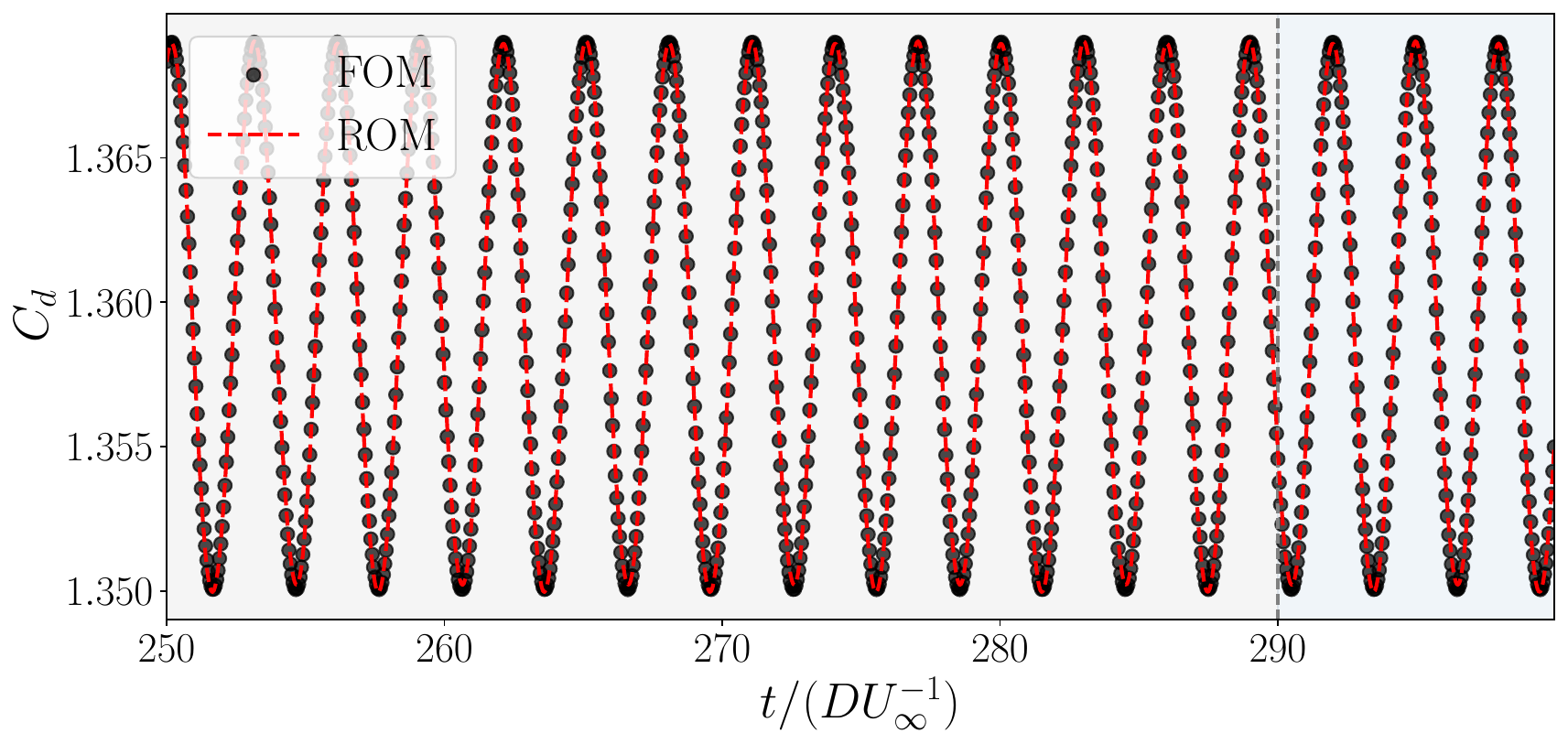}
			\put(2,120){\small (a)} 
		\end{overpic}
		\phantomcaption
		\label{subfig:case1_rom_fom_cd_time} 
	\end{subfigure}
	\hfill
	\begin{subfigure}[t]{0.36\textwidth} 
		\centering
		\begin{overpic}[height=4cm]{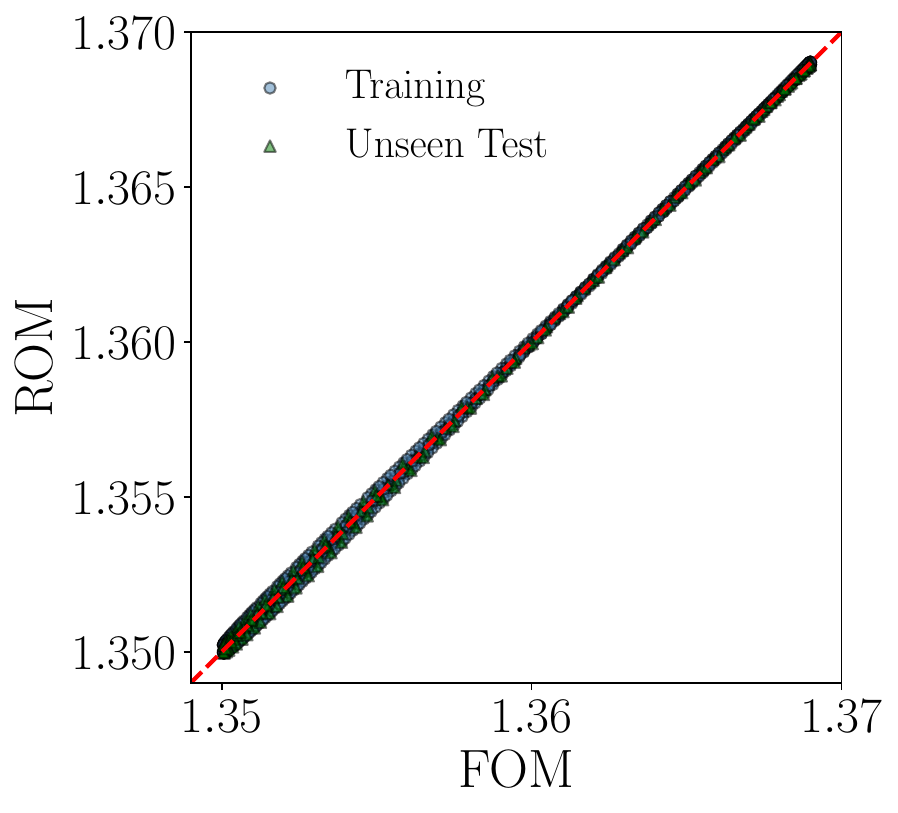}
			\put(2,120){\small (b)}
		\end{overpic}
		\phantomcaption
		\label{subfig:case1_rom_fom_cd_scatter} 
	\end{subfigure}

	\vspace{4mm}

	\begin{subfigure}[t]{0.57\textwidth} 
		\centering
		\begin{overpic}[height=4cm]{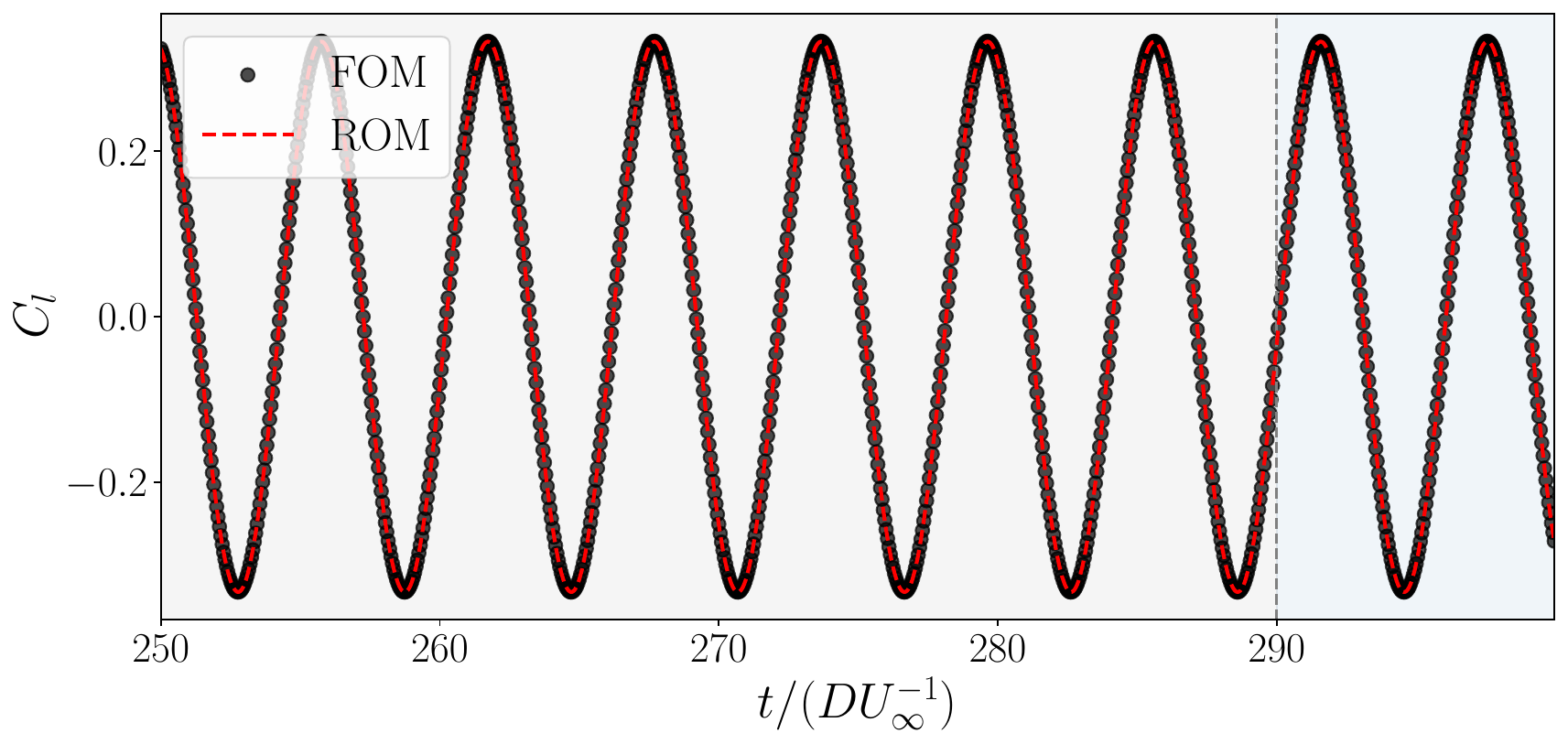} 
			\put(2,120){\small (c)} 
		\end{overpic}
        \phantomcaption
		\label{subfig:case1_rom_fom_cl_time} 
	\end{subfigure}
	\hfill
	\begin{subfigure}[t]{0.36\textwidth} 
		\centering
		\begin{overpic}[height=4cm]{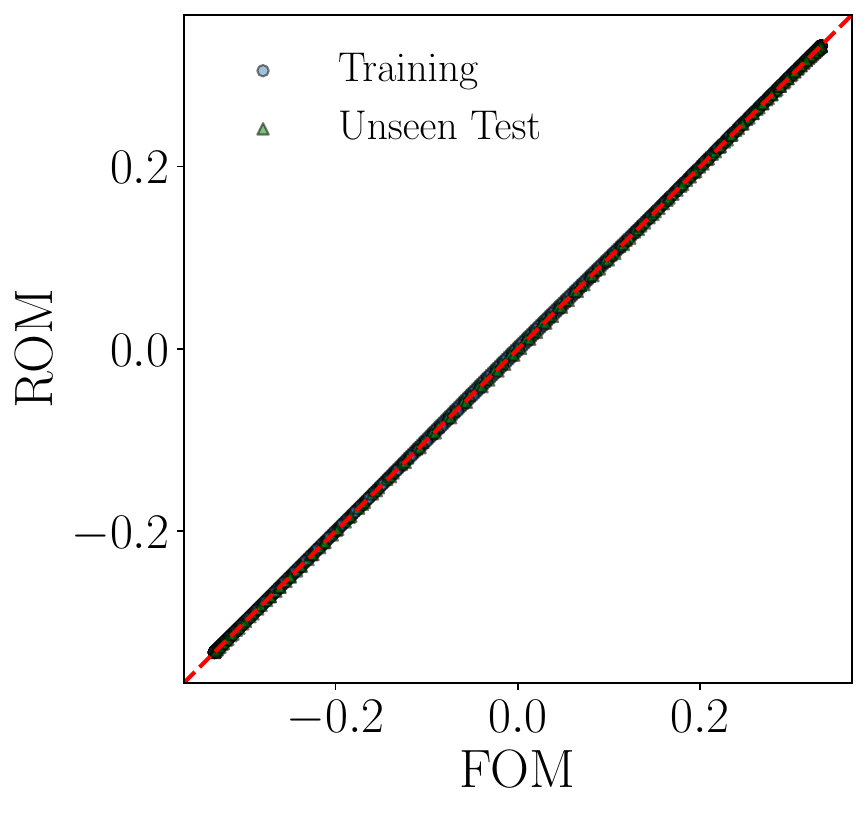} 
			\put(2,120){\small (d)}
		\end{overpic}
        \phantomcaption
		\label{subfig:case1_rom_fom_cl_scatter} 
	\end{subfigure}

	\caption{Comparison of ROM results and FOM for (a, b) $C_d$ and (c, d) $C_l$. (a, c) Temporal evolution over normalized time, where the gray-shaded and steelblue-shaded regions represent the training and test sets, respectively; FOM is represented by discrete black points and ROM reconstructions by red dashed lines. (b, d) Scatter plots of ROM results against FOM for both training (circles) and test (triangles) datasets, illustrating the high fidelity and generalization capability of the POD--AS--PRS framework.}
	\label{fig:case1_rom_fom}
\end{figure*}

The pre-constructed low-dimensional surrogate model is employed to evaluate the aerodynamic coefficient, $C_d$ and $C_l$, leveraging the two-dimensional active subspace to compress the input space from the full POD coefficient to the identified sensitivity-dominant directions. Fig.~\ref{fig:case1_rom_fom} provides a comprehensive comparison between the ROM results and the FOM ground truth, covering both the training interval and the independent chronological test interval. The temporal evolution of $C_d$ and $C_l$, shown in Figs.~\ref{subfig:case1_rom_fom_cd_time} and \ref{subfig:case1_rom_fom_cl_time}, respectively, demonstrates that the ROM closely tracks the FOM across both the training and independent test intervals, capturing the dynamic behavior with high accuracy throughout the entire time domain. Furthermore, the parity plot in Figs.~\ref{subfig:case1_rom_fom_cd_scatter} and \ref{subfig:case1_rom_fom_cl_scatter} show near-perfect alignment along the diagonal for both training and test data points, confirming the strong correlation between the sensitivity-driven surrogate and the full-order simulations.

Quantitative performance metrics for both training and test datasets are summarized in Table~\ref{tab:case1_rom_performance}. The $R^2$ values for both coefficients are remarkably high across both datasets, confirming that the surrogate model explains nearly all the variance in the FOM data and generalizes well to time instants not seen during model construction, while the low MAE and RMSE values across both intervals underscore the precision and robustness of the framework. It is worth noting that while $C_l$ exhibits higher MAE and RMSE values compared to $C_d$, its $R^2$ is notably closer to unity. This is attributed to the larger fluctuation amplitude of the lift coefficient relative to the drag coefficient. The higher absolute errors in $C_l$ are a direct consequence of its greater operational range, yet the near-perfect $R^2$ across both training and test datasets demonstrates the exceptional ability of the POD--AS--PRS framework to capture the full scaled dynamics of the lift oscillations. These results demonstrate that for periodic bluff-body flows, identifying the sensitivity-dominant parameters enables the construction of highly robust ROMs that bypass the redundancies of purely energy-based modal selection.

\begin{table*}[h]
\color{black}
\centering
\small
\begin{tabular}{cccccc}
\toprule
Model & QoI & Dataset & $R^2$ & MAE & RMSE \\
\midrule
\multirow{4}{*}{POD--AS--PRS} & \multirow{2}{*}{$C_d$} & Train & 0.999597 & $9.40 \times 10^{-5}$ & $1.35 \times 10^{-4}$ \\
\cmidrule{3-6}
                              &                        & Test  & 0.999557 & $1.01 \times 10^{-4}$ & $1.43 \times 10^{-4}$ \\
\cmidrule{2-6}
                              & \multirow{2}{*}{$C_l$} & Train & 0.999980 & $7.23 \times 10^{-4}$ & $1.04 \times 10^{-3}$ \\
\cmidrule{3-6}
                              &                        & Test  & 0.999976 & $7.67 \times 10^{-4}$ & $1.09 \times 10^{-3}$ \\
\bottomrule
\end{tabular}
\caption{Performance metrics of the POD--AS--PRS framework on training and test datasets for the cylinder flow at $Re=100$.}
\label{tab:case1_rom_performance}
\end{table*}

\subsection{Case~2: Airfoil flow with intermittent and chaotic nature}
\label{subsec:naca4412}

This part examines the application of the POD--AS--PRS framework to a more challenging flow configuration characterized by intermittent and chaotic dynamics: flow over a NACA 4412 airfoil. In contrast to the periodic vortex shedding observed in the cylinder flow, this case exhibits quasi-periodic behavior with intermittent fluctuations in aerodynamic coefficients, presenting significant challenges for reduced-order modeling due to its higher-dimensional dynamics and complex coherent structures. The investigation focuses on the framework’s efficacy in identifying sensitivity-dominant parameters despite the significantly slower energy decay of the POD spectrum. Particular emphasis is placed on the method's performance in capturing extreme events and heavy-tailed statistics in the lift coefficient, demonstrating its robustness in reconstructing both typical and rare flow regimes. 

\subsubsection{Dataset description}

In contrast to the previously discussed periodic case involving vortex shedding behind a circular cylinder, the current study examines the flow over a wing profile characterized by intermittent and chaotic behavior. Specifically, we consider a NACA 4412 airfoil geometry with a chord length of $C=1\text{m}$ and an angle of attack $\alpha=5^\circ$. The chord-based Reynolds number is $Re=U_{\infty}C/\nu=1.75\times10^{4}$. Similar to the cylinder case, length and velocity are non-dimensionalised using the chord length $C$ and $U_\infty$, respectively, while time is expressed in non-dimensional convective units normalized by $CU_\infty^{-1}$. 

The computational domain spans $(x,y)\in\left[-2,6.75\right]\times\left[-2.5,2.5\right]$, with the origin positioned at the centroid of the airfoil. The mesh comprises 4644 elements with a polynomial order of 7. Consistent with the cylinder configuration, time integration adopts the same scheme~\cite{maday1990operator}, maintaining a constant timestep of $\Delta t=0.001$. In addition, the filtering technique of~\cite{fischer2001filter} was applied to stabilize the simulation further.

\begin{figure*}[t!]
\centering
\includegraphics[width=1\textwidth]{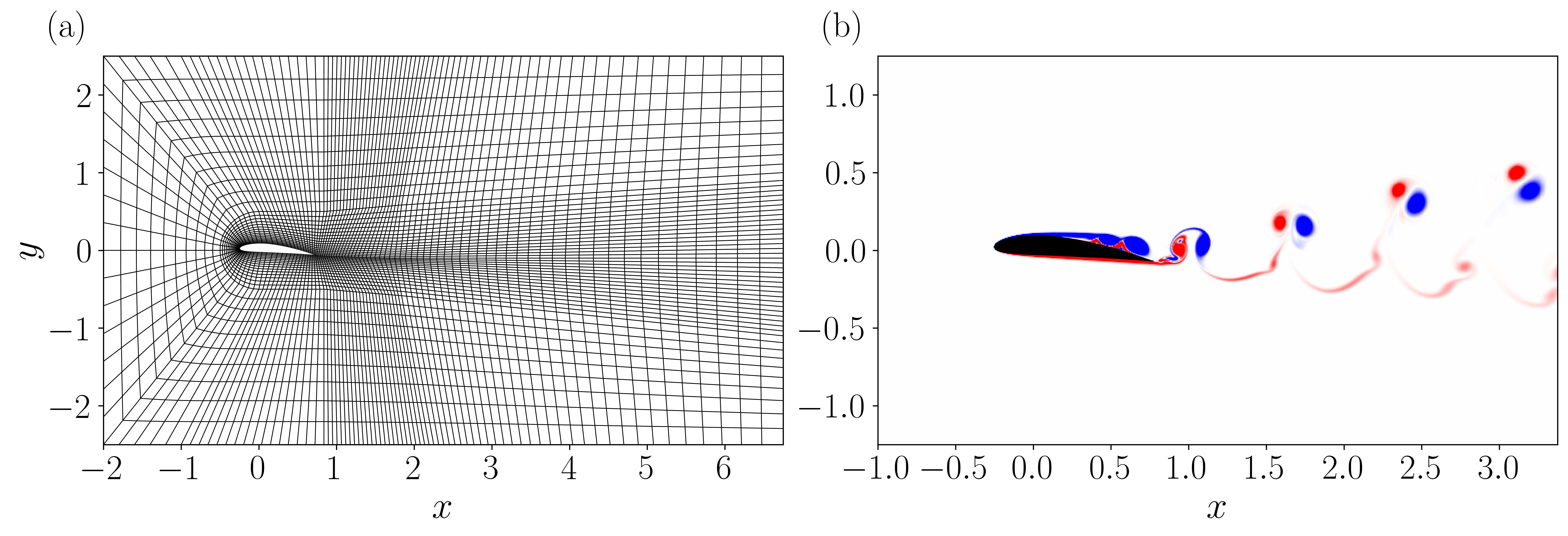}
\caption{(a) Computational domain for the flow past a NACA 4412 airfoil displaying outlines of elements without internal interpolation points. (b) Snapshot of vorticity.}
\label{fig:airfoil_mesh}
\end{figure*}

Boundary conditions mirror those of the previous case with the exception of the outlet, where a convective outflow boundary condition~\cite{DONG2015300} is utilized to facilitate the smooth departure of intense vortical structures. Fig.~\ref{fig:airfoil_mesh} illustrates the element mesh (excluding GLL interpolation points) alongside a typical vorticity field snapshot. 

In this case, snapshots from $t=600$ to $t=800$ are collected every 40 steps, yielding a total of 5000 snapshots used as inputs. 

\begin{figure*}[t!]
    \centering
    \color{black}
    \begin{subfigure}[t]{1\textwidth} 
    \centering
        \begin{overpic}[width=0.95\textwidth]{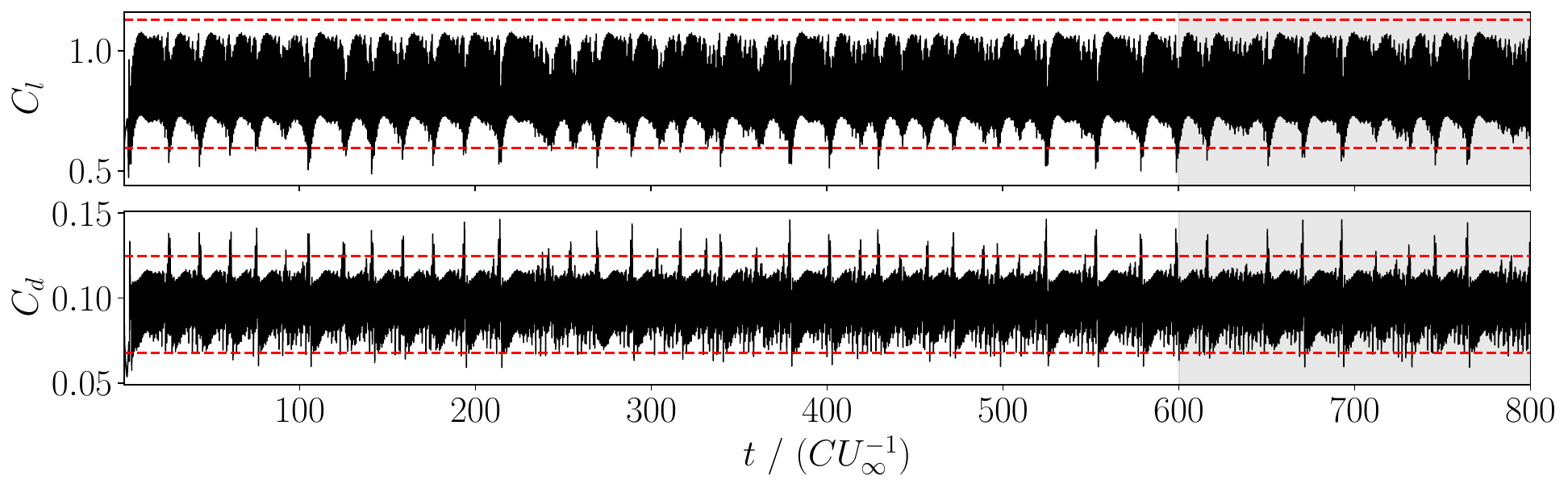}
            \put(2,110){\small{(a)}} 
        \end{overpic}
        \phantomcaption
        \label{subfig:case2_coeff} 
    \end{subfigure}
    \vspace{4mm}
    \begin{subfigure}[t]{1\textwidth}
    \centering
        \begin{overpic}[width=0.95\textwidth]{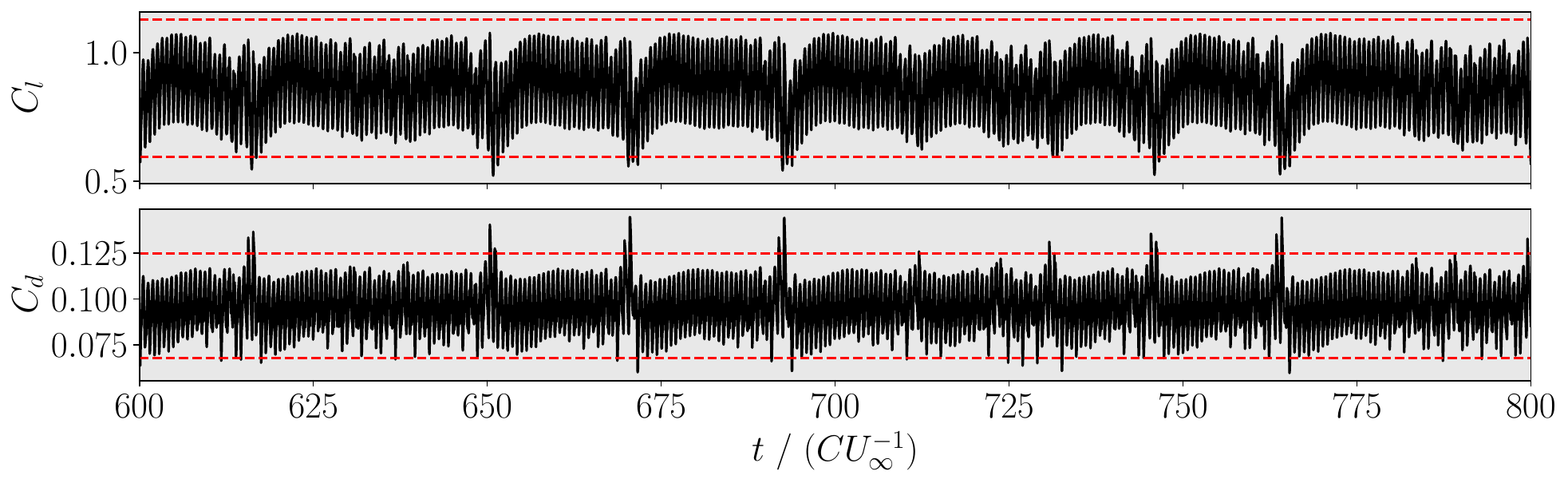}
            \put(2,110){\small{(b)}} 
        \end{overpic}
        \phantomcaption
        \label{subfig:case2_zoomedcoeff} 
    \end{subfigure}
\caption{(a)Temporal evolution of the lift coefficient $C_l$ and drag coefficient $C_d$, with the grey-shaded region denoting the snapshot sampling interval adopted for constructing the ROM; (b) zoomed view of $C_l$ (top) and $C_d$ (bottom) over the sampling interval. Red dashed lines indicate $\pm2\sigma$ away from the mean.}
\label{fig:case2_qoi}
\end{figure*}

\subsubsection{The quantity of interest}

In this study, the POD--AS--PRS framework is applied to enable efficient QoI reconstruction and to pinpoint influential parameters. For the airfoil configuration, the QoI is designated as the lift coefficient $C_l$ and the drag coefficient $C_d$, which are defined analogously to the cylinder case (see Eq.~\eqref{eq:aerodynamic_coefficients} with the characterised length being the chord length). The two-dimensional simulation yields a quasi-stable behavior in which intermittent fluctuations are observed in both aerodynamic coefficients, as depicted in Fig.~\ref{fig:case2_qoi}. This intermittent regime persists throughout the simulation duration considered herein, and the analysis focuses solely on this behavior. 

\subsubsection{Results and discussion}

\paragraph{Dimensionality reduction of POD state-space} 

In this case, the snapshot matrices were constructed from the fully converged FOM results for POD analysis. As illustrated in Fig.~\ref{subfig:case2_pod_spatial}, similar to the cylinder flow case, the spatial modes manifest in pairs, reflecting the underlying periodic structures of the flow. Differing from the cylinder flow case, however, the corresponding temporal coefficients exhibit quasi-periodic behavior with intermittently clustered patterns,  indicative of increased flow complexity. Phase diagrams Fig.~\ref{subfig:case2_pod_phase} reveal that each paired mode traces multiple intertwined trajectories, rather than single closed loops, indicating high-dimensional chaotic dynamics. The PDFs of the temporal coefficients show that the first four modes retain a bimodal distribution, while higher-order modes progressively transition to unimodal distributions, highlighting the increasing complexity of the flow structures captured by the higher modes. These observations confirm that POD effectively extracts the dominant coherent structures and temporal dynamics, providing suitable low-dimensional coefficients for subsequent ResNet-based QoI prediction and AS gradient computation, despite the enhanced complexity of the flow.

\begin{figure*}[t]
	\centering
    \begin{subfigure}[t]{0.48\columnwidth}
    \centering
        \begin{overpic}[height=4.5cm]{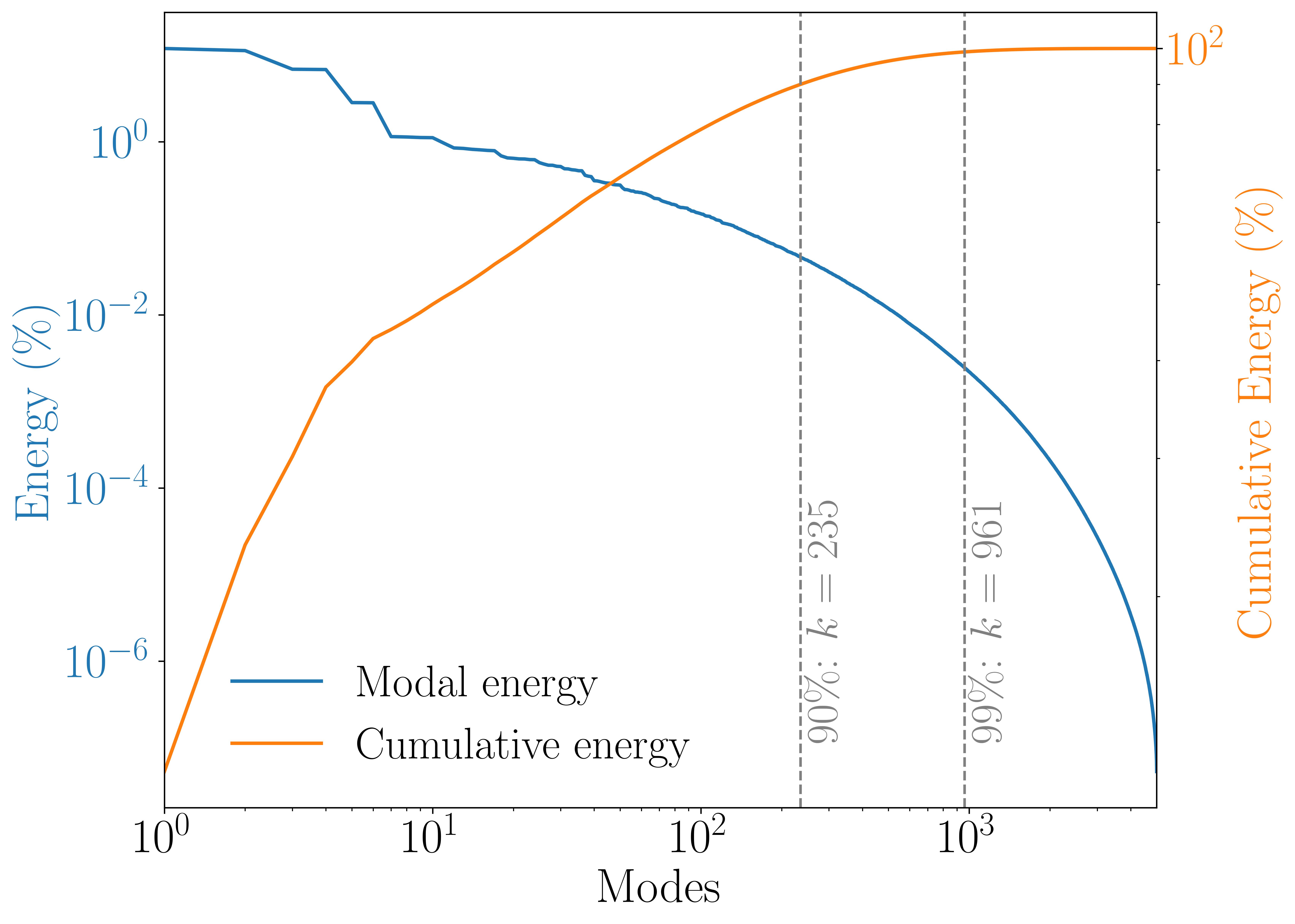}
            \put(2,130){\small{(a)}} 
        \end{overpic}
		\phantomcaption
		\label{subfig:case2_energy} 
	\end{subfigure}
	\hfill
	\begin{subfigure}[t]{0.48\columnwidth}
    \centering
	    \begin{overpic}[height=4.5cm]{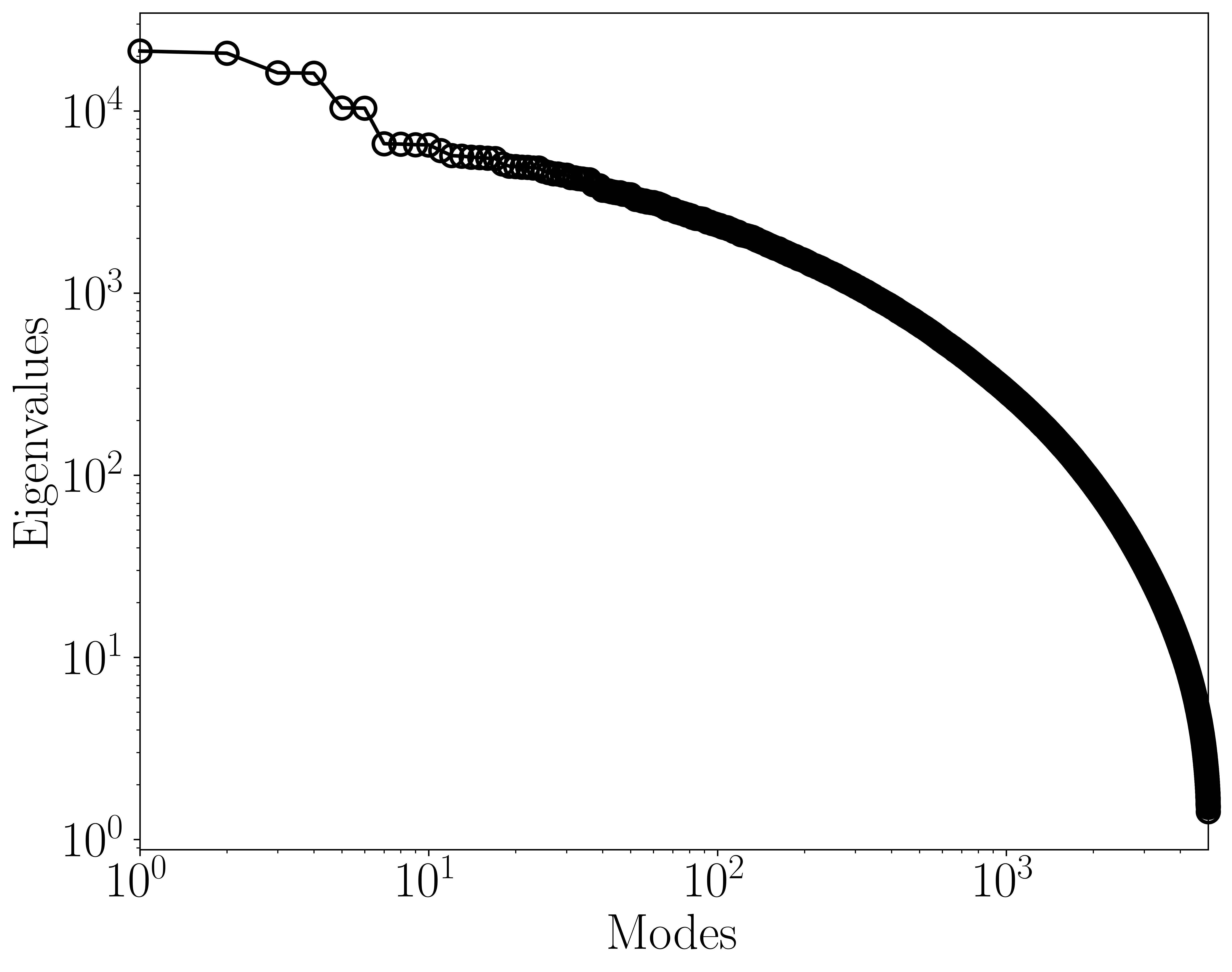}
            \put(2,130){\small{(b)}} 
        \end{overpic}
		\phantomcaption
		\label{subfig:case2_eigenval} 
	\end{subfigure}
	\caption{(a) The kinetic energy explained by each mode and the corresponding cumulative energy, and (b) the eigenvalue distribution for the test case~2.}
	\label{fig:case2_pod_energy}
\end{figure*}

\begin{table}[h]
\centering
\small
\begin{tabularx}{\textwidth}{>{\centering\arraybackslash}X >{\centering\arraybackslash}X}
\toprule
Minimum retained energy [\%]&$N_{\text{POD}}$ \\
\midrule
90&235\\
95&405\\
99&961\\
99.5&1255\\
99.9&2002\\
99.99&3097\\
\bottomrule
\end{tabularx}
\caption{Cumulative retained energy for different numbers of POD modes for the test case~2.}\label{tab:case2_pod_energy_retained}
\end{table}

The kinetic energy $\sigma_{k}^{2}$ of the POD modes and the corresponding cumulative energy are presented in Fig.\ref{subfig:case2_energy}. To capture 99\% of the total kinetic energy, as many as 961 modes are required, as detailed in Table~\ref{tab:case2_pod_energy_retained}, indicating a notably slower energy concentration among the leading modes. The eigenvalue spectrum, shown in Fig.\ref{subfig:case2_eigenval}, further illustrates that although the flow retains low-rank characteristics, the decay rate of the eigenvalues is significantly slower than that observed in the periodic bluff-body wake. This slower decay reflects the increased complexity and higher-dimensional nature of the flow, necessitating the retention of a large number of modes for accurate representation. Such insights provide critical guidance for selecting modes in the POD--AS--PRS framework, ensuring that the subsequent construction of surrogate models from AS-derived active variables adequately captures the dominant dynamics of the quasi-periodic, intermittently clustered flow.

\paragraph{Training of linear-layer ResNet and gradient computation} 

The linear-layer ResNet model was employed to map the POD coefficients to both aerodynamic coefficients, $C_l$ and $C_d$, following the same training procedure as described for the previous case. The training process for both QoIs exhibited robust convergence, and the model maintained stable performance across the validation dataset.

\begin{figure}[t]
\color{black}
    \centering

    \begin{subfigure}[t]{0.48\columnwidth} 
       \centering
        \begin{overpic}[height=5cm]{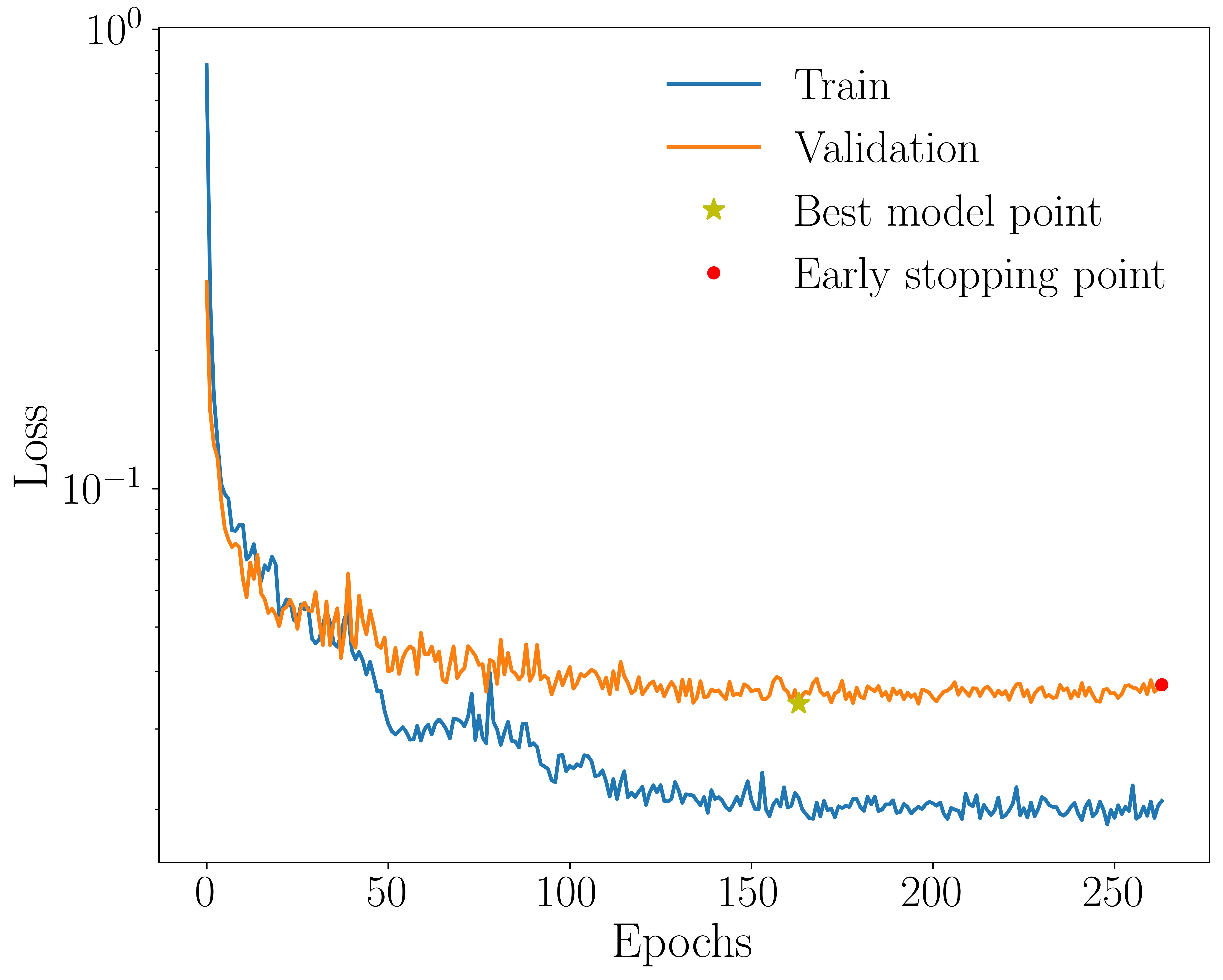}
            \put(2,140){\small{(a)}} 
        \end{overpic}
        \phantomcaption
        \label{subfig:case2_loss_cl} 
    \end{subfigure}
    \hfill
    \begin{subfigure}[t]{0.48\columnwidth}
        \centering
        \begin{overpic}[height=5cm]{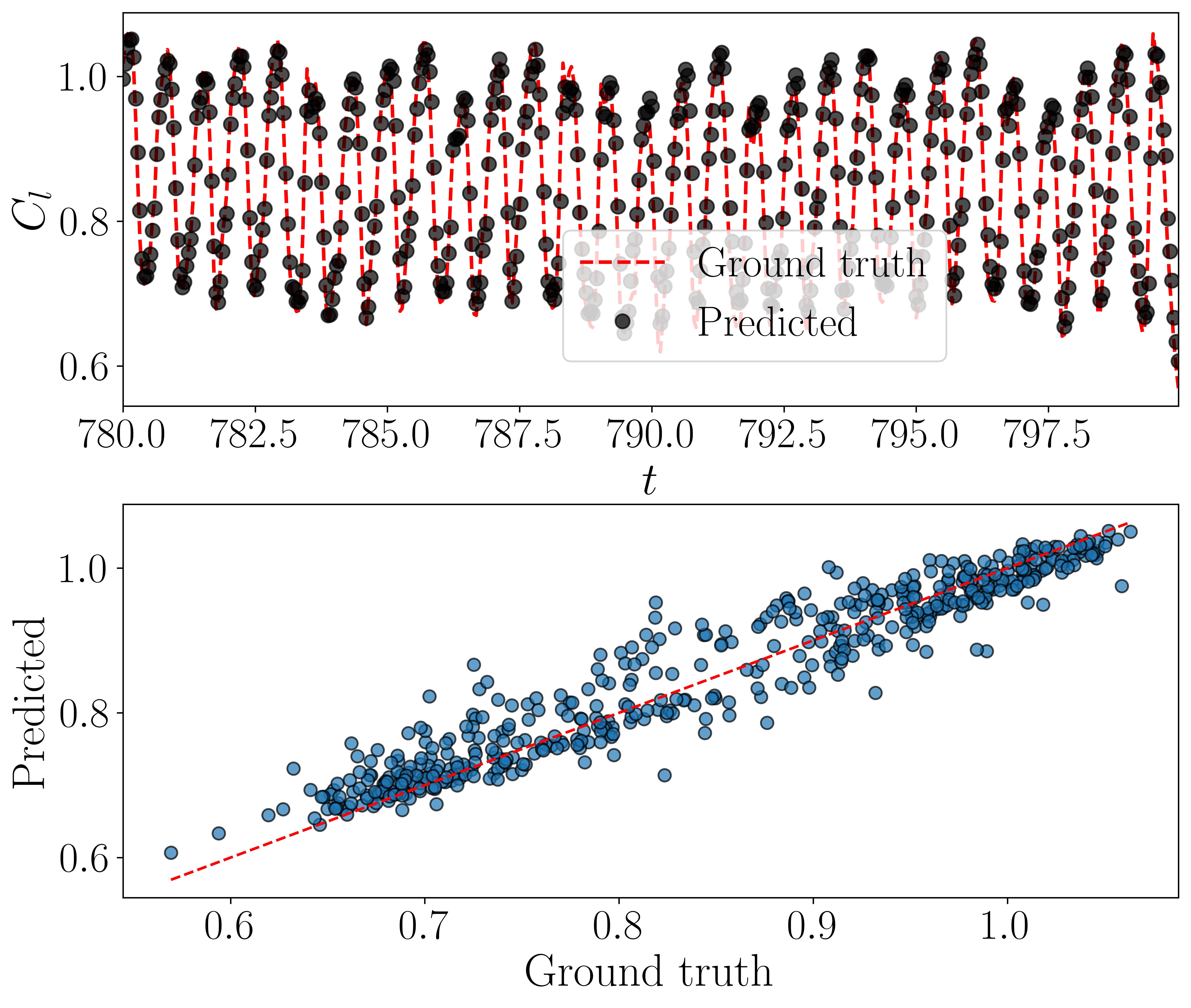}
            \put(2,140){\small{(b)}} 
        \end{overpic}
        \phantomcaption
        \label{subfig:case2_predicted_cl} 
    \end{subfigure}

    \vspace{4mm} 

    \begin{subfigure}[t]{0.48\columnwidth} 
       \centering
        \begin{overpic}[height=5cm]{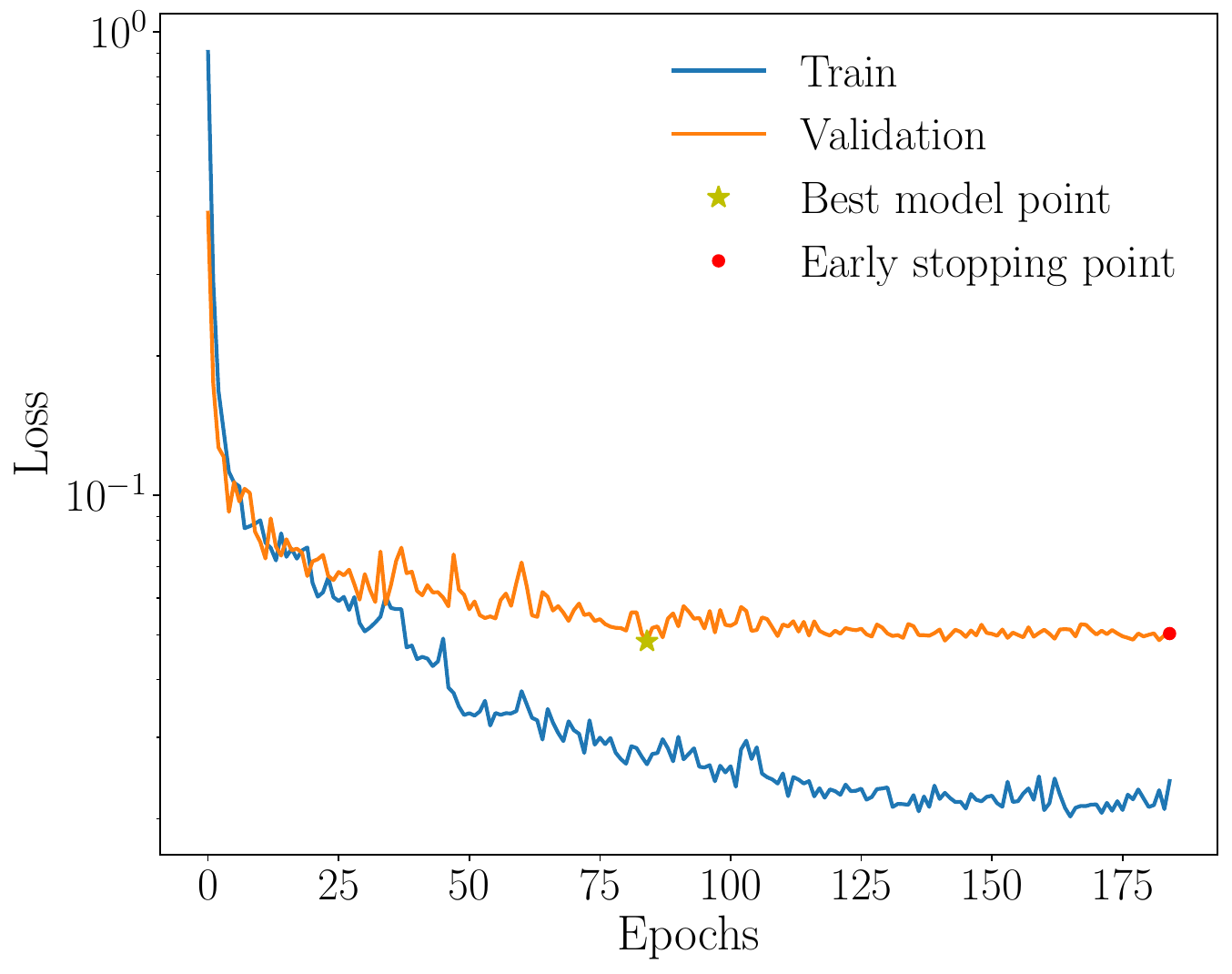}
            \put(2,140){\small{(c)}} 
        \end{overpic}
        \phantomcaption
        \label{subfig:case2_loss_cd} 
    \end{subfigure}
    \hfill
    \begin{subfigure}[t]{0.48\columnwidth}
        \centering
        \begin{overpic}[height=5cm]{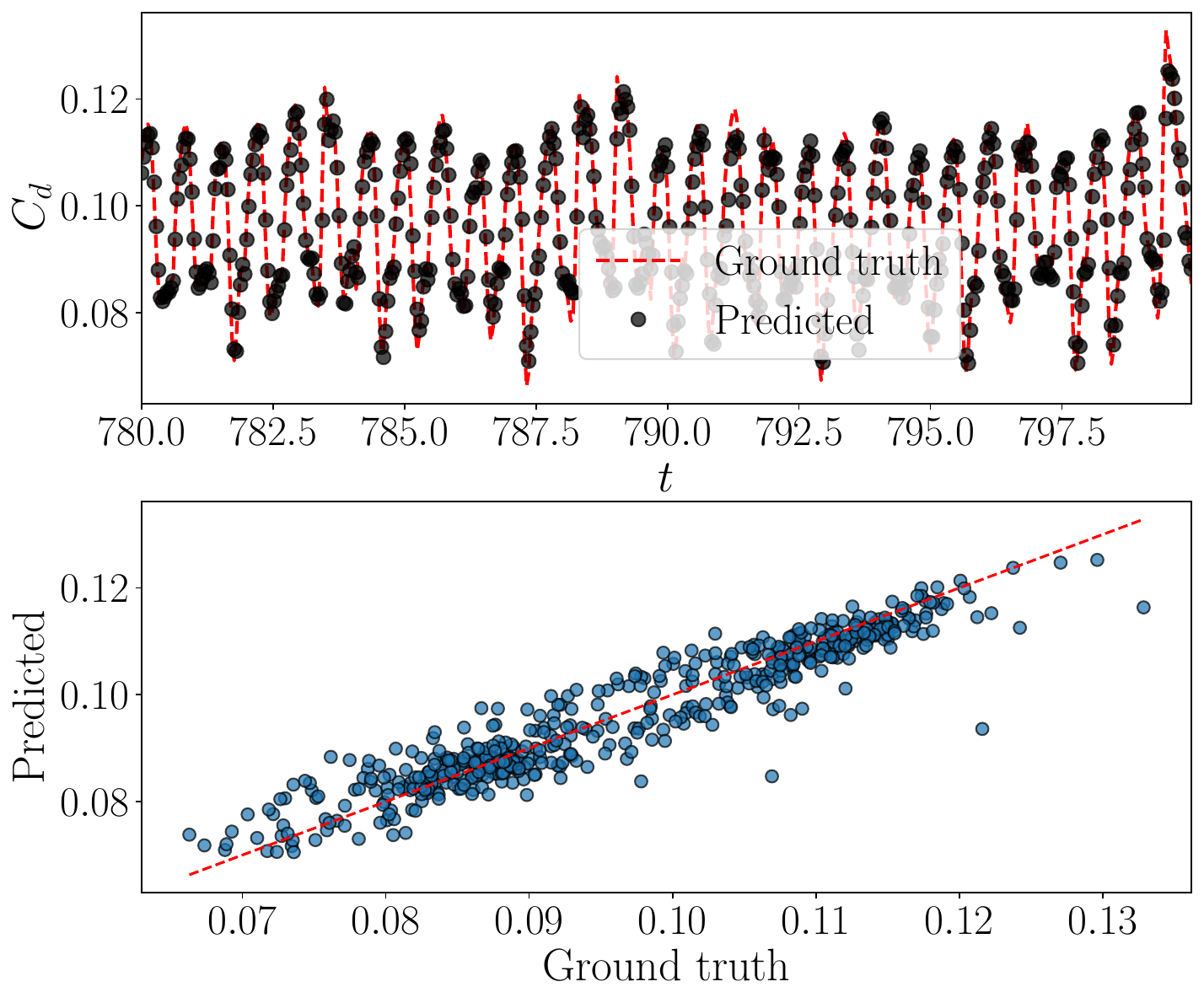}
            \put(2,140){\small{(d)}} 
        \end{overpic}
        \phantomcaption
        \label{subfig:case2_predicted_cd} 
    \end{subfigure}

    \caption{(a, c) Loss function evolution for both training and validation regarding $C_l$ and $C_d$ respectively, the yellow star (\textcolor{goldenyellow}{\ding{72}}) indicates the best model point where the validation loss ceases to decrease, while the red dot (\textcolor{red}{$\bullet$}) marks the early stopping point determined by the stopping criterion(Eq.~\eqref{eq:early_stopping}). (b, d) Model performance on the test dataset for $C_l$ and $C_d$: the upper panel compares the predicted and ground truth values over time, and the lower panel shows the parity plot between the predicted and ground truth values.}
    \label{fig:case2_resnet_traning}
\end{figure}

The training dynamics for $C_l$ and $C_d$ are illustrated in Fig.~\ref{subfig:case2_loss_cl} and Fig.~\ref{subfig:case2_loss_cd}, respectively. In both cases, the loss function decreases steadily and eventually stabilizes, indicating effective learning without signs of overfitting. The predictive performance is further assessed on an independent test dataset, as shown in Fig.~\ref{subfig:case2_predicted_cl} for $C_l$ and Fig.~\ref{subfig:case2_predicted_cd} for $C_d$. The time-series comparison demonstrates close agreement between predicted and reference values, successfully capturing the key transient features. The corresponding parity plot further confirms that the predictions for both coefficients closely follow the ground truth with minor deviations, thereby underscoring the high accuracy of the model.

\begin{table*}[b]
\color{black}
\centering
\small
\begin{tabular}{ccccc}
\toprule
QoI & Metric & Training set & Validation set & Test set \\
\midrule
\multirow{3}{*}{$C_l$} 
    & RMSE & $8.26\times 10^{-3}$ & $2.52\times 10^{-2}$ & $3.67\times 10^{-2}$ \\
    & MAE  & $6.79\times 10^{-3}$ & $2.04\times 10^{-2}$ & $2.79\times 10^{-2}$ \\
    & $R^2$ & 0.996 & 0.965 & 0.917 \\
\midrule
\multirow{3}{*}{$C_d$} 
    & RMSE & $9.98\times 10^{-4}$ & $3.12\times 10^{-3}$ & $4.38\times 10^{-3}$ \\
    & MAE  & $8.10\times 10^{-4}$ & $2.44\times 10^{-3}$ & $3.30\times 10^{-3}$ \\
    & $R^2$ & 0.995 & 0.953 & 0.898 \\
\bottomrule
\end{tabular}
\caption{Performance metrics of the linear layer ResNet for test case~2.}
\label{tab:case2_resnet_performance}
\end{table*}

Quantitative performance metrics are summarized in Table~\ref{tab:case2_resnet_performance}, including RMSE, MAE, and $R^2$ values evaluated over the training, validation, and test datasets. While the $R^2$ scores for the training and validation sets exceed 0.95, a relative decrease is observed on the test dataset. This discrepancy is not an indication of overfitting, as training and validation losses converge consistently; rather, it stems from the chronological data splitting applied to this inherently intermittent and quasi-periodic flow. The non-stationary nature of the intermittent regime makes late-time forecasting significantly more challenging than interpolation within the training window. Crucially, within the POD--AS--PRS framework, the ResNet is primarily utilized as a differentiable ``gradient engine" to extract sensitivity information rather than as a standalone forecasting tool. The high training accuracy confirms that the models have effectively captured the underlying nonlinear mapping and its derivatives within the training manifold, providing a reliable foundation for the subsequent AS analysis.

\begin{figure}[t!]
\color{black}
    \centering

    \begin{subfigure}[t]{0.81\textwidth} 
        \centering
        \begin{overpic}[height=3.5cm]{figures/Figure_17a}
            \put(2,100){\small{(a)}} 
        \end{overpic}
        \phantomcaption
        \label{subfig:case2_grad_diff_cd} 
    \end{subfigure}
    \hfill
    \begin{subfigure}[t]{0.18\textwidth}
        \centering
        \begin{overpic}[height=3.5cm]{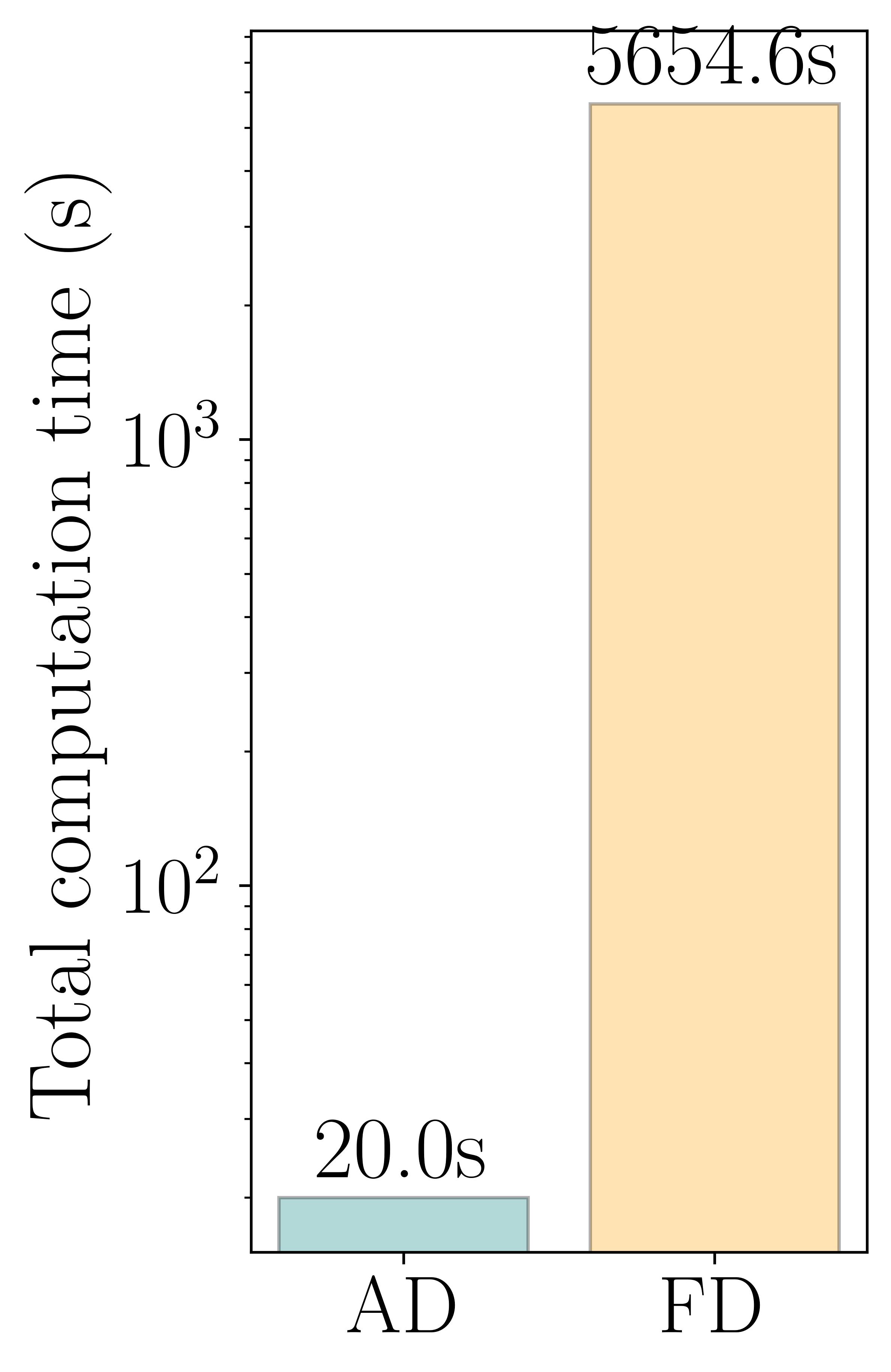}
            \put(2,100){\small{(b)}} 
        \end{overpic}
        \phantomcaption
        \label{subfig:case2_grad_time_cd} 
    \end{subfigure}

    \vspace{4mm}

    \begin{subfigure}[t]{0.81\textwidth} 
        \centering
        \begin{overpic}[height=3.5cm]{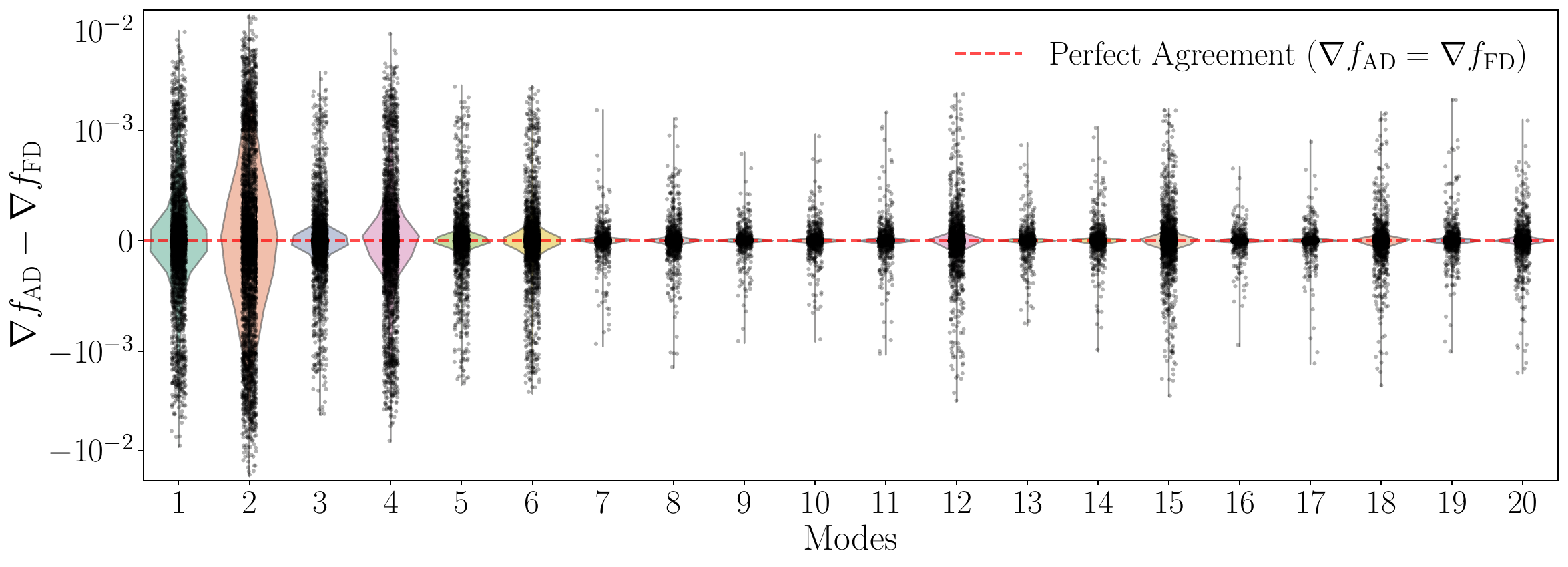} 
            \put(2,100){\small{(c)}} 
        \end{overpic}
        \phantomcaption
        \label{subfig:case2_grad_diff_cl} 
    \end{subfigure}
    \hfill
    \begin{subfigure}[t]{0.18\textwidth}
        \centering
        \begin{overpic}[height=3.5cm]{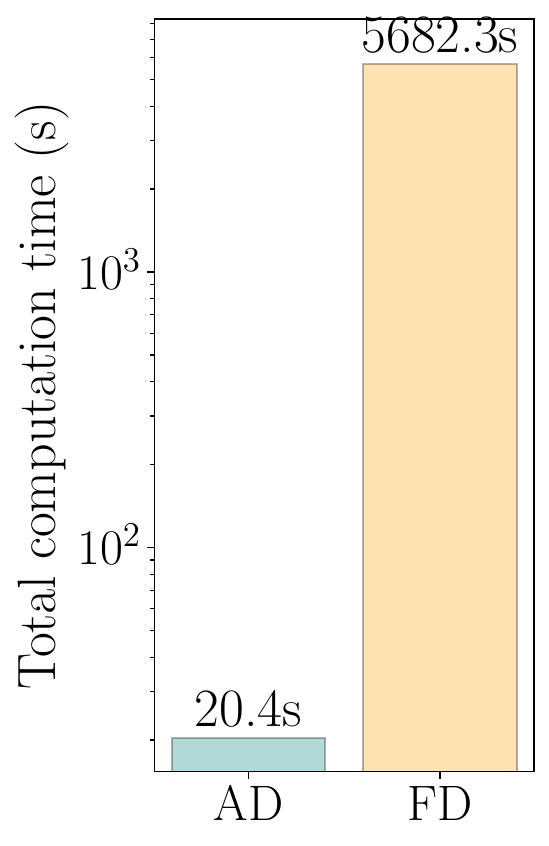} 
            \put(2,100){\small{(d)}} 
        \end{overpic}
        \phantomcaption
        \label{subfig:case2_grad_time_cl} 
    \end{subfigure}

    \caption{Gradient differences of the first 20 POD modes across all samples for (a) $C_l$ and (c) $C_d$, shown as violin plots with boxplot and strip overlays; (b, d) total gradient computation times for AD and FD for (b) $C_l$ and (d) $C_d$ in test case~2.}
    \label{fig:case2_gradient}
\end{figure}

Gradients of the predicted aerodynamic coefficients, $C_l$ and $C_d$, with respect to the POD coefficients were computed via reverse-mode AD through the trained network. Similar to the cylinder case, the first 500 POD modes were retained for each sample to provide a sufficiently rich candidate space for identifying all potentially sensitive physical directions. To evaluate computational efficiency, the AD-based gradients were compared against FD approximations, as shown in Fig.~\ref{fig:case2_gradient}. The AD-based gradient computation required a total of 20.0s for $C_l$ and 20.4s for $C_d$, whereas FD took 5654.6s and 5682.3s, achieving significant speedups of 282.73-fold and 278.5-fold, respectively. Relative to the previous cylinder wake case, these results demonstrate that, even for the more complex quasi-periodic and intermittently clustered flow, AD preserves both high computational efficiency and machine-precision accuracy in gradient evaluation.

\paragraph{Dimensionality reduction of active subspaces parameter-space}

\begin{figure}[t!]
\color{black}
    \centering
      \begin{subfigure}[t]{0.48\textwidth} 
        \centering
        \begin{overpic}[height=5cm]{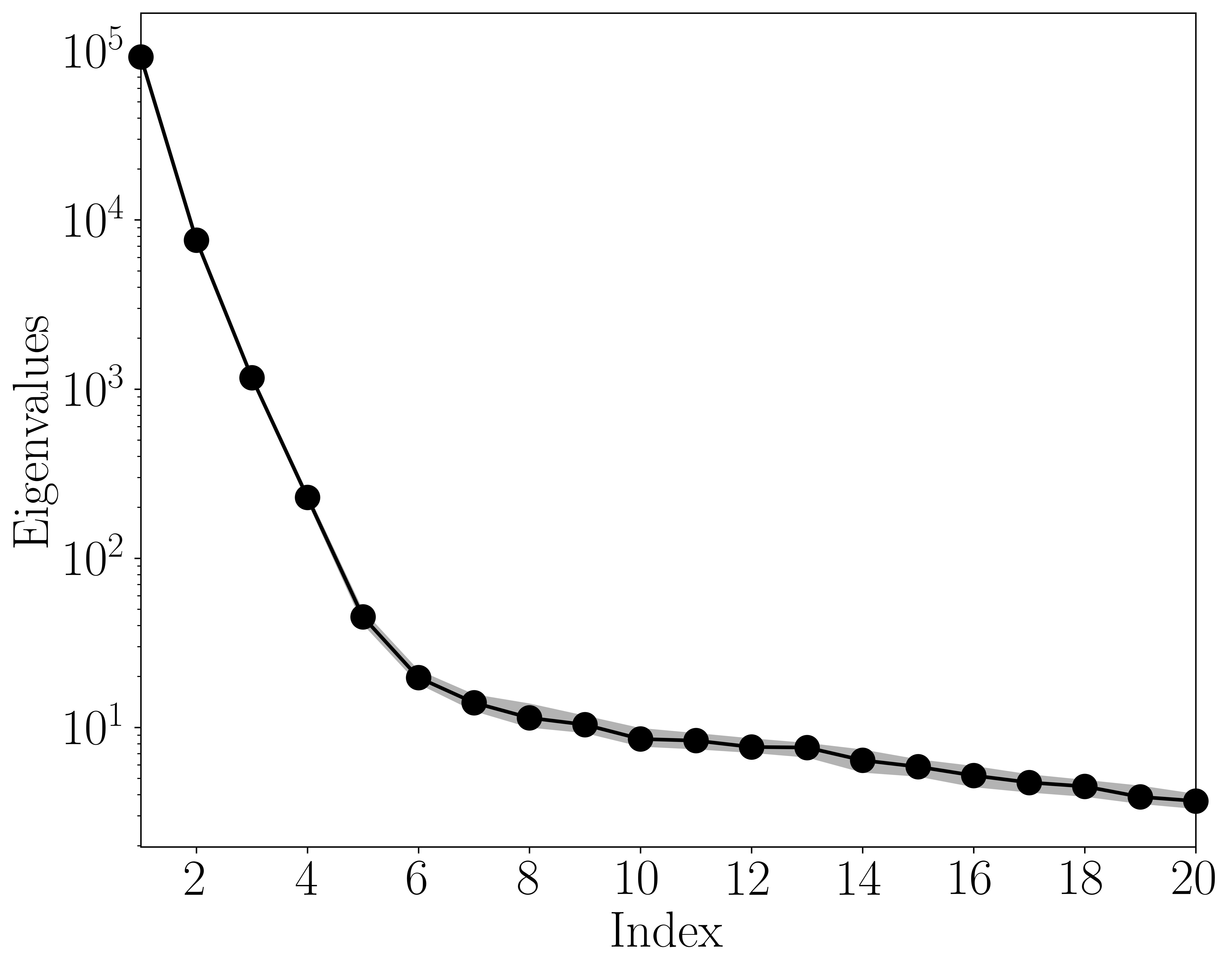}
            \put(2,140){\small (a)} 
        \end{overpic}
        \phantomcaption
        \label{subfig:case2_as_eigenval_cl} 
    \end{subfigure}
    \hfill
    \begin{subfigure}[t]{0.48\textwidth} 
        \centering
        \begin{overpic}[height=5cm]{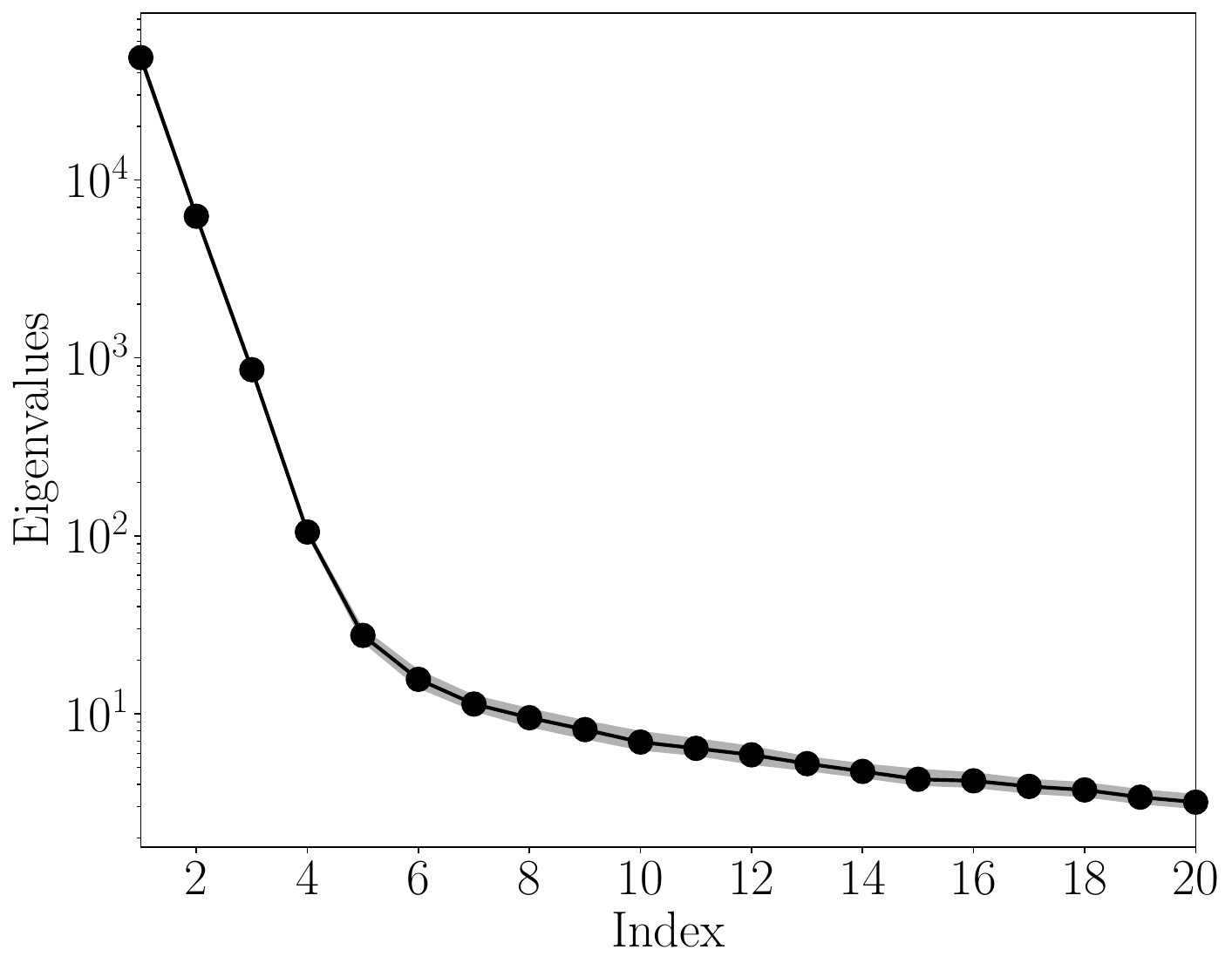} 
            \put(2,140){\small (b)}
        \end{overpic}
        \phantomcaption
        \label{subfig:case2_as_eigenval_cd} 
    \end{subfigure}

    \vspace{4mm}

    \begin{subfigure}[t]{0.48\textwidth} 
        \centering
        \begin{overpic}[height=5cm]{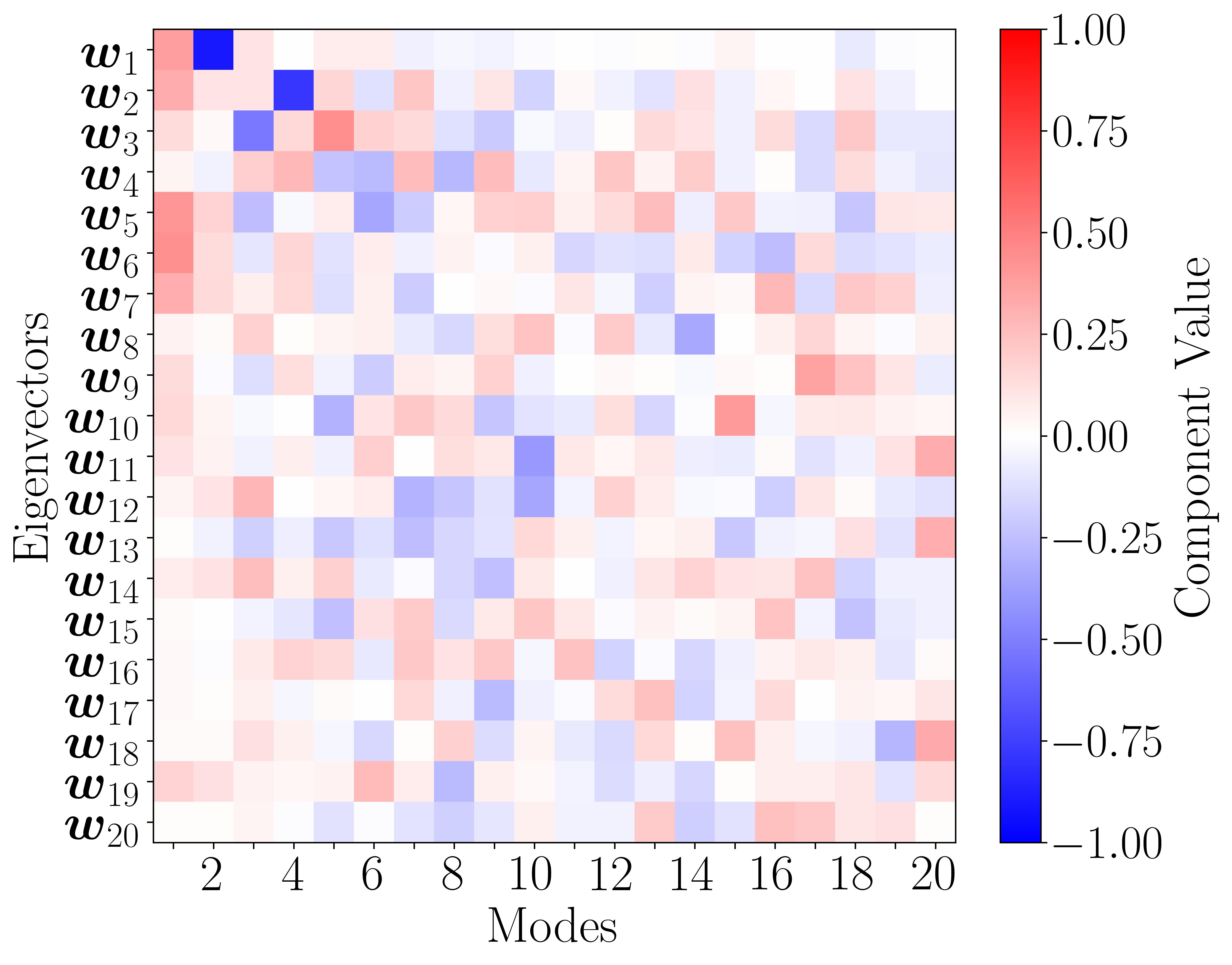} 
            \put(2,140){\small (c)} 
        \end{overpic}
        \phantomcaption
        \label{subfig:case2_as_eigenvec_cl} 
    \end{subfigure}
    \hfill
    \begin{subfigure}[t]{0.48\textwidth} 
        \centering
        \begin{overpic}[height=5cm]{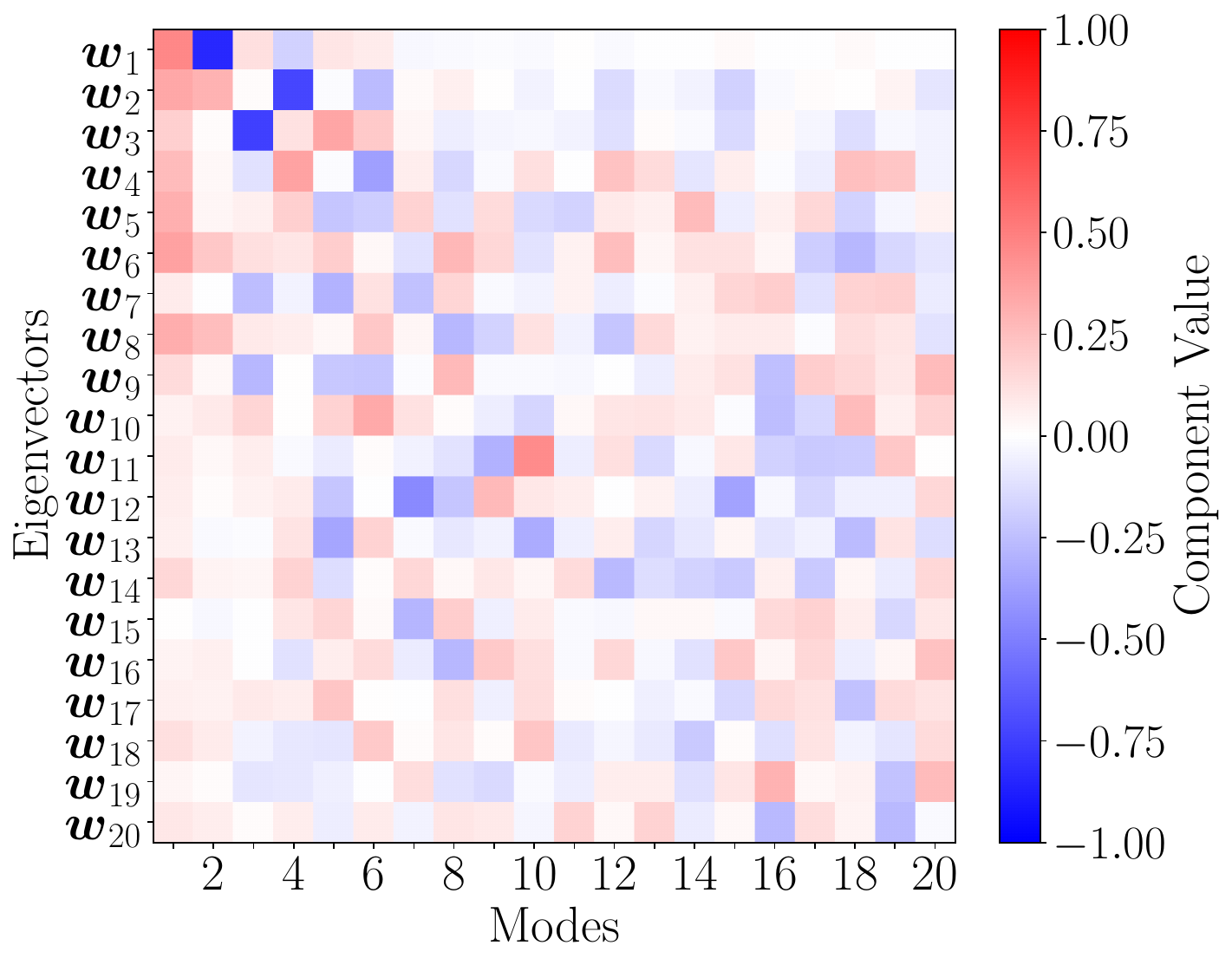}
            \put(2,140){\small (d)}
        \end{overpic}
        \phantomcaption
        \label{subfig:case2_as_eigenvec_cd} 
    \end{subfigure}

    \caption{The first 20 eigenvalues of the covariance matrix with bootstrap confidence intervals (grey regions) for (a) $C_l$ and (b) $C_d$ and components of the first 20 eigenvectors for (c) $C_l$ and (d) $C_d$ in test case~2.}
    \label{fig:case2_as_eigen}
\end{figure}

Figs.~\ref{subfig:case2_as_eigenval_cl} and \ref{subfig:case2_as_eigenval_cd} present the distribution of the first 20 eigenvalues of the covariance matrix $\hat{\boldsymbol{C}}$ derived from the gradients of the aerodynamic coefficients, $C_l$ and $C_d$, with respect to the POD coefficients. Although the leading eigenvalues exhibit significant gaps, multiple AS dimensions are retained to adequately capture the variability of QoIs, Specifically, $N_{\text{AS}}=22$ and $N_{\text{AS}}=24$ are identified as optimal for $C_l$ and $C_d$, respectively, via a grid search on the AS dimension and the polynomial order for the response surface. Figs.~\ref{subfig:case2_as_eigenvec_cd} and \ref{subfig:case2_as_eigenvec_cl} depict the components of the corresponding eigenvectors, where color indicates the sign of contribution to the active variables. Remarkably, the sensitivity rankings for both $C_l$ and $C_d$ exhibit a high degree of consistency across the leading eigenvectors, which stands in contrast to the previous circular cylinder case. In the cylinder flow, the inherent two-to-one frequency relationship between $C_d$ and $C_l$, whereby the drag coefficient oscillates at twice the vortex shedding frequency ($2f_s$) while the lift coefficient 
oscillates at the fundamental shedding frequency ($f_s$)~\citep{marzouk2026extended, zheng2008frequency}, necessarily causes the two force coefficients to be governed by distinct POD modes, those capturing the fundamental frequency and those capturing its first harmonic, independently of the degree of flow nonlinearity. However, for the NACA4412 airfoil, the presence of non-zero mean lift, arising from the camber and the effective angle of attack, fundamentally breaks this reflective symmetry. As a result, the drag and lift responses share the same dominant frequency, and the same POD modes govern both $C_l$ and $C_d$ simultaneously. Examination of the leading three eigenvectors elucidates the comparative sensitivities: for both QoIs, the primary eigenvector is dominated by the largest absolute component in Mode 2 (negative sign indicating an inverse coupling), followed by Mode 1 (positive); the secondary eigenvector exhibits primacy in Mode 4 (negative); and the tertiary eigenvector displays the peak magnitude in Mode 3 (negative), with subsequent positive influences from Mode 5. This suggests that in non-symmetric lifting configurations with non-zero mean $C_l$, the broken wake symmetry eliminates the two-to-one frequency relationship between $C_d$ and $C_l$, such that both force coefficients oscillate at the same dominant frequency and are consequently governed by the same set of dominant POD modes, as identified by the active subspace.

\begin{figure}[!ht]
\color{black}
	\centering

	\begin{subfigure}[t]{0.48\textwidth} 
		\centering
		\begin{overpic}[height=5.5cm]{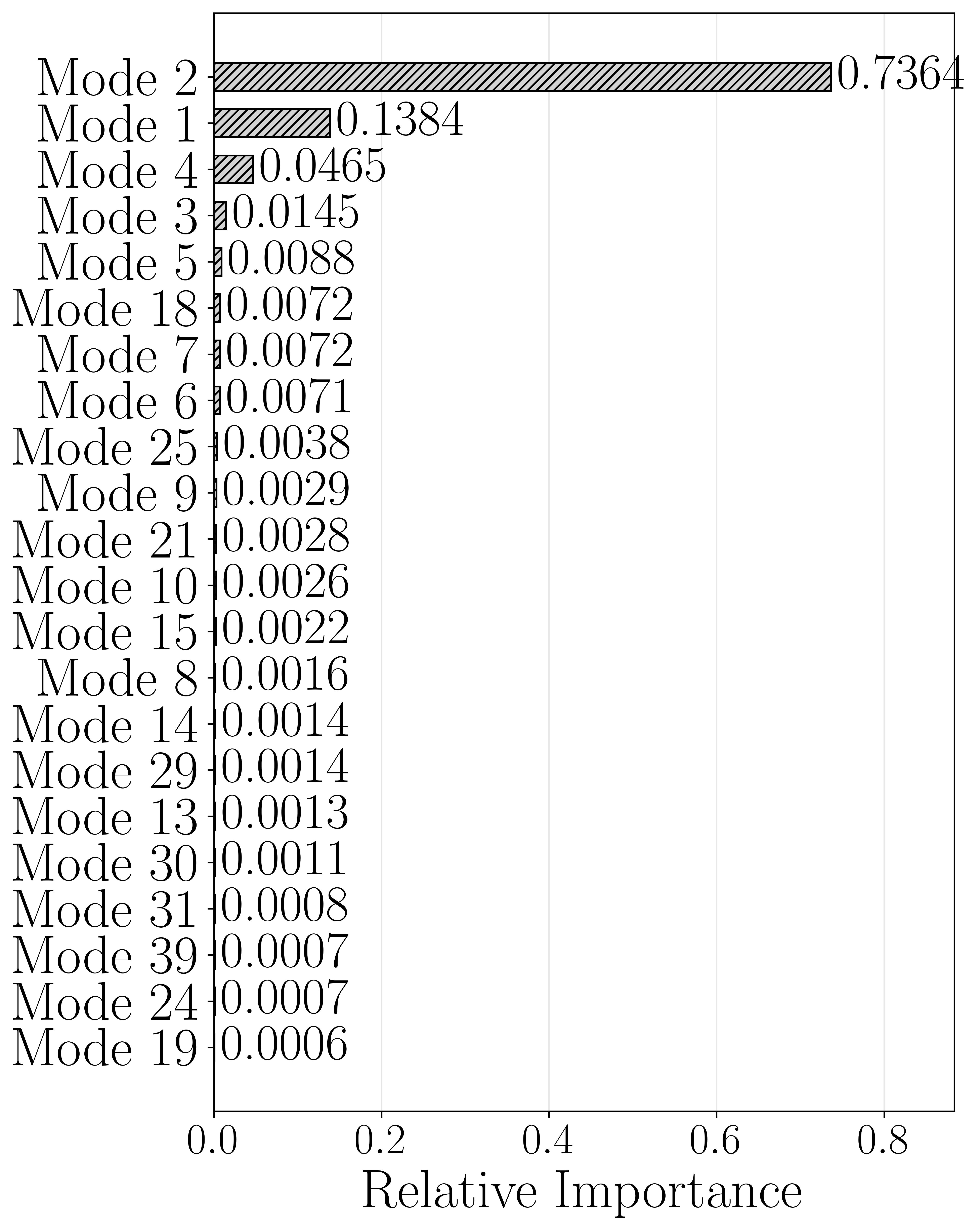}
			\put(2,150){\small (a)} 
		\end{overpic}
		\phantomcaption
		\label{subfig:case2_importance_cl} 
	\end{subfigure}
	\hfill
	\begin{subfigure}[t]{0.48\textwidth} 
		\centering
		\begin{overpic}[height=5.5cm]{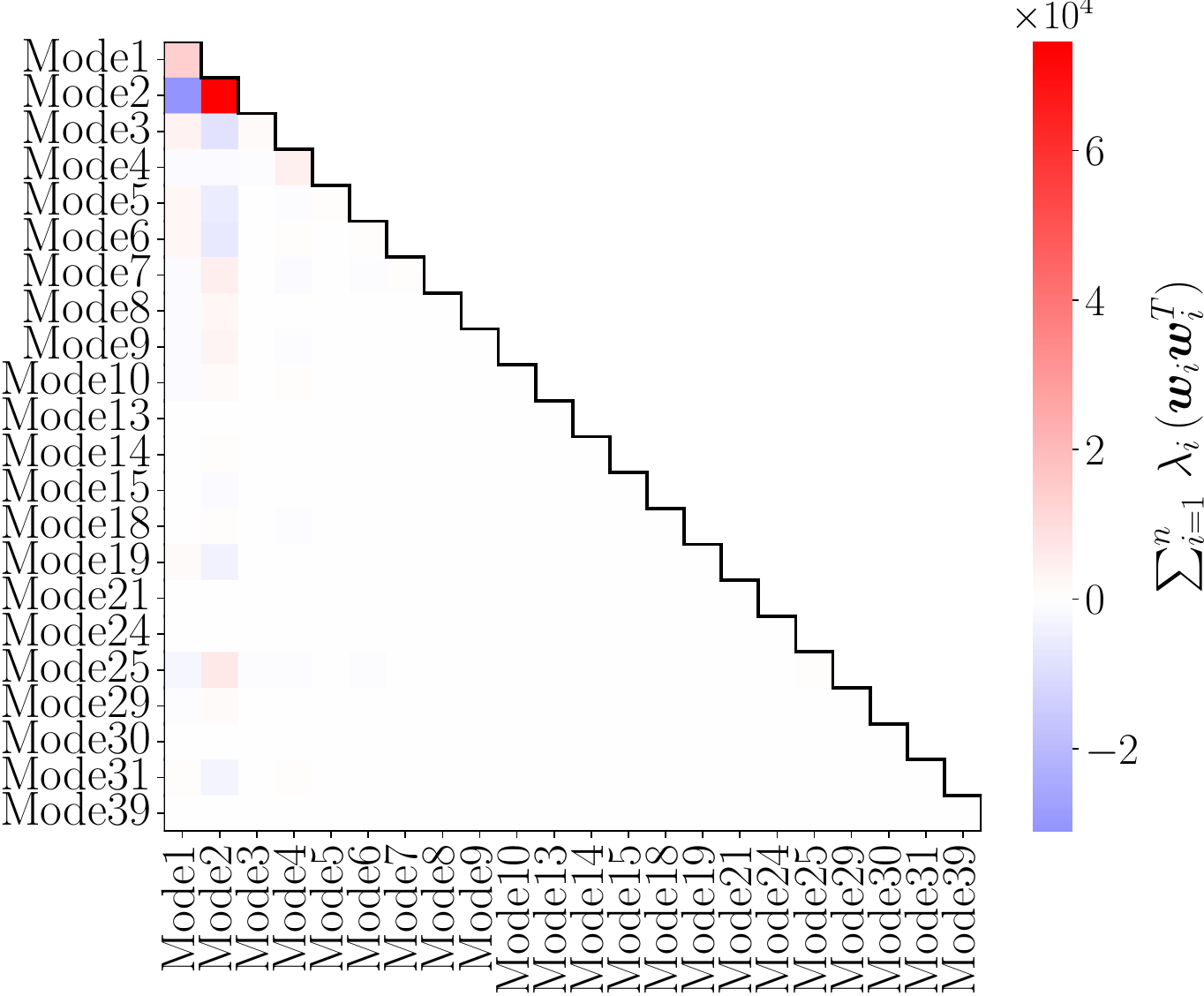}
			\put(-10,150){\small (b)}
		\end{overpic}
		\phantomcaption
		\label{subfig:case2_importance_heat_cl} 
	\end{subfigure}

	\vspace{0mm}

	\begin{subfigure}[t]{0.48\textwidth} 
		\centering
		\begin{overpic}[height=5.5cm]{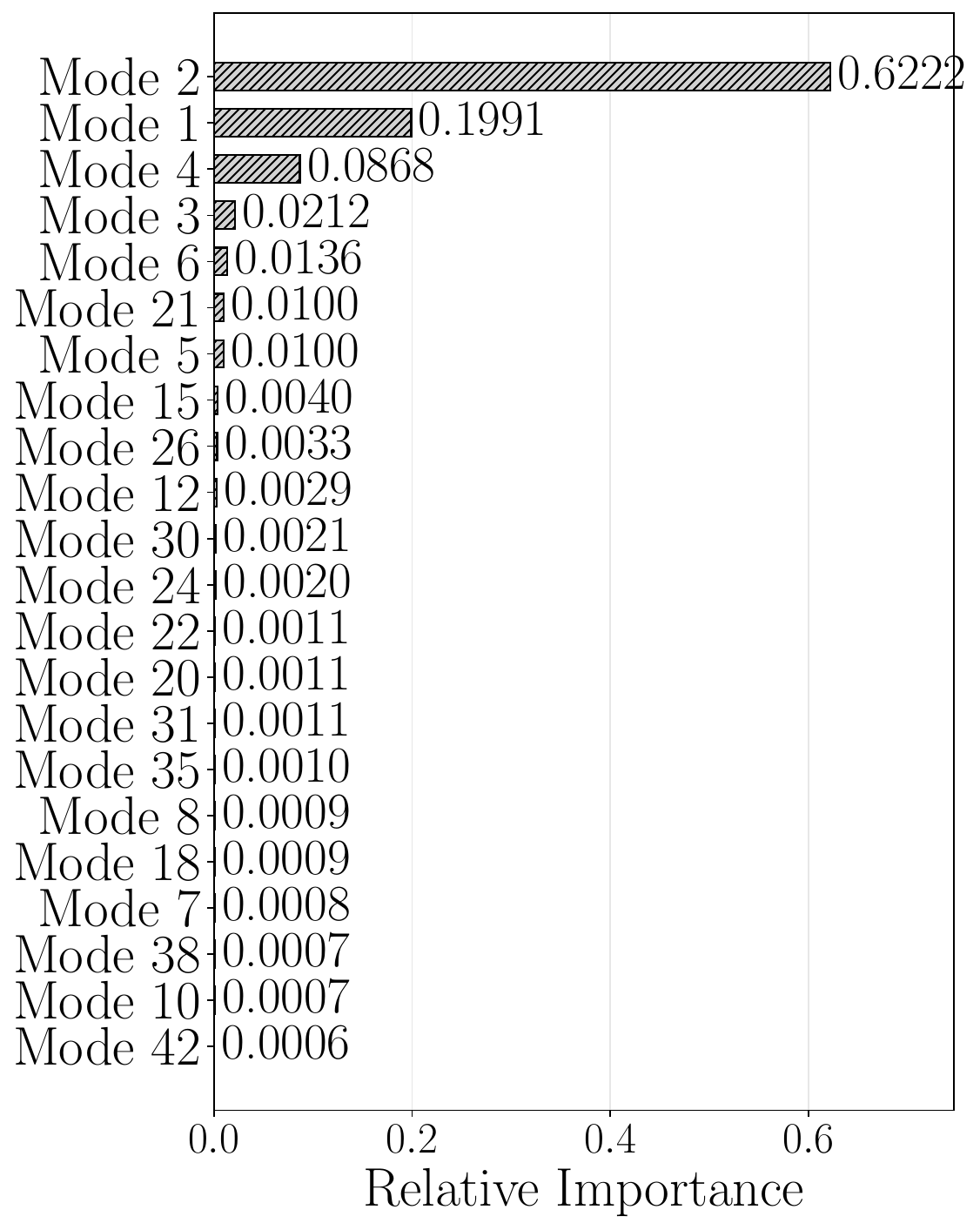} 
			\put(2,150){\small (c)} 
		\end{overpic}
		\phantomcaption
		\label{subfig:case2_importance_cd} 
	\end{subfigure}
	\hfill
	\begin{subfigure}[t]{0.48\textwidth} 
		\centering
		\begin{overpic}[height=5.5cm]{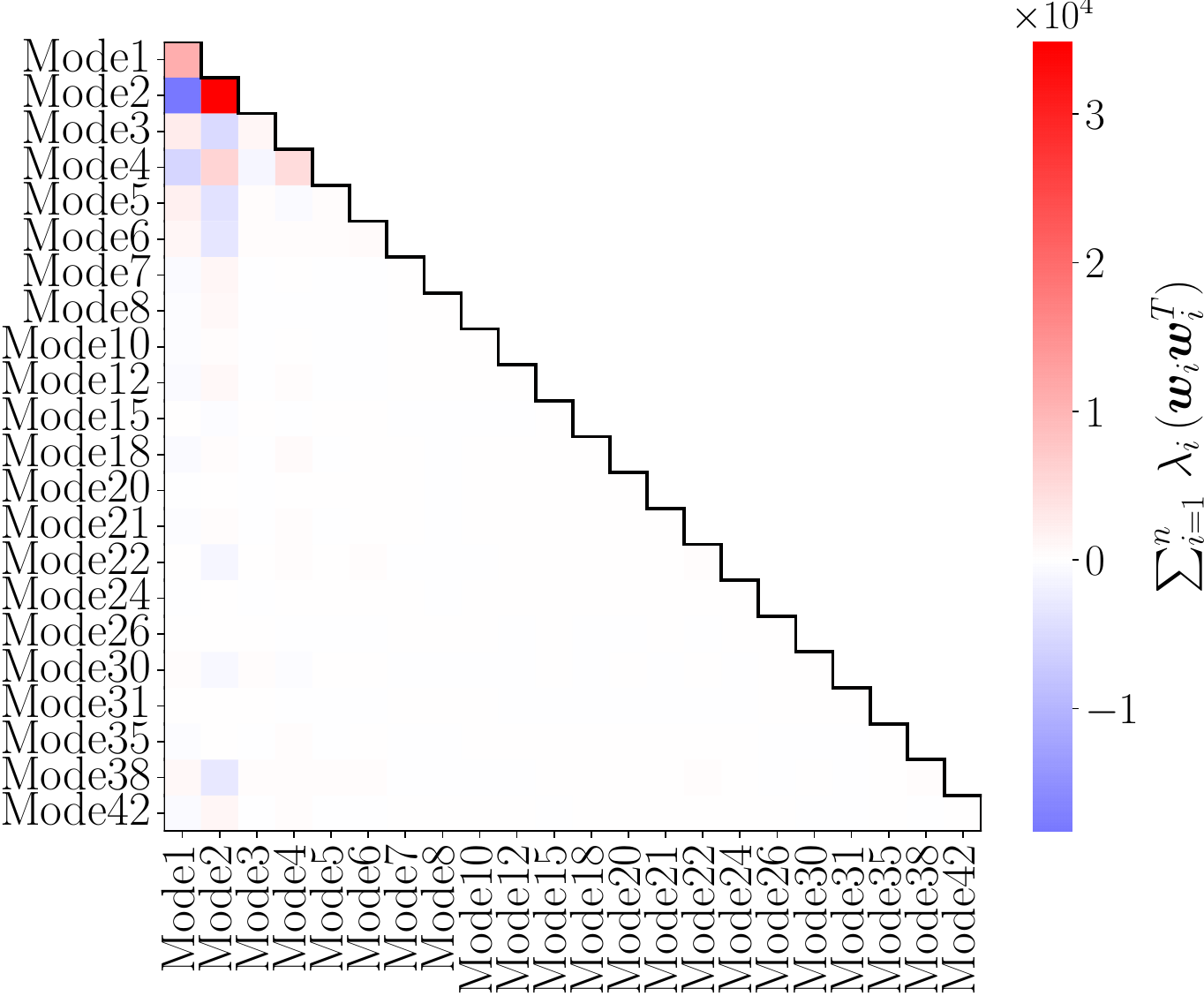} 
			\put(-10,150){\small (d)}
		\end{overpic}
		\phantomcaption
		\label{subfig:case2_importance_heat_cd} 
	\end{subfigure}

	\caption{Global sensitivity analysis of POD coefficients with respect to (a, b) $C_l$ and (c, d) $C_d$ for test case~2. (a, c) Normalized activity scores for the POD coefficients whose cumulative contributions reach 99\%; (b, d) pairwise interaction matrices of the leading active subspace directions, where diagonal entries show individual sensitivities and off-diagonal entries indicate correlations (red for synergistic amplification, blue for antagonistic reduction).}
	\label{fig:case2_importance}
\end{figure}

\newcolumntype{Y}{>{\centering\arraybackslash}X}

\begin{table}[h]
    \centering
    \small
    
    \begin{tabularx}{\textwidth}{YYY}
        \toprule
        & \multicolumn{2}{c}{$N_{\text{POD}}$} \\
        \cmidrule(lr){2-3} 
        Minimum retained influence [\%] & {$C_l$} & {$C_d$} \\
        \midrule
        95    & 6  & 6   \\
        96    & 7  & 7   \\
        97    & 9  & 9   \\
        98    & 13 & 15  \\
        99    & 22 & 32  \\
        99.99 & 95 & 115 \\
        \bottomrule
    \end{tabularx}
    \caption{Cumulative retained influence of POD modes on $C_l$ and $C_d$ for test case~2.}
    \label{tab:case2_as_influence_retained}
\end{table}

To further elucidate the influence of the POD coefficients on the aerodynamic performance, global sensitivity was assessed using normalized activity scores for both $C_l$ and $C_d$. Figs.~\ref{subfig:case2_importance_cl} and \ref{subfig:case2_importance_cd} present the scores for the leading POD modes. To facilitate a direct comparison between the two QoIs, both figures display the top 22 modes—the number required for $C_l$ to reach a 99\% cumulative influence—even though $C_d$ necessitates 32 modes to achieve the same threshold. For both coefficients, Mode 2 is identified as the sensitivity-dominant parameter, followed by Modes 1, 4, 3, and 5. This ranking is highly consistent with the relative magnitudes of the components in the leading active subspace eigenvectors (Figs.~\ref{subfig:case2_as_eigenvec_cl} and \ref{subfig:case2_as_eigenvec_cd}), further corroborating that higher-magnitude eigenvector components indicate relative sensitivity to the corresponding parameters. Table~\ref{tab:case2_as_influence_retained} details the cumulative influence of these modes on QoI, illustrating the number of modes required to capture specified percentages of the total influence. Figs.~\ref{subfig:case2_importance_heat_cl} and \ref{subfig:case2_importance_heat_cd} depict the pairwise interactions of the active variables through the outer product of the eigenvectors. Notably, for both coefficients, the interaction between Modes 1 and 2 exhibits a relatively pronounced negative value, indicating antagonistic effects that attenuate their combined impact on aerodynamic coefficients. Overall, these results demonstrate that the variability of both aerodynamic forces is governed by a shared set of sensitivity-dominant parameters and their intermodal couplings, despite the slightly higher modal requirements for $C_d$ to account for the same level of global influence. This finding underscores the efficiency of the active subspace in isolating the low-dimensional mechanisms that drive complex aerodynamic response.

\paragraph{Construction of response surface surrogate models} 

\begin{figure}[t]
\color{black}
    \centering
      \begin{subfigure}[t]{0.46\textwidth} 
        \centering
        \begin{overpic}[height=5.5cm]{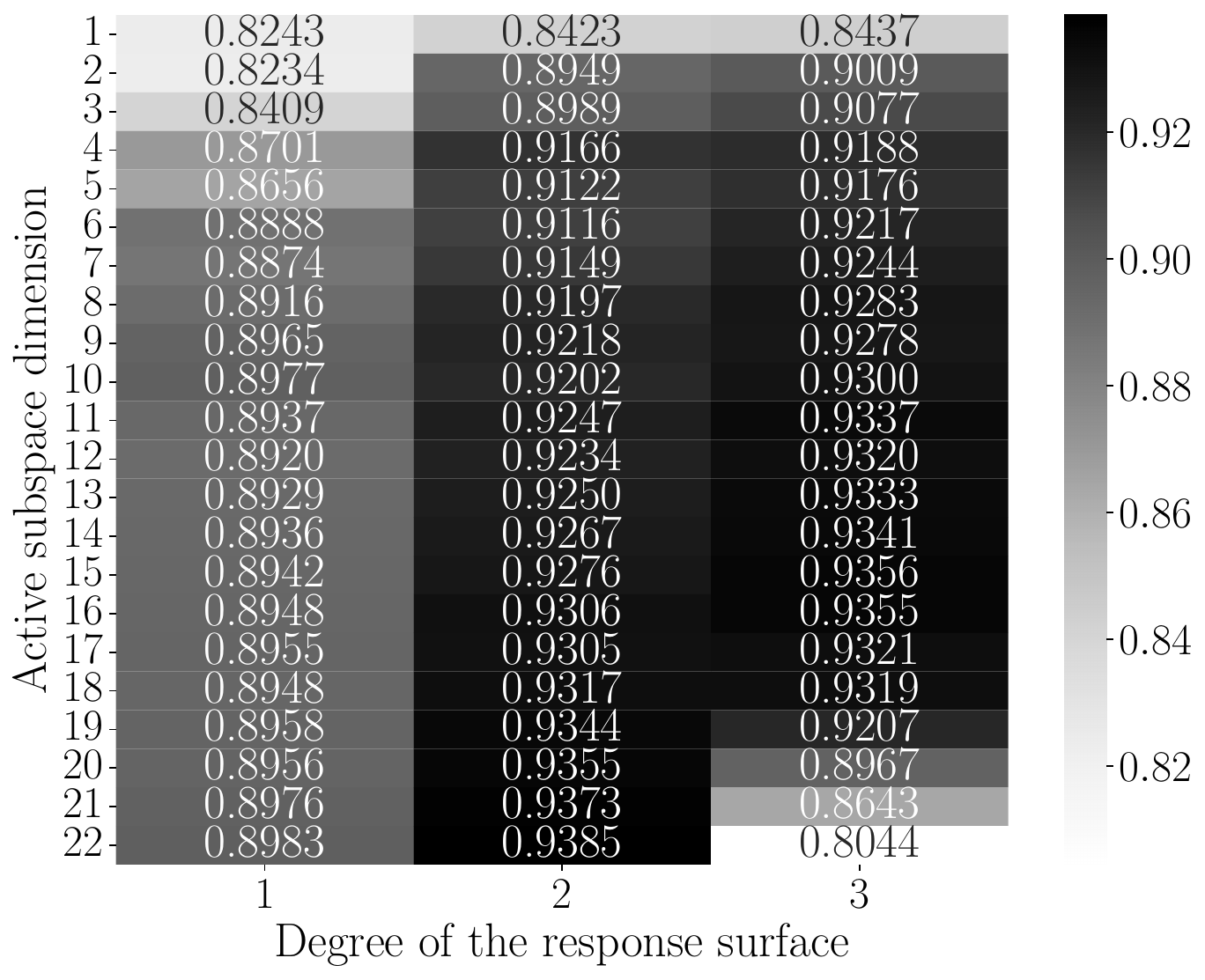}
            \put(2,160){\small (a)} 
        \end{overpic}
        \phantomcaption
        \label{subfig:case2_rs_cl} 
    \end{subfigure}
    \hfill
    \begin{subfigure}[t]{0.46\textwidth} 
        \centering
        \begin{overpic}[height=5.5cm]{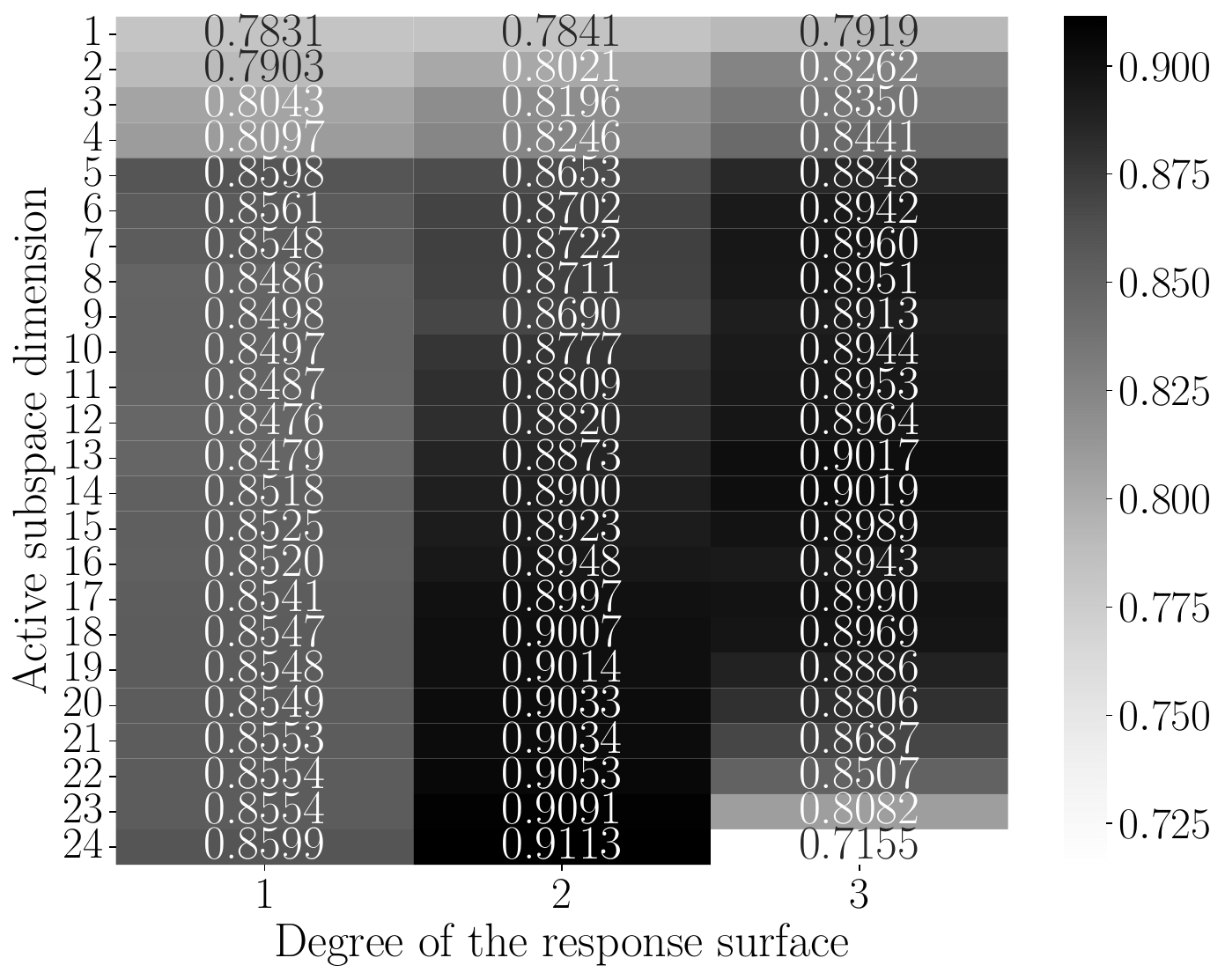} 
            \put(2,160){\small (b)}
        \end{overpic}
        \phantomcaption
        \label{subfig:case2_rs_cd} 
    \end{subfigure}
	\caption{Construction and assessment of the response surface for evaluating the generalization capability of the surrogate for (a) lift coefficient $C_l$ and (b)drag coefficient $C_d$. The heatmap presents the $R^2$ scores on the test dataset corresponding to various active subspace dimensions and polynomial orders.}
	\label{fig:case2_rs}
\end{figure}

Based on the identified sensitivity-dominant parameters, the response surface was constructed using the active variables derived from the leading POD modes that collectively account for over 99\% of the total influence.
By employing the AS-based response surface construction strategy described previously, a grid search over different active subspace dimensions and polynomial orders was conducted to evaluate the generalization capability of the surrogate models on the independent test set. Fig.~\ref{fig:case2_rs} depicts the $R^2$ values corresponding to the tested combinations of active subspace dimension and polynomial order. For $C_d$, the grid search is presented up to $N_{\text{AS}}=24$, beyond which the generalization performance begins to decline due to overfitting; the remaining dimensions are omitted for clarity of visualization. The analysis reveals that for both coefficients, the generalization performance improves consistently with increasing active subspace dimension, with the optimal balance between accuracy and model complexity achieved at $(N_{\text{AS}},p)=(22,2)$ for $C_l$ and $(N_{\text{AS}},p)=(24,2)$ for $C_d$. This configuration ensures sufficient representation of the QoI variability while maintaining computational tractability for subsequent analyses.

\paragraph{Accuracy assessment of the POD--AS--PRS framework}

\begin{figure}[t!]
	\centering
\begin{subfigure}[t]{0.95\textwidth}
		\centering
		\begin{overpic}[width=1\textwidth]{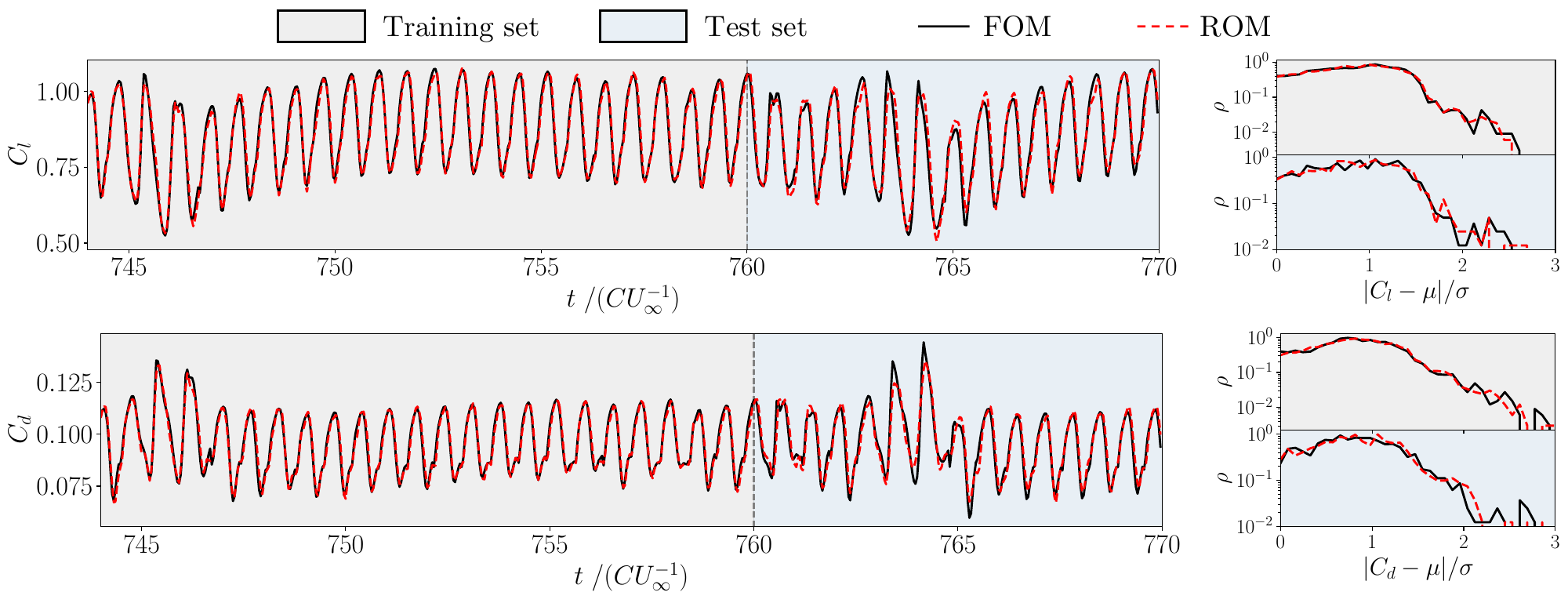}
			\put(2,130){\small (a)} 
		\end{overpic}
		
		\phantomcaption
		\label{subfig:case2_temporal_density}
	\end{subfigure}
	\hfill
\vspace{0mm}
	\begin{subfigure}[t]{1\textwidth}
		\centering
		\begin{overpic}[width=0.95\textwidth]{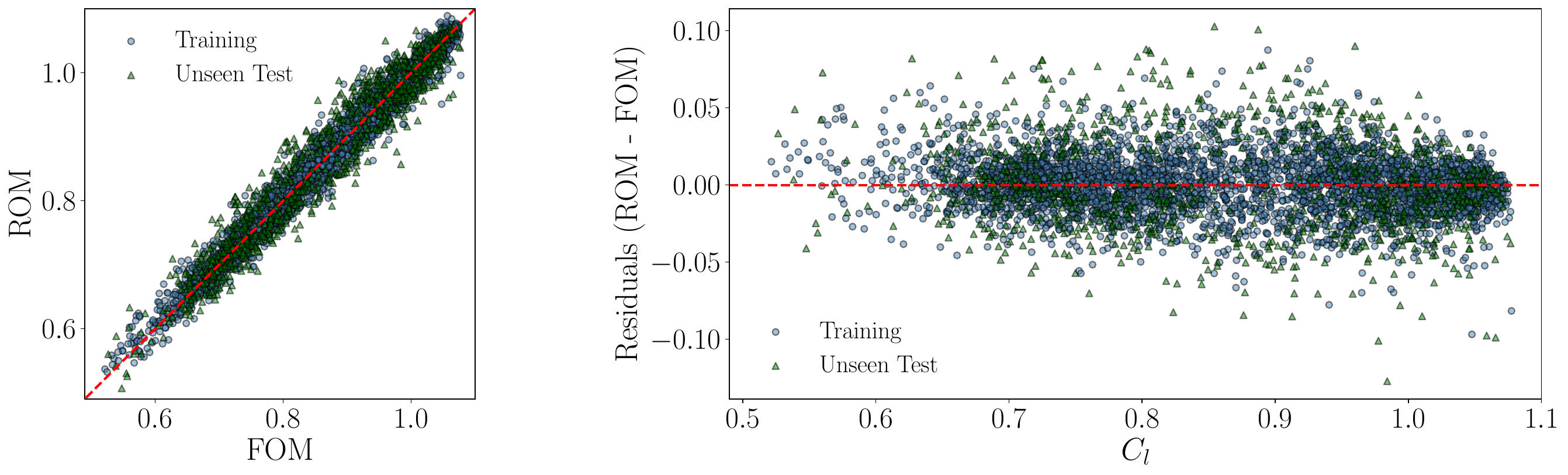}
			\put(2,110){\small (b)}
		\end{overpic}
		\vspace{0.2em} \\
		\begin{overpic}[width=0.95\textwidth]{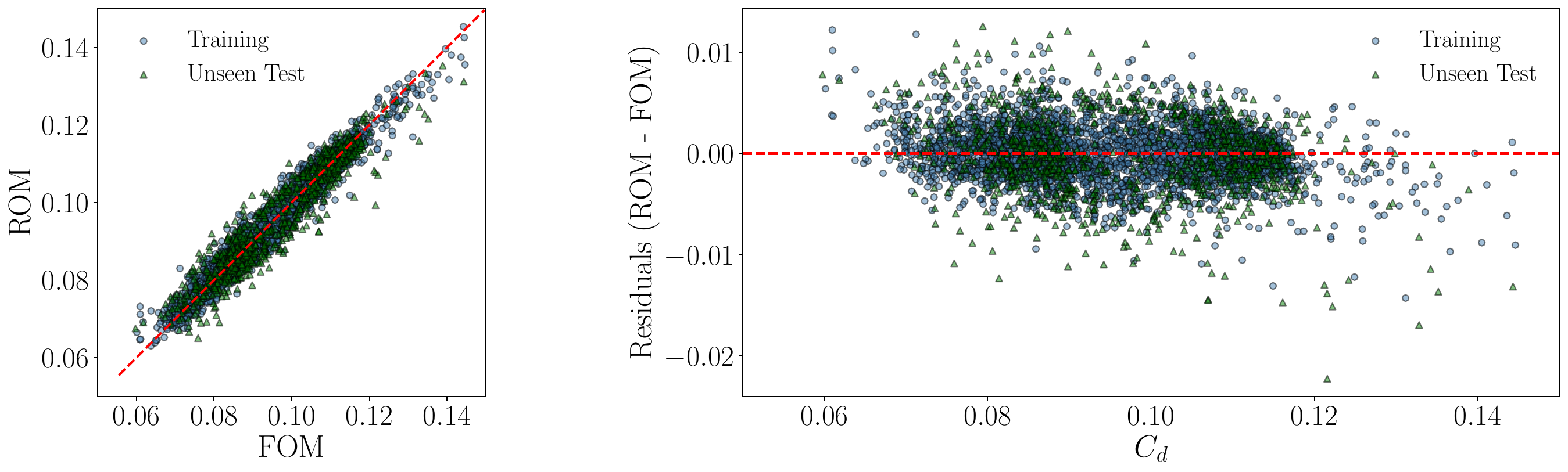}

		\end{overpic}
		\phantomcaption
		\label{subfig:case2_scatter_residual}
	\end{subfigure}

	\caption{Comparison of ROM results and FOM for aerodynamic coefficients ($C_l$ and $C_d$), where the upper and lower panels in each subfigure correspond to $C_l$ and $C_d$, respectively. 
    (a) Time-series comparison of ROM and FOM: the left panels show the temporal evolution, where the gray-shaded and steelblue-shaded regions represent the training (reconstruction) and independent test (generalization) intervals, respectively. FOM values are represented as black solid lines and ROM results as red dashed lines. The right panels present the density functions corresponding to the histograms of the absolute deviations, $|\mathrm{QoI}-\mu|/\sigma$, where $\mu$ and $\sigma$ denote the mean and standard deviation of QoI, for both training and test phases, highlighting the statistical consistency. (b) Scatter plots of ROM results against FOM (left) and residual analysis (right) for both training (circles) and test (triangles) datasets, illustrating the model fidelity and the error distributions across both aerodynamic coefficients.}
	\label{fig:case2_rom_fom}
\end{figure}

To evaluate the performance of the POD--AS--PRS framework in a more complex quasi-periodic and chaotic flow, the pre-constructed low-dimensional surrogate models were employed to perform reconstruction analysis within the training interval and generalization assessment within the independent test interval, covering both the lift coefficient $C_l$ and the drag coefficient $C_d$. Fig.~\ref{fig:case2_rom_fom} provides a comprehensive comparison between ROM results and FOM ground truth for both aerodynamic coefficients. The time-series comparisons (left panel of Fig.~\ref{subfig:case2_temporal_density}) demonstrate that the ROM accurately reproduces the temporal evolution of both $C_l$ and $C_d$ in the training set and exhibits robust generalization to the intermittent fluctuations characteristic of this flow regime in the test set. The right panel illustrates the density functions of the absolute deviations of the ROM results from the FOM mean for both training and test phases. For the training interval, the ROM successfully captures the heavy-tail characteristics associated with intermittent extreme events~\cite{sapsis2021}, confirming the high reconstruction fidelity of the identified active subspace. For the independent test interval, although a moderate deviation in the tail region is observed, the surrogate still demonstrates stable generalization performance. Fig.~\ref{subfig:case2_scatter_residual} further confirms that the ROM results are in strong alignment with the FOM, with minimal errors and no evident bias. 

As summarized in Table~\ref{tab:case2_rom_performance}, the surrogate model achieves $R^2$ values of 0.9726 and 0.9577 on the training set for $C_l$ and $C_d$, respectively, confirming that the identified low-dimensional active subspace enables efficient and high-fidelity reconstruction of the QoI within the reduced manifold. The independent test set yields $R^2$ values of 0.9385 and 0.9113, respectively. The moderate reduction in $R^2$ from training to test intervals, as well as the tail deviation observed in the density functions, are consistent with the increased complexity of the chaotic flow regime, where intermittent extreme events in the test interval may exhibit statistical properties not fully represented in the training data. Nevertheless, despite the inherent complexity of the chaotic flow regime, the high $R^2$ values across both QoIs on the independent test set confirm that the identified active subspace generalizes well to unseen flow states, demonstrating the robustness of the framework under challenging conditions. Overall, these results demonstrate the applicability and reliability of the POD--AS--PRS framework in identifying and modeling sensitivity-dominant mechanisms that govern both typical and extreme flow conditions.

\begin{table*}[h]
\color{black}
\centering
\small
\begin{tabular}{cccccc}
\toprule
Model & QoI & Dataset & $R^2$ & MAE & RMSE \\
\midrule
\multirow{4}{*}{POD--AS--PRS} & \multirow{2}{*}{$C_l$} & Train & 0.9726 & $1.60 \times 10^{-2}$ & $2.17 \times 10^{-2}$ \\
\cmidrule{3-6}
                              &                        & Test  & 0.9385 & $2.59 \times 10^{-2}$ & $3.27 \times 10^{-2}$ \\
\cmidrule{2-6}
                              & \multirow{2}{*}{$C_d$} & Train & 0.9577 & $2.07 \times 10^{-3}$ & $2.87 \times 10^{-3}$ \\
\cmidrule{3-6}
                              &                        & Test  & 0.9113 & $3.16 \times 10^{-3}$ & $4.18 \times 10^{-3}$ \\
\bottomrule
\end{tabular}
\caption{Performance metrics of the POD--AS--PRS framework on training and test datasets for the NACA 4412 airfoil flow at $Re = 1.75 \times 10^4$.}
\label{tab:case2_rom_performance}
\end{table*}

\section{Conclusions and perspectives}
\label{sec:conclusion}

In this work, we propose a novel reduced-order modeling framework, termed POD--AS--PRS, which synergistically integrates AS-based parameter-space sensitivity analysis with the non-intrusive reduced-order modeling capability of POD and PRS surrogate modeling. The framework extracts low-dimensional POD coefficients from high-fidelity CFD snapshots and maps them to the QoI using a linear-layer ResNet. Gradients of the QoI with respect to the POD coefficients are computed via reverse-mode AD and are subsequently employed in AS analysis to identify the sensitivity-dominant parameters that exert the most significant functional influence on the QoI. Based on these identified active directions, a compact PRS is constructed in the reduced active subspace, yielding a surrogate model built in the physically meaningful active subspace. Compared to conventional approaches, the innovations of the POD--AS--PRS framework include:
\begin{enumerate}[
    label=(\arabic*),
    leftmargin=3em,
    itemsep=0pt,
    parsep=0pt,
    topsep=0pt
]
    \item The introduction of gradient-based AS analysis into non-intrusive ROM: By integrating AS analysis, the framework can efficiently identify and rank the influential POD modes that govern the variations of QoI, thereby bridging the gap between black-box data-driven modeling and physical interpretability. The PRS surrogate, constructed over the identified active subspace, further reinforces this interpretability: its explicit polynomial structure directly reveals how the QoI depends on the sensitivity-dominant directions and their interactions.
    \item Efficient gradient evaluation via neural network backpropagation: The framework utilizes a linear-layer ResNet to map POD coefficients to the QoI, enabling the computation of exact gradients via reverse-mode AD. This differentiable architecture achieves machine-precision accuracy and significant speedup compared to finite difference methods, all without the need for complex adjoint solvers or explicit analytical formulas.
    \item Simultaneous reduction in both state and parameter spaces: Unlike traditional methods that only reduce the state dimension, this framework achieves dual-dimensionality reduction. By identifying a low-dimensional active subspace within the POD coefficient space, it further compresses the input dimension of the surrogate model, thereby significantly reducing the computational costs of both model training and online prediction.
    \item Influence-based truncation for high-fidelity QoI reconstruction: The framework establishes a novel truncation criterion that prioritizes modes based on their sensitivity scores rather than modal energy. This ensures that even with only a few key POD modes, the surrogate model can capture the essential dynamics required for precise QoI reconstruction, ensuring reliable and precise surrogate modeling.
\end{enumerate}

Overall, the proposed POD--AS--PRS framework advances reduced-order modeling by identifying sensitivity-dominant parameters through the synergistic unification of modal decomposition and active subspace analysis, establishing a robust methodology that optimizes computational efficiency while ensuring high reconstruction fidelity, generalizability to unseen flow states, and structural insight.

Two numerical examples are employed to demonstrate the robustness and versatility of the framework across diverse flow regimes. To assess both the reconstruction capability and generalization capability of the surrogate models, the dataset in each case is partitioned chronologically, with the first 80\% of snapshots used for training and the remaining 20\% reserved as an independent test set. For both flow configurations, AD achieves approximately two orders of magnitude acceleration in gradient evaluation over FD while maintaining machine-precision accuracy, and the framework demonstrates high reconstruction fidelity on the training datasets and satisfactory generalizability on the independent test sets across both cases. For the cylinder flow, accurate drag and lift coefficients reconstructions were achieved using only a two-dimensional active subspace and two POD coefficients. In the airfoil flow, although 961 modes were required to capture 99\% of the system energy, the framework identifies only 22 and 32 sensitivity-dominant POD modes for $C_l$ and $C_d$, respectively, to reliably capture 99\% of their functional influence. Within a 22-dimensional ($C_l$) and 24-dimensional ($C_d$) active subspace. Notably, the analysis reveals that for this non-symmetric lifting configuration, the primary sensitivity-dominant POD modes and the resulting structural composition of the leading active subspace eigenvectors for both $C_l$ and $C_d$ exhibit a high degree of consistency. This contrasts with the symmetric cylinder case, where the two-to-one frequency relationship between $C_d$ and $C_l$ enforces a spectral separation that causes the two force coefficients to be governed by distinct modes. The elimination of this frequency constraint in the NACA~4412 lifting configuration thus allows the active subspace to identify a shared set of primary flow structures concurrently governing multiple aerodynamic performance metrics, even though $C_d$ exhibits a more distributed sensitivity profile in the higher-order modes. Furthermore, AD enables end-to-end differentiable gradient computation through the nonlinear mapping, enhancing the robustness of active subspace identification and sensitivity analysis. The results confirm that the sensitivity ranking of POD modes does not strictly follow their kinetic energy ordering, underscoring the necessity of identifying specific flow structures most relevant to the target QoI. 

In conclusion, the POD--AS--PRS framework demonstrates that shifting the reduction criterion from modal energy to QoI sensitivity ensures that the resulting ROMs capture the essential dynamics governing system performance. The present study focuses on aerodynamic force coefficients as representative performance metrics that are widely used across a broad range of engineering applications. By identifying sensitivity-dominant parameters, the framework provides an efficient, high-fidelity, and physically interpretable approach to reduced-order modeling in fluid dynamics. While the present framework is demonstrated for these critical engineering objectives, it is inherently adaptable to other QoI. In particular, higher-order metrics constitute more demanding and informative targets for sensitivity analysis, as they cannot be trivially correlated with individual POD modes. Future extensions will address higher-order flow-field metrics, compressible three-dimensional flows, and multiple parameter variations, further broadening the framework’s applicability to complex industrial design scenarios.

\backmatter

\bmhead{Acknowledgements}
The authors warmly thank the reviewers for their careful reading of the manuscript and for their constructive comments and suggestions, which have significantly contributed to improving the quality and clarity of the present work.

\bmhead{Author contribution}
Writing – original draft, D.Y.; Writing – review \& editing, D.Y., R.W., F.W., H.X.; Visualization, D.Y., H.X.; Validation, D.Y., H.X.; Software, D.Y., J.W.; Methodology, D.Y., R.W., P.L., H.X.; Data curation, D.Y.; Conceptualization, D.Y., R.W., F.W., H.X.; Supervision, R.W., H.X.; Project administration, H.X.; and funding acquisition, H.X.

\bmhead{Data \& Code availability}
Data and code are publicly available at \url{https://github.com/Dewu-Yang/POD-AS-PRS}.

\section*{Declarations}

\bmhead{Competing interests}
The authors declare that they have no financial or personal conflicts of interest that could have appeared to influence the research reported in this paper.

\begin{appendices}

\section{Derivation of gradient computation via reverse-mode AD}
\label{app:ad_derivation}

The goal is to compute the gradient $\nabla_{\mathbf{x}}f$, which represents the sensitivities of the QoI with respect to the POD coefficients. Since $f$ is a one-dimensional objective function, its Jacobian $\mathcal{J}$ reduces to a $1\times m$ row vector, equivalent to the gradient:
\begin{equation}
	\mathcal{J}_p=\frac{\partial f}{\partial x_p}, \quad p = 1, \dots, m.
\end{equation}

The function $f$ is implemented as a composition of $l$ residual blocks in the ResNet architecture:
\begin{equation}
	f=f^{l+2} \circ f^{l+1} \circ \ldots \circ f^1,\label{eq:f-l}
\end{equation}
where $f^1$ denotes the initial projection layer mapping the input to the hidden dimension, $f^{\ell}$ for $\ell=2,\dots, l+1$ represents the mapping implemented by the $({\ell}-1)$-th residual block operating on the hidden dimension, and $f^{l+2}$ is the final linear layer producing the scalar output. Each residual block incorporates two FC layers, BN, ReLU activation, and a skip connection that adds the block input to its output. 
The corresponding Jacobian matrix satisfies the chain rule:
\begin{equation}
	\mathcal{J}=\mathcal{J}_{l+2} \cdot \mathcal{J}_{l+1} \cdot \ldots \cdot \mathcal{J}_1,\label{eq:Jacobian_verifies}
\end{equation}
where $\mathcal{J}_{\ell}=\frac{\partial{f^\ell}}{\partial{f^{\ell-1}}}$ is the Jacobian of the $\ell$-th operation, evaluated with respect to its input. For a composite function $f(\mathbf{x})=h\circ g(\mathbf{x})=h(g(\mathbf{x}))$, the Jacobian is given by:
\begin{equation}
	\mathcal{J}=\mathcal{J}_{h \circ g}=\mathcal{J}_h(g(\mathbf{x})) \cdot \mathcal{J}_g(\mathbf{x}),
\end{equation}
where elements:
\begin{equation}
	\mathcal{J}_{p}=\frac{\partial f}{\partial x_p}=\sum_{r=1}^q \frac{\partial h}{\partial g_{r}} \frac{\partial g_{r}}{\partial x_p},
\end{equation}
where $q$ is the dimension of the intermediate output $g(\mathbf{x})$, and the summation reflects the chain rule applied across the composite mapping.

In reverse-mode AD, the computational graph $\mathcal{G}$ represents the ResNet architecture as a directed acyclic graph (DAG), where nodes correspond to intermediate variables and edges denote functional dependencies. The graph begins with the input POD coefficients $\mathbf{x} \in \mathbb{R}^m$, progresses through the initial projection layer, $l$ residual blocks, and the final linear layer, culminating in the scalar QoI output $f(\mathbf{x})$.

To compute the gradient $\nabla_{\mathbf{x}} f$, reverse-mode AD initializes an adjoint at the scalar output with $\bar{f} = 1$ (where the overbar denotes adjoints) and propagates it backward. The process recursively computes adjoints for each intermediate variable by applying the transpose of the local Jacobians, accumulating sensitivities via the chain rule in reverse.

Let $\mathbf{v}_{\ell}$ denotes the output of the $\ell$-th operation, with $\mathbf{v}_0=\mathbf{x}$ and $\mathbf{v}_{l+2}=f(\mathbf{x})$. The initial projection computes $\mathbf{v}_1=f^1(\mathbf{v}_0)$, each residual block computes $\mathbf{v}_\ell=f^{\ell}(\mathbf{v}_{\ell-1})=\mathcal{F}^{\ell-1}(\mathbf{v}_{\ell-1})+\mathbf{v}_{\ell-1}$, where $\mathcal{F}^{\ell-1}$ represents the residual mapping, and the final layer computes $\mathbf{v}_{l+2}=f^{l+2}(\mathbf{v}_{l+1})$. The local Jacobian for the initial and final layers is the corresponding derivative matrices, while for each residual block: 
\begin{eqnarray}
	\mathcal{J}_{\ell}=\frac{\partial{\mathbf{v}_\ell}}{\partial{\mathbf{v}_{\ell-1}}}=\frac{\partial\mathcal{F}^{\ell-1}}{\partial\mathbf{v}_{\ell-1}}+\mathbb{I}.
\end{eqnarray}
Starting from the output adjoint $ \bar{\mathbf{v}}_{l+2} = \bar{f} = 1$ (treated as a one-dimensional vector for consistency), the adjoints are propagated backward as:
\begin{equation}
	\overline{\mathbf{v}}_{\ell-1}=\mathcal{J}_{\ell}^T \cdot \overline{\mathbf{v}}_{\ell}, \quad \ell=l+2, l+1, \ldots, 1 .
\end{equation}

This recursion yields $ \bar{\mathbf{v}}_0 = \nabla_{\mathbf{x}} f$, the gradient of the QoI with respect to the POD coefficients.

Within each residual block, the residual mapping is
\begin{equation}
	\mathcal{F}(\mathbf{h})=\sigma\left(\mathrm{BN}_2\left(\mathcal{W}_2 \cdot \sigma\left(\mathrm{BN}_1\left(\mathcal{W}_1 \cdot \mathbf{h}\right)\right)\right)\right),
\end{equation}
where $ \mathbf{h} = \mathbf{v}_{\ell-1} $, $ \mathcal{W}_1 $ and $ \mathcal{W}_2 $ are weight matrices of the FC layers, $ \mathrm{BN}_i $ denotes batch normalization layers, and $ \sigma $ is the ReLU linear activation function, the adjoint propagation bifurcates across the skip connection and the residual path. Specifically, the adjoint for the block input is given by:
\begin{equation}
	\overline{\mathbf{h}}=\overline{\mathbf{v}}_{\ell}+\left(\frac{\partial \mathcal{F}}{\partial \mathbf{h}}\right)^T \cdot \overline{\mathbf{v}}_{\ell},
\end{equation}
where the first term arises from the identity skip (direct passthrough of sensitivities), and the second term originates from the residual mapping layers. The adjoints subsequently propagate backward through the individual operations in $\mathcal{F}$. Specifically, within a linear layer inside $\mathcal{F}$, let $\textbf{u}$ denote the input to the linear layer, which is the output of the previous sub-layer within $\mathcal{F}$ (e.g., a ReLU or BN layer). And $\textbf{r} = \mathcal{W}\cdot\mathbf{u}$ denotes the output of this linear layer. The corresponding adjoint is then computed as:
\begin{equation}
	\bar{\mathbf{u}} = \mathcal{W}^T \cdot \bar{\mathbf{r}}.
\end{equation}
This procedure continues recursively through all layers of $\mathcal{F}$, involving the computation of derivatives for the ReLU activation (a diagonal matrix with entries of 1 where the input is positive and 0 otherwise), batch normalization (incorporating adjustments for mean and variance normalization along with learnable parameters). This propagation persists until the adjoints reach the input $\mathbf{x}$, thereby yielding the gradient $\nabla_{\mathbf{x}} f$.

The computational complexity of reverse-mode AD satisfies $\mathrm{OPS}(f, \nabla_{\mathbf{x}} f) \leq 4 \times \mathrm{OPS}(f)$,  rendering reverse-mode AD particularly efficient for scalar-valued outputs~\cite{griewank2008evaluating}, where OPS is fused multiply-add (FMA), serving as a metric for computational complexity. In contrast, forward-mode AD requires $m$ forward passes for an $m$-dimensional input, rendering it inefficient for large $m$.

\section{POD Modal Structures and Temporal Evolution}
\label{app:pod_figures}
\clearpage
\begin{sidewaysfigure}
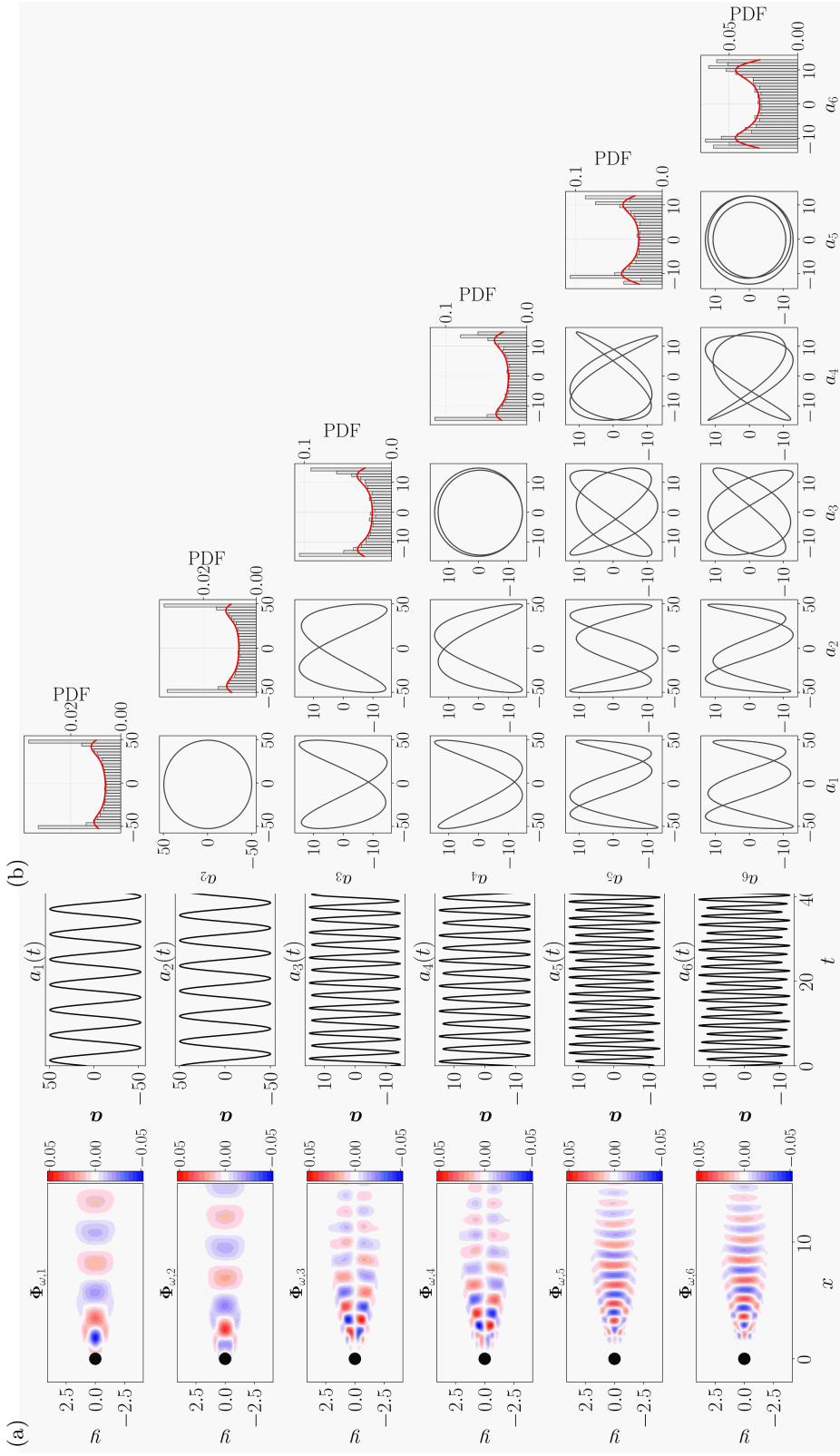

    \vspace*{0cm}
	\begin{subfigure}[t]{0.4\textwidth} 
    \centering
        \begin{overpic}[height=12cm]{figures/Figure_B22a.jpg}
            \put(2,340){\small{(a)}} 
        \end{overpic}
		\phantomcaption
		\label{subfig:case1_pod_spatial} 
	\end{subfigure}
	\hfill
	\begin{subfigure}[t]{0.58\textwidth}
    \centering
        \begin{overpic}[height=12cm]{figures/Figure_B22b.jpg}
            \put(2,340){\small{(b)}} 
        \end{overpic}
		\phantomcaption
		\label{subfig:case1_pod_phase} 
	\end{subfigure}
	\caption{(a) The first 6 POD spatial modes (left) and their corresponding temporal coefficients (right). (b) Phase-space trajectories between pairs of the first 6 POD coefficients (upper triangle) and their PDFs (diagonal), with red curves representing KDE-smoothed distributions and bars indicating histograms, highlight the dominant spatiotemporal features of the flow for test case~1.}
	\label{fig:case1_pod}
\end{sidewaysfigure}

\clearpage

\begin{sidewaysfigure}
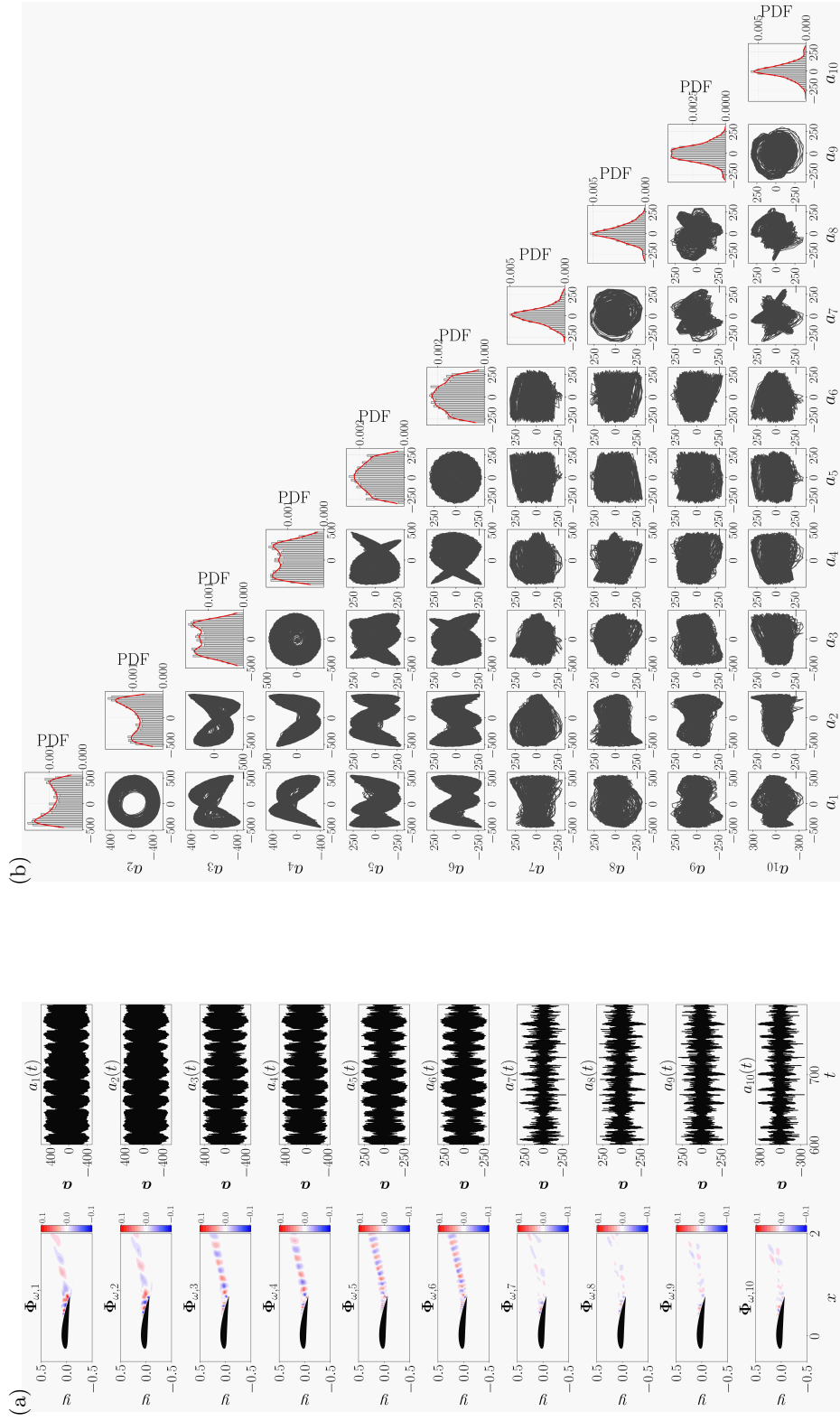

    \vspace*{0cm}
	\begin{subfigure}[t]{0.3\textwidth} 
		\centering
		\begin{overpic}[height=12cm]{figures/Figure_B23a.jpg}
			\put(-2,340){\small (a)} 
		\end{overpic}
		\phantomcaption
		\label{subfig:case2_pod_spatial} 
	\end{subfigure}
	\hfill
	\begin{subfigure}[t]{0.6\textwidth}
		\centering
		\begin{overpic}[height=12cm]{figures/Figure_B23b.jpg}
			\put(-2,340){\small (b)}
		\end{overpic}
		\phantomcaption
		\label{subfig:case2_pod_phase} 
	\end{subfigure}
	\caption{(a)  The first 10 POD spatial modes (left) and their corresponding temporal coefficients (right). (b) Phase-space trajectories between pairs of the first 10 POD coefficients (upper triangle) and their PDFs (diagonal), with red curves representing KDE-smoothed distributions and bars indicating histograms, highlight the dominant spatiotemporal features of the flow for test case~2.}
	\label{fig:case2_pod}
\end{sidewaysfigure}

\clearpage

\section{Comparison of AS-based Sensitivity with Simple Correlation Metrics}
\label{app:correlation_metrics}

This appendix provides a systematic comparison of the mode sensitivity rankings produced by three methods, including the AS activity score $I_i$, the Pearson correlation coefficient $|r_i|$, and the standardized regression coefficient $|\beta_i|$. The comparison is performed for all four QoIs considered in this work, namely the lift and drag coefficients ($C_l$ and $C_d$) for both the circular cylinder and the NACA 4412 airfoil, with the aim of demonstrating that, as flow nonlinearity increases, the AS activity scores progressively reveal inter-mode coupling contributions that are structurally inaccessible to any linear projection-based metric.
 
\paragraph{Pearson correlation coefficient.}
The Pearson correlation coefficient~\cite{pearson1895vii} between each POD coefficient $a_i$ and the scalar QoI $f$ is defined as
\begin{equation}
  r_i = \frac{\mathrm{cov}(a_i,\,f)}{\sigma_{a_i}\,\sigma_{f}}
      = \frac{\displaystyle\sum_{k=1}^{N}
        \bigl(a_i^{(k)}-\bar{a}_i\bigr)\bigl(f^{(k)}-\bar{f}\bigr)}
       {\displaystyle
        \sqrt{\sum_{k=1}^{N}\bigl(a_i^{(k)}-\bar{a}_i\bigr)^{2}}\;
        \sqrt{\sum_{k=1}^{N}\bigl(f^{(k)}-\bar{f}\bigr)^{2}}},
  \label{eq:pearson}
\end{equation}
where $\mathrm{cov}()$ in the numerator indicates the covariance between both variables, while the sigmas in the denominator indicate the standard deviation of the variable. Here, $N$ is the number of snapshots, and $\bar{a}_i$, $\bar{f}$ denote the sample means of the $i$-th POD coefficient and the QoI, respectively. 
 
\paragraph{Standardized regression coefficient.}
The standardized regression coefficient $\beta_i$ is computed as~\citep{constantine2017global,saltelli2008global}
\begin{equation}
  \beta_i = \frac{b_i}{\sqrt{3}\;\hat{\sigma}_{f}},
  \label{eq:beta}
\end{equation}
where $b_i$ is the $i$-th coefficient of the least-squares linear fit $f \approx b_0 + \mathbf{b}^T\mathbf{a}$, and the factor $\sqrt{3}$ arises from scaling by the standard deviation of a uniformly distributed variable on $[-1,1]$.  As noted by~\citet{constantine2017global},  since this coefficient is reliable only when $f$ is smooth and monotonic along each input direction, its sensitivity to nonlinear contributions is structurally limited.

\begin{figure}[h]
\centering
\includegraphics[width=0.7\textwidth]{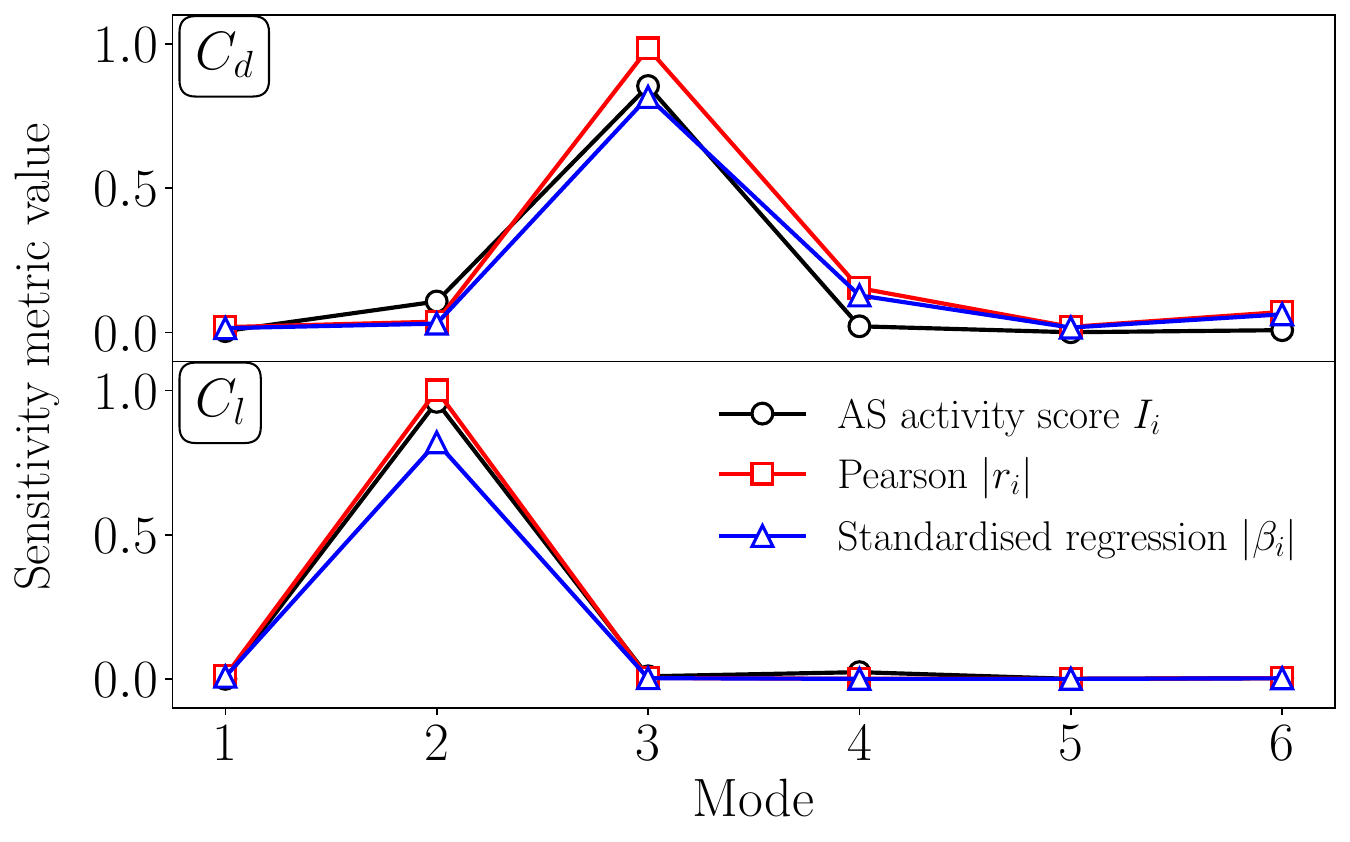}
\caption{Comparison of sensitivity rankings for the first six POD modes of the cylinder, produced by three methods: AS activity score $I_i$, Pearson correlation coefficient $|r_i|$, and standardized regression coefficient $|\beta_i|$.}
\label{fig:sens_cyl}
\end{figure}

Among the first six POD modes, the sensitivity rankings for the circular cylinder $C_l$ and $C_d$ are presented in Tables~\ref{tab:cyl_cl} and~\ref{tab:cyl_cd}, respectively, and illustrated in Fig.~\ref{fig:sens_cyl}. For the circular cylinder $C_l$, the AS rankings are broadly consistent with
both linear methods. Mode~2 dominates all three metrics ($I_2 = 0.960$, $|r_2| = 0.9999$, $|\beta_2| = 0.820$). The only notable discrepancy concerns Mode 4, which is ranked 2nd by the AS activity score ($I_4=0.024$) but only 5th by both $|r_4|=0.001$ and $|\beta_4|=0.001$. This is consistent with the observation of Constantine \& Diaz~\cite{constantine2017global} that different sensitivity metrics measure different characteristics of the function, and with the noted limitation that the standardized regression coefficient $\beta_i$ is only valid when $f$ is smooth and monotonic along each component. In the present aerodynamic problem, Mode 2 alone accounts for over 96\% of the QoI variance, effectively suppressing the detectable linear signal of the remaining modes below the reliability threshold of both the Pearson correlation coefficient and the standardized regression coefficients. The AS activity scores, being derived from the gradient covariance matrix rather than linear projections, remain sensitive to the nonlinear and coupled contributions of Mode 4 that are invisible to both linear methods in this regime. For the circular cylinder $C_d$, however, the divergence between AS and the linear methods increases substantially.  Mode~3 dominates all three metrics ($I_3 = 0.854$, $|r_3| = 0.984$, $|\beta_3| = 0.816$). Mode~2 is ranked 2nd by AS ($I_2 = 0.108$) but only 4th by both $|r_2|=0.038$ and $|\beta_2|=0.031$, Mode~4 is ranked 3rd by AS ($I_4 = 0.022$) yet 2nd by both $|r_4|=0.154$ and $|\beta_4|=0.129$. This increased divergence relative to the $C_l$ case is physically consistent with the fact that the drag coefficient of a symmetric bluff body oscillates at \emph{twice} the vortex shedding frequency ~\citep{marzouk2026extended,zheng2008frequency}: the two-to-one frequency of $C_d$ relative to $C_l$ introduces nonlinear inter-mode interactions that the gradient-based AS detects, but that linear projection metrics are structurally unable to resolve. 

\begin{table*}[h]
\centering
\small
\caption{Sensitivity rankings for the first six POD modes with respect to the circular cylinder lift coefficient $C_l$.}
\label{tab:cyl_cl}
\begin{tabular}{l
                S[table-format=1.6] c
                S[table-format=1.6] c
                S[table-format=1.6] c}
\toprule
\multirow{2}{*}{Mode}
  & \multicolumn{2}{c}{AS activity score $I_i$}
  & \multicolumn{2}{c}{Pearson $|r_i|$}
  & \multicolumn{2}{c}{Std.\ regression $|\beta_i|$} \\
\cmidrule(lr){2-3}\cmidrule(lr){4-5}\cmidrule(lr){6-7}
  & {Value} & Rank & {Value} & Rank & {Value} & Rank \\
\midrule
Mode 1 & 0.001023 & 5 & 0.010496 & 2 & 0.008534 & 2 \\
Mode 2 & 0.960367 & 1 & 0.999931 & 1 & 0.819924 & 1 \\
Mode 3 & 0.009782 & 3 & 0.003904 & 3 & 0.003237 & 3 \\
Mode 4 & 0.024136 & 2 & 0.001152 & 5 & 0.000964 & 5 \\
Mode 5 & 0.000857 & 6 & 0.000974 & 6 & 0.000853 & 6 \\
Mode 6 & 0.003753 & 4 & 0.003182 & 4 & 0.002819 & 4 \\
\bottomrule
\end{tabular}
\end{table*}

\begin{table*}[h]
\centering
\small
\caption{Sensitivity rankings for the first six POD modes with respect to the circular cylinder drag coefficient $C_d$.}
\label{tab:cyl_cd}
\begin{tabular}{l
                S[table-format=1.6] c
                S[table-format=1.6] c
                S[table-format=1.6] c}
\toprule
\multirow{2}{*}{Mode}
  & \multicolumn{2}{c}{AS activity score $I_i$}
  & \multicolumn{2}{c}{Pearson $|r_i|$}
  & \multicolumn{2}{c}{Std.\ regression $|\beta_i|$} \\
\cmidrule(lr){2-3}\cmidrule(lr){4-5}\cmidrule(lr){6-7}
  & {Value} & Rank & {Value} & Rank & {Value} & Rank \\
\midrule
Mode 1 & 0.005220 & 5 & 0.019005 & 6 & 0.015452 & 6 \\
Mode 2 & 0.107782 & 2 & 0.037990 & 4 & 0.031151 & 4 \\
Mode 3 & 0.854081 & 1 & 0.984342 & 1 & 0.816161 & 1 \\
Mode 4 & 0.021915 & 3 & 0.154046 & 2 & 0.128930 & 2 \\
Mode 5 & 0.001341 & 6 & 0.019961 & 5 & 0.017482 & 5 \\
Mode 6 & 0.008544 & 4 & 0.071556 & 3 & 0.063374 & 3 \\
\bottomrule
\end{tabular}
\end{table*}

\begin{figure}[h]
\centering
\includegraphics[width=0.7\textwidth]{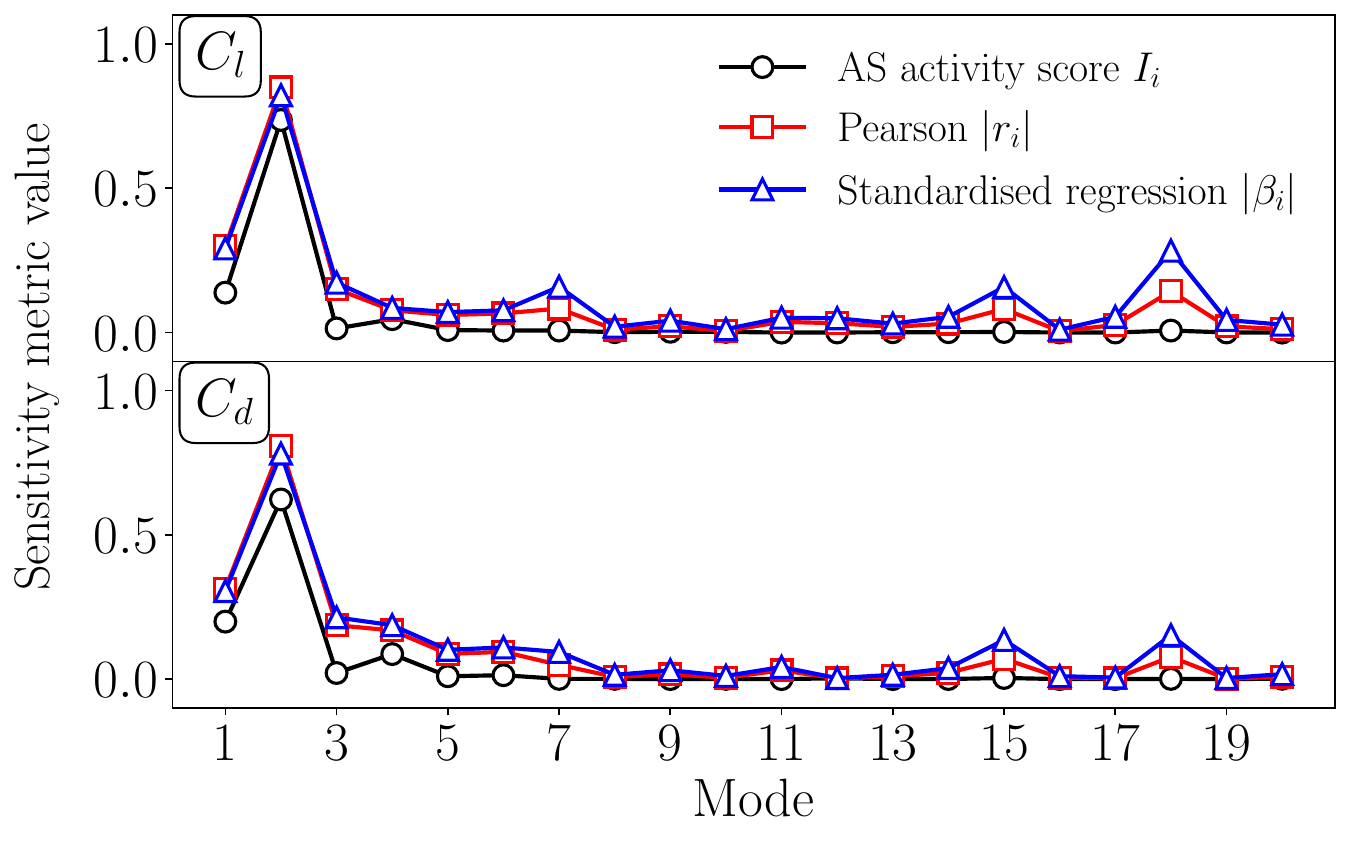}
\caption{Comparison of sensitivity rankings for the first 20 POD modes of the NACA~4412 airfoil.}
\label{fig:sens_naca}
\end{figure}

This divergence is further amplified for the NACA 4412 airfoil, where the inherent aerodynamic nonlinearities are more significant than in the canonical cylinder flow. The rankings for the airfoil $C_l$ and $C_d$ are presented in Tables~\ref{tab:naca_cl} and~\ref{tab:naca_cd} and illustrated in Fig.~\ref{fig:sens_naca}. For the airfoil $C_l$, Mode~2 and Mode~1 retain a consistent top-two ranking across all three metrics ($I_2 = 0.736$, $|r_2| = 0.849$, $|\beta_2| = 0.822$; $I_1 = 0.138$, $|r_1| = 0.299$, $|\beta_1| = 0.292$). Mode~4 is ranked 3rd by AS ($I_4 = 0.047$) yet only 10th by $|r_4| = 0.077$ and 15th by $|\beta_4| = 0.085$; Mode~5 and Mode~6, ranked 5th ($I_5 = 0.009$) and 8th ($I_6 = 0.007$) by AS, fall to 13th/18th and 12th/17th by the two linear methods, respectively. The Mode~18 presents the converse failure: it receives rank 5th under $|r_{18}| = 0.144$ and rank 3rd under $|\beta_{18}| = 0.284$, yet is placed only 6th by AS ($I_{18} = 0.007$). The linear metrics amplify the strong marginal projection of Mode~18 onto $C_l$ without accounting for the moderating effect of inter-mode coupling. In contrast, the AS, evaluated through the full gradient covariance matrix, correctly weights this contribution within the global nonlinear response. For the airfoil $C_d$, an analogous pattern is observed, with the divergence between AS and the linear metrics more pronounced still. Mode~2 is identified as the most dominant mode across all three metrics ($I_2 = 0.622$, $|r_2| = 0.807$, $|\beta_2| = 0.781$). Mode~1, ranked 2nd by both AS ($I_1 = 0.199$) and Pearson ($|r_1| = 0.313$), already falls to 3rd under the standardized regression coefficient ($|\beta_1| = 0.305$), indicating that the discrepancy between AS and the linear metrics emerges at the very head of the sensitivity spectrum and diverges further for the sub-dominant modes. Mode~4 is ranked 3rd by AS ($I_4 = 0.087$) but only 4th by $|r_4| = 0.169$ and 6th by $|\beta_4| = 0.187$; Mode~6, ranked 5th by AS ($I_6 = 0.014$), falls to 7th under $|r_6| = 0.095$ and 14th under $|\beta_6| = 0.110$. The bidirectional nature of the failure is most starkly illustrated by Modes~12 and~18. Mode~12, ranked 10th by AS ($I_{12} = 0.003$) is assigned rank 138th by $|r_{12}| = 0.002$ and rank 142nd by $|\beta_{12}| = 0.004$, rendering it effectively negligible to both linear methods. Mode~18 exhibits the opposite behaviour: ranked 10th by $|r_{18}| = 0.077$ and 8th by $|\beta_{18}| = 0.152$, it is placed only 18th by AS ($I_{18} = 0.001$). This confirms that its high linear sensitivity is primarily an apparent correlation, one that fades once inter-mode coupling is properly resolved. These results reveal that linear metrics prioritize modes with strong marginal projections but systematically miss those whose influence stems from nonlinear coupling. This discrepancy underscores the fundamental insufficiency of univariate correlation- and regression-based metrics for nonlinear aerodynamic sensitivity analysis and provides direct empirical support for adopting the AS framework in the present work.

\begingroup
\small
\setlength{\tabcolsep}{5pt}

\begin{longtable}{l S[table-format=1.6] c S[table-format=1.6] c S[table-format=1.6] c}

    \caption{Sensitivity rankings for the first 20 POD modes with respect to the NACA 4412 airfoil lift coefficient $C_l$.} \label{tab:naca_cl} \\
    \toprule
    \multirow{2}{*}{Mode} & \multicolumn{2}{c}{AS activity score $I_i$} & \multicolumn{2}{c}{Pearson $|r_i|$} & \multicolumn{2}{c}{Std.\ regression $|\beta_i|$} \\
    \cmidrule(lr){2-3} \cmidrule(lr){4-5} \cmidrule(lr){6-7}
    & {Value} & {Rank} & {Value} & {Rank} & {Value} & {Rank} \\
    \midrule
    \endfirsthead 
    \endhead

    \midrule
    \multicolumn{7}{r}{\textit{Continued on next page}} \\
    \endfoot

    \bottomrule
    \endlastfoot

    Mode 1  & 0.138442 & 2  & 0.299396 & 2   & 0.291676 & 2   \\
    Mode 2  & 0.736419 & 1  & 0.849197 & 1   & 0.821857 & 1   \\
    Mode 3  & 0.014488 & 4  & 0.150959 & 3   & 0.172917 & 8   \\
    Mode 4  & 0.046527 & 3  & 0.077257 & 10  & 0.085499 & 15  \\
    Mode 5  & 0.008838 & 5  & 0.061079 & 13  & 0.071060 & 18  \\
    Mode 6  & 0.007139 & 8  & 0.066518 & 12  & 0.077143 & 17  \\
    Mode 7  & 0.007164 & 7  & 0.083446 & 8   & 0.158921 & 9   \\
    Mode 8  & 0.001564 & 14 & 0.010612 & 80  & 0.020397 & 87  \\
    Mode 9  & 0.002891 & 10 & 0.021894 & 37  & 0.041348 & 42  \\
    Mode 10 & 0.002586 & 12 & 0.005540 & 107 & 0.011911 & 112 \\
    Mode 11 & 0.000156 & 38 & 0.037284 & 19  & 0.051498 & 33  \\
    Mode 12 & 0.000439 & 23 & 0.032504 & 21  & 0.049708 & 34  \\
    Mode 13 & 0.001290 & 17 & 0.019941 & 41  & 0.031201 & 57  \\
    Mode 14 & 0.001442 & 15 & 0.031389 & 24  & 0.054672 & 31  \\
    Mode 15 & 0.002226 & 13 & 0.081998 & 9   & 0.157721 & 10  \\
    Mode 16 & 0.000370 & 26 & 0.005142 & 112 & 0.010125 & 117 \\
    Mode 17 & 0.000349 & 27 & 0.026457 & 29  & 0.054821 & 30  \\
    Mode 18 & 0.007166 & 6  & 0.143640 & 5   & 0.283847 & 3   \\
    Mode 19 & 0.000607 & 22 & 0.021540 & 38  & 0.043933 & 40  \\
    Mode 20 & 0.000150 & 40 & 0.012167 & 72  & 0.027411 & 68  \\
\end{longtable}

\vspace{1cm}

\begin{longtable}{l S[table-format=1.6] c S[table-format=1.6] c S[table-format=1.6] c}
    \caption{Sensitivity rankings for the first 20 POD modes with respect to the NACA 4412 airfoil drag coefficient $C_d$.} \label{tab:naca_cd} \\
    \toprule
    \multirow{2}{*}{Mode} & \multicolumn{2}{c}{AS activity score $I_i$} & \multicolumn{2}{c}{Pearson $|r_i|$} & \multicolumn{2}{c}{Std.\ regression $|\beta_i|$} \\
    \cmidrule(lr){2-3} \cmidrule(lr){4-5} \cmidrule(lr){6-7}
    & {Value} & {Rank} & {Value} & {Rank} & {Value} & {Rank} \\
    \midrule
    \endfirsthead

    \endhead 

    \midrule
    \multicolumn{7}{r}{\textit{Continued on next page}} \\
    \endfoot

    \bottomrule
    \endlastfoot

    Mode 1  & 0.199137 & 2  & 0.313147 & 2   & 0.305072 & 3   \\
    Mode 2  & 0.622167 & 1  & 0.806902 & 1   & 0.780924 & 1   \\
    Mode 3  & 0.021194 & 4  & 0.186637 & 3   & 0.213785 & 5   \\
    Mode 4  & 0.086762 & 3  & 0.168659 & 4   & 0.186651 & 6   \\
    Mode 5  & 0.009953 & 7  & 0.087295 & 8   & 0.101560 & 16  \\
    Mode 6  & 0.013627 & 5  & 0.094603 & 7   & 0.109714 & 14  \\
    Mode 7  & 0.000807 & 19 & 0.049360 & 15  & 0.094005 & 19  \\
    Mode 8  & 0.000907 & 17 & 0.007711 & 104 & 0.014821 & 111 \\
    Mode 9  & 0.000247 & 35 & 0.016241 & 65  & 0.030672 & 77  \\
    Mode 10 & 0.000693 & 21 & 0.005124 & 119 & 0.011015 & 119 \\
    Mode 11 & 0.000101 & 68 & 0.030363 & 33  & 0.041938 & 59  \\
    Mode 12 & 0.002894 & 10 & 0.002289 & 138 & 0.003500 & 142 \\
    Mode 13 & 0.000099 & 70 & 0.009596 & 94  & 0.015015 & 108 \\
    Mode 14 & 0.000348 & 30 & 0.021884 & 48  & 0.038116 & 63  \\
    Mode 15 & 0.004025 & 8  & 0.070435 & 11  & 0.135481 & 11  \\
    Mode 16 & 0.000090 & 73 & 0.005186 & 118 & 0.010211 & 121 \\
    Mode 17 & 0.000090 & 74 & 0.002579 & 135 & 0.005344 & 136 \\
    Mode 18 & 0.000877 & 18 & 0.077144 & 10  & 0.152445 & 8   \\
    Mode 19 & 0.000350 & 28 & 0.001820 & 142 & 0.003712 & 141 \\
    Mode 20 & 0.001100 & 14 & 0.007634 & 105 & 0.017199 & 104 \\
\end{longtable}
\endgroup

\end{appendices}

\bibliography{hyrom}

\end{document}